\documentclass[aps,onecolumn,nofootinbib,nobibnotes,superscriptaddress,preprintnumbers]{revtex4}
\pdfoutput=1

\usepackage{amsmath,amssymb,graphicx,color} 
\usepackage{epsf}
\usepackage[letterpaper, total={6.5in, 9.5in}]{geometry}

\usepackage[utf8]{inputenc}
\usepackage{youngtab}
\usepackage{multirow}
\usepackage{dcolumn}
\usepackage{bm}
\usepackage{slashed}
\usepackage{hyperref}
\usepackage{dsfont}
\usepackage{booktabs}
\usepackage{cleveref}
\usepackage{ulem}
\usepackage{titlesec}
\usepackage{cleveref}

\newcommand{\avg}[1]{\left\langle#1\right\rangle}

\def\beq{\begin{equation}}
\def\eeq#1{\label{#1}\end{equation}}
\def\eeqn{\end{equation}}

\def\beqna{\begin{eqnarray}}
\def\eeqna#1{\label{#1}\end{eqnarray}}
\def\eeqnan{\end{eqnarray}}

\newcommand{\GeV}{\,\mathrm{GeV}}

\begin{document}

\preprint{KIAS-P17049}

\title{On the Validity of the Effective Potential and the Precision of Higgs Self Couplings}

\author{Bithika Jain}
\affiliation{ICTP South American Institute for Fundamental
	Research \& Instituto de F\'isica Te\'orica \\ 
	Universidade Estadual  Paulista, S\~ao Paulo, Brazil}
\affiliation{School of Physics, Korea Institute for Advanced Study, Seoul 130-722, Republic of Korea}
\author{Seung~J.~Lee}
\affiliation{Department of Physics, Korea University, Seoul 136-713, Republic of Korea}
\affiliation{School of Physics, Korea Institute for Advanced Study, Seoul 130-722, Republic of Korea}
\author{Minho Son}
\affiliation{Department of Physics, Korea Advanced Institute of Science and Technology,
	291 Daehak-ro, Yuseong-gu, Daejeon 34141, Republic of Korea}



\begin{abstract}
The global picture of the Higgs potential in the bottom-up approach is still unknown. A large deviation as big as $\mathcal{O}(1)$ fluctuations of the Higgs self couplings is still a viable option for the New Physics. An interesting New Physics scenario which can be linked to a large Higgs self coupling is the baryogenesis based on the strong first order phase transition. We revisit the strong first order phase transition in two classes of Beyond the Standard Models, namely the Higgs portal with the singlet scalar under the Standard Model gauge group with  $\mathds{Z}_2$ symmetry and the effective field theory approach with higher-dimensional operators. We numerically investigate a few important issues in the validity of the effective potential, caused by the breakdown of the high-temperature approximation, and in the criteria for the strong first order phase transition. We illustrate that these issues can lead to $\mathcal{O}(1)$ uncertainties in the precision of the Higgs self couplings, which are relevant when discussing sensitivity limits of different future colliders. 
We also find that the quartic coupling of the above two classes of scenarios compatible with the strong first order electroweak phase transition where the cubic coupling is not negligible, can achieve a $2\sigma$ sensitivity at the 100 TeV $pp$-collider. From this novel observation, we show that the correlation between the Higgs cubic coupling and the quartic coupling will be useful for differentiating various underlying New Physics scenarios and discuss its prospect for the future colliders. Throughout our numerical investigation, the contribution from Goldstone boson is not included.
\end{abstract}

\maketitle


\section{Introduction}

The baryon asymmetry of the universe remains a challenging mystery.  An explanation of the baryon asymmetry based on the strong first order electroweak phase transition (SFOEPT) of the Higgs potential is a commonly explored option. A motivation for pursuing this idea is partly related to the fact that a large deviation of the Higgs self coupling with respect to the Standard Model (SM) value is still phenomenologically allowed~\cite{Aad:2015xja,Aad:2015uka,Khachatryan:2016sey}. A reconstruction of the global picture of the Higgs potential via the Higgs self coupling measurement is of utmost importance for a better understanding of the nature of the electroweak phase transition as well as the dynamics of the electroweak symmetry breaking (EWSB).

The electroweak baryogenesis (EWBG) based on the SFOEPT is an attractive way to explain the baryon asymmetry~\cite{Kuzmin:1985mm}. Today, electroweak symmetry is broken but in the early universe it was electroweak symmetric~\cite{Kirzhnits:1972iw,Kirzhnits:1972ut,Dolan:1973qd,Weinberg:1974hy}. About $10^{-11}$ seconds after the Big Bang, the universe undergoes a phase transition from the electroweak symmetric (unbroken) phase to asymmetric (broken) phase.  This leads to the formation and expansion of bubbles of the true vacuum configuration into the false one if the phase transition is the first order~\cite{Kirzhnits:1972iw,Kirzhnits:1972ut}. If there exists CP violation, particle interactions with expanding bubble may create a baryon asymmetry in the vicinity of the bubble via baryon number violating process induced by the sphaleron. The generated baryon asymmetry near the bubble will enter into the bubble as it expands.  However, the baryon asymmetry would have been washed out if the sphaleron process inside the bubbles is not sufficiently suppressed. This requires that the phase transition needs to be strongly first order for the successful EWBG based on the SFOEPT. While a realistic EWBG scenario requires a large enough CP violation as one of the Sakharov conditions for the EWBG to be realized~\cite{Sakharov:1967dj}, we will focus in this work only on the plausibility for having a strong first order electroweak phase transition in the Higgs potential~\cite{Shaposhnikov:1987tw,Shaposhnikov:1987pf,Shaposhnikov:1991cu}(for a recent review, see Ref~\cite{Morrissey:2012db}).

A tight correlation between the Higgs cubic coupling and the dynamics of electroweak phase transition has been explored in several beyond the SM (BSM) scenarios in the context of the EWBG and it has been shown that these models can be tested at future colliders via the cubic coupling measurement~\cite{Noble:2007kk,Huang:2015tdv,Katz:2014bha,Curtin:2014jma}. The effective potential at a finite temperature is the main theoretical tool to check the compatibility of BSM scenarios with the SFOEPT, but there have been some variations in the form of the effective potential which has been used in their analyses. For a successful EWBG, the phase transition must be strongly first order.  The criterion for baryon number preservation which requires the suppression of the sphaleron process is approximated by the threshold value of $v_c/T_c$. Here $v_c$ is the critical Higgs vacuum expectation value (VEV) at critical temperature, $T_c$. The uncertainty in the precision calculation of baryon number asymmetry is reflected in the range of $v_c/T_c$ $\gtrsim 0.6 - 1.4$ used in literature instead of a unique threshold value (for example, see~\cite{Patel:2011th}).  This approximate ratio has been used to claim that the phase transition is strongly first order\footnote{As will be discussed in Section~\ref{sec:SFOPTnSelfcoupling}, a similar ambiguity exists in the determination of the nucleation temperature and the Higgs VEV, denoted by $T_N$ and $v_N$, in a more sophisticated treatment of the criteria for the SFOEPT (see~\cite{Kurup:2017dzf} for a related work).}. In this work, we aim to study in detail the impact on the precision of the Higgs self couplings caused by the ambiguity in the form of the effective potential and by the finite range of $v_c/T_c$ values. We will show that the uncertainty on the Higgs self coupling due to the various ambiguities can be as big as $\mathcal{O}(1)$ and that there can also be a dramatic impact on the prospect for the future colliders. To this end, we will consider two classes of BSM scenarios that have been extensively considered in literature: a Higgs portal with a singlet scalar\footnote{Note that we are interested in exploring the case which is difficult to falsify. i.e. a nightmare scenario of a $Z_2$ symmetric singlet scalar extension of SM is of interest for us as it is the scenario of which we need to take care in order to cover the whole possible scenarios relevant for the SFOEPT.}~\cite{Espinosa:2007qk,Barger:2007im,Espinosa:2008kw,Espinosa:2011ax,Cline:2012hg,Cline:2013gha,Alanne:2014bra,Curtin:2014jma,Vaskonen:2016yiu,Marzola:2017jzl,Kurup:2017dzf} and an effective field theory (EFT) approach with higher dimension operators~\cite{Grojean:2004xa,Bodeker:2004ws,Delaunay:2007wb,Huang:2015tdv,Gan:2017mcv}. For the EFT approach, we will consider not only the case only with dimension-six operator, $|H|^6$, but also extend to the scenario where all higher dimensional operators, $|H|^{4+2n}$ ($n \geq 1$), are re-summed up to the infinite order in the Higgs field~\cite{Huang:2015tdv}\footnote{Unlike the case in~\cite{Huang:2015tdv}, we keep the universal Willson coefficient as the free parameter while taking into account the power counting in the SILH basis~\cite{Giudice:2007fh} (see Section~\ref{sec:EFT}).}. In particular, universal Wilson coefficients are assumed and the EFT description will be valid even when coefficients deviate largely, provided the energy is well below the cutoff scale.

The quartic coupling in the context of EWBG based on the SFOEPT has not been well studied. This is mainly because the production channels which directly accesses this coupling have a very small production cross-section~\cite{Maltoni:2014eza,Plehn:2005nk,Binoth:2006ym,Papaefstathiou:2015paa,Chen:2015gva,Fuks:2015hna,Contino:2016spe}. We make the novel observation that there is a large parameter space for SFOEPT in the commonly explored BSM scenarios, where the quartic coupling can deviate with respect to the SM one by a factor of $\mathcal{O}(1-10)$. We found it very illuminating to study the strength of the electroweak phase transition in the cubic versus quartic Higgs self coupling plane as it highlights the utility of the quartic coupling as a way to disentangle various BSM scenarios. In particular, if the deviation of the cubic coupling with respect to the SM one is not negligible,
then the measurement of Higgs quartic coupling is palpable in the future colliders~\cite{Papaefstathiou:2015paa,Contino:2016spe,Fuks:2017zkg}. This can serve as a discriminator among different underlying models responsible for the electroweak phase transition.

Our paper is organized in the following way. In Section~\ref{sec:finiteT:effpot}, we define two prescriptions for the effective potential at a finite temperature differ by the use of the high-temperature approximation of the thermal potential. We then briefly discuss about the thermal potential in various temperature limits. We first focus our attention to the benefit of using the series-sum of Bessel functions of the second kind for thermal potential computation. This approximation reproduces the exact evaluation for the entire temperature range when a large number of terms are included -- we choose $n=50$ in this work. In Section~\ref{sec:BS}, we introduce two types of benchmark scenarios: the Higgs portal with a singlet scalar with $\mathds{Z}_2$ symmetry and the effective field theory approach with higher-dimensional operators. In Section~\ref{sec:SFOPTnSelfcoupling}, we explore the commonly used criteria for the SFOEPT and review related issues. We scan the variables of the benchmark BSM scenarios to identify the compatible parameter space with SFOEPT. In each benchmark scenario, we examine the relationship between the cubic and quartic couplings. Finally, we present the prospects for the Higgs self couplings at the future colliders. In the course of our discussion in Section~\ref{sec:validityEffpot}, we comment on the issue regarding the validity of effective potential, mainly caused by the high-temperature approximation of the thermal potential and $v_c/T_c$ values in a finite range. We conclude in Section~\ref{sec:conclusion}. 

\section{Effective Potential in Finite Temperature}
\label{sec:finiteT:effpot}
The dynamics of the electroweak phase transition is governed by the finite-temperature effective action, $S_{eff}(T)$, where $T$ is the temperature. The $S_{eff}(T)$ reduces to an integral over the effective potential $V_{eff}(\phi_{j},T)$, which is the free energy density  for 
fields, $\phi_i$ (where $i \geq 1$ ). An one-loop effective potential, that we explore in this work, is~\cite{Gross:1980br,Parwani:1991gq,Arnold:1992rz,Carrington:1991hz} (see~\cite{Curtin:2016urg} for related discussion)
\begin{equation}
\label{eq:TFD}
V_{eff}(\phi_{i},T) \equiv V_{tree}(\phi_{i}) +V_{CW}(m^2_i(\phi) + \Pi_i) +V_{T}(m^2_i(\phi) + \Pi_i,T)~,
\end{equation}
where $\Pi_i$ is thermal masses (or Debye masses). It was pointed out in~\cite{Boyd:1992xn,Dine:1992wr} that the effective potential in Eq.~(\ref{eq:TFD}) is not theoretically consistent in that it miscounts the two-loop daisy diagrams.
The first term $V_{tree}$ is the tree-level SM Higgs potential augmented by BSM features which we will discuss in detail later. The second term $V_{CW}$ is the one-loop Coleman-Weinberg potential~\cite{Coleman:1973jx}. Using on-shell renormalization scheme in the Landau gauge, it is given by 
\begin{equation}
V_{CW}(m^2_i(\phi) + \Pi_i) =\sum_i  (-1)^{F_i}\frac{g_i}{64 \pi^2} \bigg[ m_i^4(\phi) \left ( \log \frac{m_i^2(\phi) + \Pi_i}{m_i^2(v) + \Pi_i} -\frac{3}{2} \right ) + 2 \left ( m_i^2(\phi)+ \Pi_i \right ) \left ( m_i^2(v)+ \Pi_i \right ) \bigg] \label{eq:A_VCW}~,
\end{equation}
where the sum runs over the SM particles including the Goldstone bosons and BSM particles. The degrees of freedom for each particle is $g_i$ with $F_i$ being the fermion number. The expression for $V_{T}$, the thermal potential at finite temperature,  is defined as
\begin{equation}
V_{T}(m^2_i (\phi) + \Pi_i,T) = \sum_i (-1)^{F_i}\frac{g_i T^4}{2 \pi^2} \int_0^{\infty} dx ~x^2 \log \bigg[1 \mp \exp \big(-\sqrt{x^2+ \left ( m_i^2(\phi)+ \Pi_i \right )/T^2} \big)    \bigg] \label{eq:A_VT}~,
\end{equation}
where the integral with ``$-/+$" sign denotes the thermal bosonic/fermionic function. 
The exact thermal mass is computed by solving finite-temperature gap equations properly, which is beyond the scope of this paper. We simply take the leading contribution in temperature to thermal mass (which amounts to the re-summation of daisy diagrams) in the high temperature limit. We call the recipe in Eq.~(\ref{eq:TFD}) with truncated thermal masses {\bf prescription A}.

In a fully consistent high temperature expansion, the thermal potential should also be approximated accordingly.  In this approximation, $V_{T}$ effectively splits into the one-loop thermal potential without thermal mass and the IR divergent piece, known as ring-term $V_{ring}$ which counts zero modes of the Debye masses~\cite{Fendley:1987ef,Carrington:1991hz}. The effective potential with this self-consistent high-temperature approximation will be referred as {\bf prescription B} which is defined as
\begin{equation}\label{eq:Veff:B}
V_{eff}(\phi,T) \equiv V_{tree}(\phi) +V_{CW}(m^2_i(\phi)) +V_{T}(m^2_i(\phi),T)+V_{ring}(m^2_i(\phi),T)~,
\end{equation}
where one-loop Coleman-Weinberg potential is the familiar expression
\begin{equation}
V_{CW}(m^2_i(\phi))=\sum_i  (-1)^{F_i}\frac{g_i}{64 \pi^2} \bigg[ m_i^4(\phi) \left ( \log \frac{m_i^2(\phi)}{m_i^2(v)} -\frac{3}{2} \right ) + 2 m_i^2(\phi) m_i^2(v) \bigg]~,
\label{eq:B_CW} 
\end{equation}
and the ring term is given by
\begin{equation}
V_{ring}(m^2_i(\phi),T) = \sum_i \frac{T}{12 \pi}  Tr \big[ m_i^3(\phi_i) - (m_i^2(\phi)+\Pi_i(0))^{3/2} \big]~.
\label{eq:B_ring}
\end{equation}
The thermal potential for fermions and bosons are written as
\begin{equation}
V_{T}(m^2_i(\phi),T) = \sum_i (-1)^{F_i}\frac{g_i T^4}{2 \pi^2} J_{B/F}\left(\frac{m_i^2(\phi)}{T^2}\right)~,
\end{equation}
where the loop functions, $J_{B/F}$ are given in the high temperature expansion, $\alpha = m/T \ll 1$~\cite{Dolan:1973qd},
\begin{equation}\label{eq:VT:highTapprox}
\begin{split}
 J_B (\alpha^2) = \int_0^\infty dx\, x^2\, {\rm ln} \left (1- e^{-\sqrt{x^2+\alpha^2}} \right )  \sim&
  \frac{\pi^2}{12} \alpha^2 - \frac{\pi}{6}\alpha^3-\frac{\pi^4}{45} - \frac{1}{32}\alpha^4\, {\rm ln} \left ( {\alpha^2 \over a_b} \right ) ~,
  \\[2.5pt]
 J_F (\alpha^2) = \int_0^\infty dx\, x^2\, {\rm ln} \left (1+ e^{-\sqrt{x^2+\alpha^2}} \right )  \sim&  
  -\frac{\pi^2}{24} \alpha^2 + \frac{7\pi^4}{360} - \frac{1}{32}\alpha^4\, {\rm ln} \left ( {\alpha^2 \over a_f} \right ) ~.
\end{split}
\end{equation}
where $a_b = 16 \pi^2\, {\rm exp}(3/2-2\gamma_E)$ and $a_f = \pi^2\, {\rm exp}(3/2-2\gamma_E)$\footnote{${\rm log} a_b \sim 5.4076$ and ${\rm log} a_f \sim 2.6350$.}. Note that $J_B$ from bosons has a $T$-dependent cubic term which can induce a first order phase transition via thermal effects~\cite{Anderson:1991zb}.  The sign of $\alpha^3$ term is undetermined and a flip of the sign can dramatically change conclusions.

Another variation of the effective potential takes the similar form to Eq.~(\ref{eq:Veff:B}) except that thermal potential $V_T$ is not expanded to be valid in a larger domain of $m/T$. We call it {\bf prescription C}

The low-temperature approximation~\cite{Anderson:1991zb}, namely $\alpha = m/T \gg 1$ yields the following analytical expressions:
\begin{equation}\label{eq:VT:lowTapprox}
\begin{split}
 J_B (\alpha^2) \equiv J_B (\alpha^2;\, n) &= - \sum^n_{k=1} \frac{1}{k^2} \alpha^2 K_2 (\alpha\, k)~,
 \\[2.5pt]
 J_F (\alpha^2) \equiv J_F (\alpha^2;\, n) &= - \sum^n_{k=1}\frac{(-1)^k}{k^2} \alpha^2 K_2 (\alpha\, k)~,
\end{split}
\end{equation}
where $K_2$ is the modified Bessel function of the second kind. The series in Eq.~(\ref{eq:VT:lowTapprox}) are uniformly convergent for all positive $\alpha^2$. It is possible to combine two approximations in Eqs.~(\ref{eq:VT:highTapprox}) and ~(\ref{eq:VT:lowTapprox}) to define a single piecewise function which covers the entire $m/T$ range for a relatively small $n$~\cite{Curtin:2016urg}. However, we find that the low-temperature approximation in Eq.~(\ref{eq:VT:lowTapprox}) with $n$ of order a few 10's  gives a very good agreement with the exact evaluation, even for near-zero $m/T$ values. This implies that the low-$T$ approximation with a sufficiently large $n$ can replace the exact evaluation for the entire range of the argument. An additional advantage of using low-temperature approximation with large $n$ is the better agreement with the exact method, even for a large negative $m^2/T^2$,  as compared to the piece-wise function. 
Throughout our simulation, we adopt the analytic expressions in Eq.~(\ref{eq:VT:lowTapprox}) with $n=50$ as the replacement of the exact thermal potential in the prescription A and C.\\
\begin{figure}[htp] 
	\centering
	\includegraphics[width=0.4\linewidth]{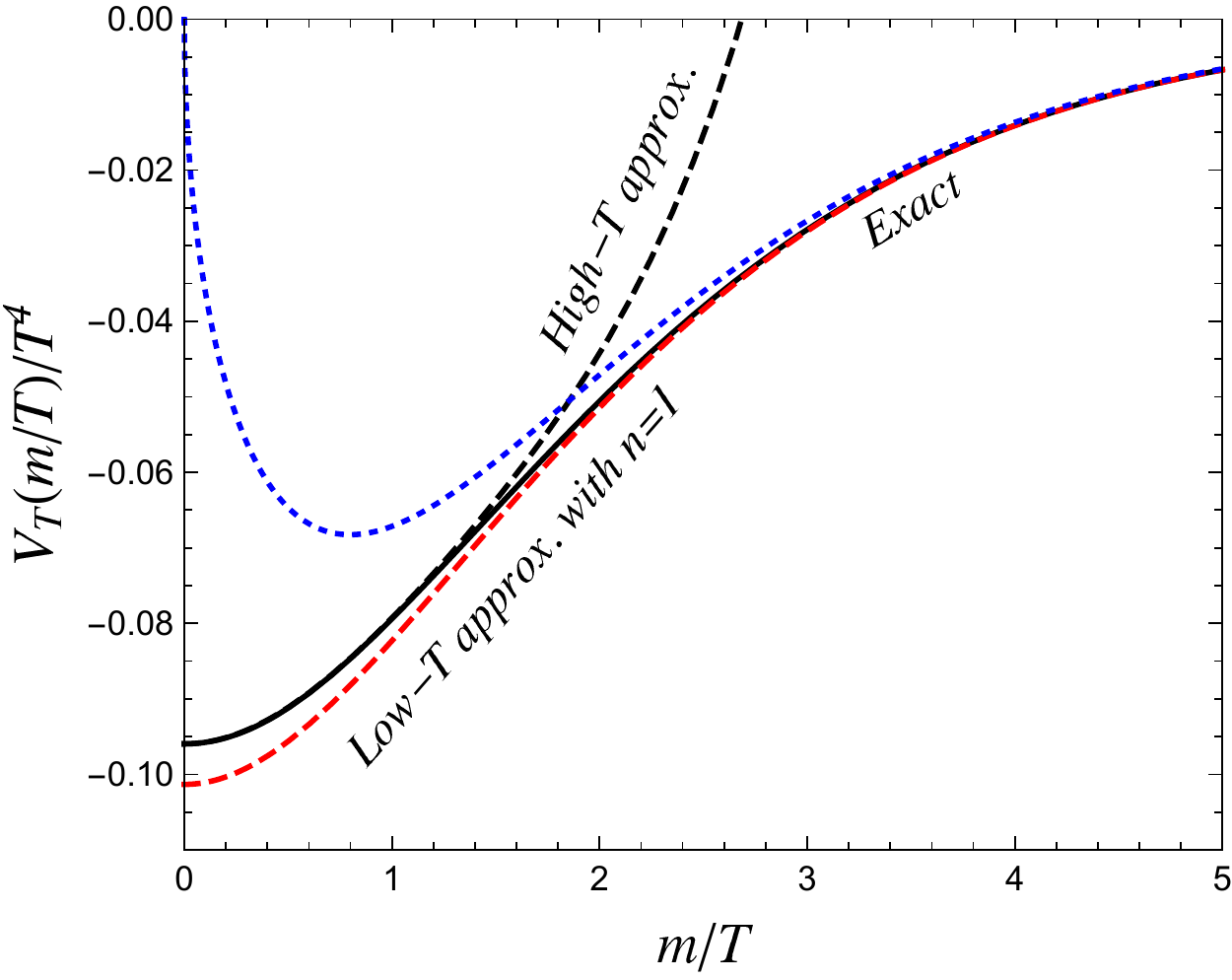} \quad
	\includegraphics[width=0.4\linewidth]{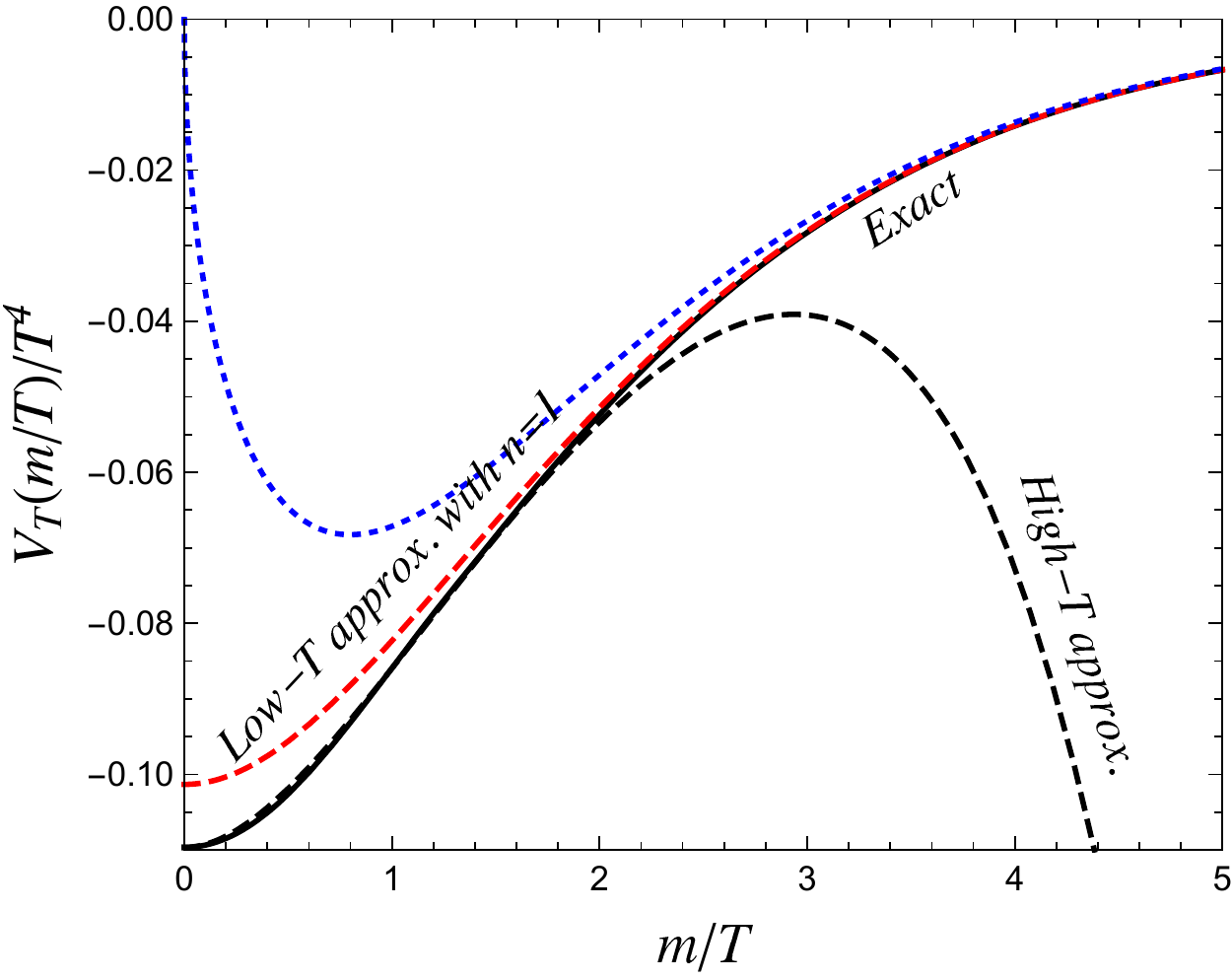}
	\caption{A contribution to the thermal potential, $V_T$, from a fermion (left) and boson (right) as a function of $m/T$, in the high-$T$ approximation (black-dashed), low-$T$ approximation in Eq.~(\ref{eq:VT:lowTapprox}) (red-dashed) with $n=1$, and in the exact form (black-solid). The dotted-blue line is the low-$T$ approximation with the approximated $K_2$ as in~\cite{Anderson:1991zb}.
}\label{fig:VTExactVSapprox}
\end{figure}
%
%
We will use the effective potential using the above mentioned three prescriptions in context of two different BSM scenarios. While these prescriptions differ in forms of the thermal potential approximations, which prescription suits the case better should depend on the typical values of $m/T$ ratios in the integrals in the domain of interest for SFOEPT. A related issue is illustrated in Fig.~\ref{fig:VTExactVSapprox} for the bosonic and fermionic thermal potentials. It is evident in Fig.~\ref{fig:VTExactVSapprox}, that the high-temperature approximation starts breaking down roughly around $m/T \gtrsim 2$ above which the low-temperature approximation with just $n =1$, matches with the exact contribution. When it comes to the SFOEPT in BSM scenarios where usually larger values of $m/T$ above the breaking point 
are likely, using high temperature approximation introduces $\mathcal{O}(1)$ uncertainties in evaluating the thermal potential (as we will show in the following sections). In this sense we see that our ``prescription A"  is more accurate as it doesn't expand the thermal potential in $m/T$ when it is not small. However, the ``prescription B" might be considered more consistent choice in terms of the consistency of the high-temperature approximation.

The gauge (in)dependence of the effective potential at finite temperature is an important issue (or source of uncertainty) which has not been firmly established (see~\cite{Patel:2011th,Garny:2012cg} for related discussion). Addressing this issue even numerically is beyond the scope of our work. While we adopt the on-shell renormalization scheme for the effective potential, the $\rm \overline{MS}$ scheme is another option. See~\cite{Curtin:2014jma} for the discussion of the scheme dependence in the Higgs portal scenario.

\section{Benchmark Scenarios}
 \label{sec:BS}
We focus on two classes of benchmark scenarios that have been extensively considered in the literature. For the first scenario, SM is extended to include a real singlet scalar with $\mathds{Z}_2$ symmetry via renormalizable interactions, and in the second, SM is extended to include higher dimensional operators. In both scenarios, the one-loop effective potentials are computed using two prescriptions introduced in Section~\ref{sec:finiteT:effpot}.

 \subsection{Higgs portal with a Singlet Scalar}
 \label{sec:singlet}
 
This scenario has been well studied in a variety of different contexts~\cite{Barger:2007im,Espinosa:2007qk,Espinosa:2011ax,Noble:2007kk,Cline:2012hg,Cline:2013gha,Alanne:2014bra,Curtin:2014jma,Chala:2016ykx,Artymowski:2016tme,Beniwal:2017eik,Chen:2017qcz}. We restrict to the case of a Higgs portal with a real singlet scalar, denoted by $S$, under the SM gauge group which respects the discrete $\mathds{Z}_2$, to avoid the mixing with Higgs field. The singlet mass is assumed to be larger than $m_h/2$ in order to evade the constraints from exotic Higgs decay searches. We take the exact $\mathds{Z}_2$ symmetric case for simplicity even though $\mathds{Z}_2$ can be softly broken in certain scenarios, allowing $S$ to decay to SM particles. Such a model provides the so called ``no-lose" theorem for testing EWBG in future colliders, as it is the most phenomenologically challenging case, i.e. it can be probed only in the future collider searches via measurements of the Higgs potential.
 Tree-level potential in the unitary gauge takes the form,
 \begin{equation}
V_{tree} =- \frac{\mu^2}{2}h^2+ \frac{\lambda}{4}h^4 +\frac{1}{2}\lambda_{HS} h^2 S^2+ \frac{1}{2}\mu_S^2 S^2 +\frac{1}{4}\lambda_S S^4~,
\label{eq:Vsinglet_tree}
\end{equation}
where $h$ is the electromagnetic neutral real component of the Higgs doublet. The EWSB occurs when $\mu^2 >0$, giving rise to EWSB minimum for the higgs, $\avg{h}=v=\mu/\sqrt{\lambda}\approx246 \GeV$. One among the three Lagrangian parameters, $\mu_S, \lambda_{HS}, \lambda_S$, related to the singlet can be traded for the physical singlet mass, $m_S$. The potential at large fields can be approximated to
 \begin{equation}
 \begin{split}
V_{tree} &\approx \frac{\lambda}{4}h^4 +\frac{1}{2}\lambda_{HS} h^2 S^2 +\frac{1}{4}\lambda_S S^4~,\\
&= \frac{1}{4} \Big [ \left (\sqrt{\lambda} h^2 - \sqrt{\lambda_S} S^2 \right )^2 + 2h^2 S^2 \left ( \lambda_{HS} + \sqrt{\lambda\, \lambda_S} \right ) \Big ]~.
\end{split}
\label{eq:Vsinglet_tree:large}
\end{equation}
The stability of the potential at large fields, or avoiding negative runaway directions in the potential, requires $\lambda>0, ~\lambda_S>0$, and $\lambda_{HS}> -\sqrt{\lambda \lambda_S}$.  

In the early universe at a very high temperature, the temperature dependent mass term dominates the effective potential, and the global minimum occurs at the electroweak symmetry preserving point, or $(\avg{h},\, \avg{S}) = (0,\, 0)$. As the universe cools down, EWSB global vacua can develop away from the symmetric point. The phase transition can proceed in two different ways in this scenario, assuming that our global vacua is always $(v,\, 0)$. First possibility is the direct transit from $(\avg{h},\, \avg{S}) = (0,\, 0)$ to the global minimum $(v,\, 0)$, called one-step phase transition. 
This occurs when $\mu_S^2>0$, and in the event $\lambda_{HS}<0$,  $\lambda_S$ gets a lower bound to prevent the negative runaways.  However, the singlet quartic, $\lambda_S$ mildly affects the effective potential as it only enters the Debye mass terms for the singlet. We parametrize one-step phase transition effectively in terms of two relevant parameters, $m_S$ and $\lambda_{HS}$, while setting $\lambda_S = 0$, in our numerical simulations. 
Alternately, it is also possible for the singlet to acquire a VEV at some point in the cosmological history while the Higgs is still in a unbroken phase, thus a local minima develops at $(\avg{h}, \avg{S})=(0,\, w)$. As the universe evolves, the EWSB finally occurs, ensuring that $(v,\, 0)$ is the global minimum. This case is called the two-step phase transition. The singlet gets a VEV when $\mu_S^2<0$. Demanding $V_{eff}(0,\, w)>V_{eff}(v,\, 0)$ at zero temperature ensures that $(v,\, 0)$ is the global vacua, and it puts a lower bound on the singlet quartic as 
\begin{equation}\label{eq:twostep:lamSmin}
\lambda_S \geq \lambda \frac{\mu_S^4}{\mu^4} \equiv \lambda_{S\, \rm min}~.
\end{equation} 
The scalar sector masses exhibit field dependence on both $h$ and $S$. The diagonalized masses of the scalar sector enter the one-loop masses,$~m_i$ where $i={h,S}$. In both possibilities of the transition history, the upper limit on  $\lambda_S$ is set by the perturbativity. 

It would be interesting to explore if an exactly $\mathds{Z}_2$-symmetric scalar singlet, $S$ in the Higgs portal can also partly account for the DM abundance~\cite{Curtin:2014jma}.

\subsection{Effective Field Theory Approach}
\label{sec:EFT}

We will consider the possibility of SFOEPT in context of the effective field theory~\cite{Noble:2007kk,Grojean:2004xa,Delaunay:2007wb,Chung:2012vg,Huang:2015tdv,Huang:2015izx}. We proceed with the assumption that Higgs belongs to the linear representation of $SU(2)$ gauge group. Furthermore, we have assumed that the Higgs is realized as a Goldstone boson. We first focus on one type of dimension-six operator $\mathcal{O}_6 \sim |H|^6$ in the SILH basis~\cite{Giudice:2007fh} with the same normalization as in Ref.~\cite{Azatov:2015oxa}
\begin{equation}\label{eq:O6}
  \Delta\mathcal{L} = \frac{c_6}{v^2}\frac{m^2_h}{2v^2} |H|^6~.
\end{equation}
The Higgs trilinear coupling, in principle, is also modified by $\mathcal{O}_H \sim (\partial_\mu |H|^2)^2$ operator which is strongly constrained by the current Higgs coupling measurements and the electroweak precision measurements~\cite{deFlorian:2016spz}. Our consideration is suitable for a New Physics scenario with $\mathcal{O}_H \ll \mathcal{O}_6$, which we will briefly discuss later in this subsection.
As the current LHC has a poor sensitivity on the Higgs self coupling, a large deviation of the $\mathcal{O}_6$ is still a viable option for the New Physics such as the SFOEPT. With $\mathcal{O}_6$ operator in Eq.~(\ref{eq:O6}) added to the SM Higgs potential, the tree-level potential in terms of $h$ is
\begin{equation}
 V_{tree} =-\frac{\mu^2}{2}h^2+ \frac{\lambda}{4}h^4 + \frac{1}{8}\frac{c_6}{v^2}\frac{m^2_h}{2v^2} h^6~.
\end{equation}
The Higgs VEV is determined by the equation,
\begin{equation}
\left .  -\mu^2 + \lambda h^2 + \frac{3c_6 m^2_h}{8 v^4} h^4 \right |_{h=v} = 0 ~.
\end{equation}
The physical Higgs mass is obtained by $d^2V_{tree}(h)/dh^2|_{h=v}$,
\begin{equation}\label{d6:eq:mh}
  m^2_h = - \mu^2 + 3 \lambda v^2  + \frac{15}{8}c_6\, m^2_h = 2\lambda v^2 + \frac{3}{2} c_6\, m^2_h~,
\end{equation}
which determines the quartic coupling as a function of the $m^2_h$, $v$, and $c_6$,
\begin{equation}\label{d6:eq:lam4}
\lambda = \frac{m^2_h}{2 v^2} \left (1-{3\over 2} c_6 \right )~.
\end{equation}
The bare mass-squared as a function of the $m^2_h$ and $c_6$ can be written as
\begin{equation}\label{d6:eq:mu2}
-\mu^2 = -\frac{m^2_h}{2} \left (1-{3\over 4} c_6 \right )~.
\end{equation}
Using Eqs.~(\ref{d6:eq:lam4}) and~(\ref{d6:eq:mu2}), the field-dependent Higgs mass term which goes into the effective potential is given by 
\begin{equation}
  m^2_h (h) = -\frac{m^2_h}{2} \left (1-{3\over 4} c_6 \right ) + \frac{3}{2} m^2_h \left (1-{3\over 2} c_6 \right ) \frac{h^2}{v^2} + \frac{15}{8}c_6\, m^2_h \frac{h^4}{v^4}~.
\end{equation}
The cubic and quartic couplings at tree-level are given by 
\begin{equation}\label{eq:34coupling:dim6}
\begin{split}
 \left . \lambda_3 = \frac{d^3V_{tree}(h)}{dh^3} \right |_{h=v} &= 6\lambda v + \frac{15}{2}\frac{c_6 m^2_h}{v} = \frac{3m^2_h}{v} ( 1+ c_6)~,\\[2.5pt]
 \left . \lambda_4 = \frac{d^4V_{tree}(h)}{dh^4} \right |_{h=v} &= 6\lambda + \frac{45}{2}\frac{c_6 m^2_h}{v^2} = \frac{3m^2_h}{v^2} ( 1+ 6\,  c_6)~.
\end{split}
\end{equation}
In our normalization, $c_6$ is literally the deviation of the cubic coupling from the SM value,
\begin{equation}
\frac{\lambda_3}{\lambda_{3\, SM}} -1 = c_6~,\quad  \frac{\lambda_4}{\lambda_{4\, SM}} -1 = 6\, c_6~.
\end{equation}
The relation between the cubic coupling and $c_6$ is linear with the choice of our normalization of $\mathcal{O}_6$ operator, and the same is true for quartic coupling. Even though the correlation between $\lambda_3$ and $\lambda_4$ is an interesting observation, this relation holds only at the level of dimension-six operators, e.g. adding a dimension-eight operator could break the relation. When assuming $T$-dependence only in the Higgs mass parameter as in~\cite{Grojean:2004xa,Huang:2015tdv}, the constraint on $c_6$ compatible with the first order (either strong or weak) phase transition can be analytically obtained~\cite{Huang:2015tdv},
\begin{equation}\label{eq:c6:crudeRange}
  \frac{2}{3} < c_6 < 2~.
\end{equation}
While we will be exploring the viable parameter space of $c_6$ with the full effective potential, the $\mathcal{O}(1)$ deviation in Eq.~(\ref{eq:c6:crudeRange}) is already alarming from the EFT viewpoint, and the truncation at the level of the dimension-six operators may not be well justified. The effect from the dimension-six operators will be order $\mathcal{O}(\mu_{EW}^2/\Lambda^2)$, as the electroweak phase transition occurs around the electroweak scale, $\mu_{EW}$. A large modification of the Higgs potential from the SM for the SFOEPT would imply a large EFT coefficient, compared to its naive dimensional analysis (NDA) estimate, or equivalently a low cutoff scale for the NDA-sized coefficient.

A large deviation of the trilinear coupling or a large size of the $\mathcal{O}_6$ operator in the linear representation, while suppressing the remaining operators to be consistent with the current Higgs data is usually not a generic feature of the EFT. However, there are a few well motivated scenarios that could give a parametric hierarchy between $\mathcal{O}_6$ and $\mathcal{O}_H$, which was discussed in~\cite{Azatov:2015oxa}. Here, we will assume that the New Physics sector is broadly characterized by one coupling $g_*$ and one mass scale $\Lambda$, associated with the new states. When the Higgs is assumed to be a generic composite state, not being a Goldstone boson, the $\mathcal{O}_6$ can be bigger than $\mathcal{O}_H$ by the factor $g^2_*/\lambda_4$. Basically, the suppression of the $\mathcal{O}_6$ operator by $\lambda_4/g^2_*$ which accounts for the shift symmetry breaking is undone. Another alternative is to couple the Higgs to a strongly coupled sector via Higgs portal, or $\lambda |H|^2 \mathcal{O}$. When the operator $\mathcal{O}$ in the strongly coupled sector is characterized by one strong coupling and one scale, it can be shown that the ratio of $\mathcal{O}_6$ to $\mathcal{O}_H$ can get enhanced by the factor $\lambda/\lambda_4$.

When the cutoff scale, for a given power counting, is dangerously low or the coefficient of the higher-dimensional operator is large for the SFOPT, the truncation at the level of dimension-six operators may cause a large uncertainty, as it is not well justified to ignore all the higher-dimensional operators. In such a situation, the EFT approach with resummed operators is a good illustration to represent a possible qualitative behavior of the scenario where the validity of EFT expansion is guaranteed~\footnote{Another aspect of the resummed EFT is that it matches to the EFT in the nonlinear basis, up to resuming over the Higgs powers and expanding them in terms of the neutral Higgs $h$, which can accommodate the large deviations from the SM values.}. While this type of the EFT might not be readily matched to a concrete UV model, this is one of few example where the EFT description is valid provided the energy is well below the cutoff scale, $E < \Lambda$. This type of EFT can be described as follows.

The generic Higgs potential in the EFT approach can be written as
\begin{equation}\label{eq:EFT:normTypeI}
 V_{tree} =-\mu^2 |H|^2+ \lambda|H|^4 + \sum_{n=1}^\infty \frac{c_{4+2n}}{v^{2n}}\frac{m^2_h}{2 v^2} |H|^{4+2n}~.
\end{equation}
We have chosen the normalization in Eq.~(\ref{eq:EFT:normTypeI}) such that the NDA estimates of the coefficients scale like $c_{4+2n} \sim (v/f)^{2n}$ where the factor $f$ is defined as $f \equiv \Lambda/g_*$.
Using the parametrization $H = (0,\, h/\sqrt{2})^T$, the above potential becomes
\begin{equation}\label{eq:Vtree:resummedTypeI}
 V_{tree} =-\frac{\mu^2}{2} h^2+ \frac{\lambda}{4} h^4 + \sum_{n=1}^\infty \frac{c_{4+2n}}{v^{2n}}\frac{m^2_h}{2 v^2} \left ( \frac{h^2}{2}\right )^{2+n}~.
\end{equation}
Assuming that all the Wilson coefficients are universal (and therefore, it would require only one counterterm for all higher-dimensional operators for the renormalization), while employing NDA scaling for all the universal coefficients, namely,  $c_{4+2n} = c\, (v/f)^{2n}$ with $c \sim \mathcal{O}(1)$, all higher-dimensional operators up to infinite order in the Higgs field can now be re-summed to give
\begin{equation}
 V_{tree} =-\frac{\mu^2}{2} h^2+ \frac{\lambda}{4} h^4 + \frac{1}{8}\frac{c}{f^2}\frac{m^2_h}{2 v^2} h^6 \frac{1}{1- \frac{h^2}{2 f^2}}~.
\end{equation}
Analogous to the case of the dimension-six operator, the Higgs VEV is determined by the equation $dV_{tree}(h)/dh|_{h=v} = 0$ of the tree-level potential and the physical Higgs mass by $d^2V_{tree}(h)/dh^2|_{h=v} \equiv m^2_h$. The bare mass parameter and the quartic coupling in terms of physical Higgs mass and Higgs VEV along with $c$ and $f$ (or $\xi \equiv (v/f)^2$) are determined to be
\begin{equation}
 -\mu^2 = -\frac{m^2_h}{2} \left ( 1- \frac{3}{4}\, c\, \frac{\xi}{1-\xi/2} - \frac{5}{8}\, c\, \frac{\xi^2}{(1-\xi/2)^2}
 - \frac{1}{8}\, c\, \frac{\xi^3}{(1-\xi/2)^3} \right )~,
\end{equation}
and
\begin{equation}
 \lambda = \frac{m^2_h}{2 v^2} \left [ 1+ c \left ( 1- \frac{1}{(1-\xi/2)^3} \right ) \right ]~.
\end{equation}
The field-dependent Higgs mass term is given by 
\begin{equation}
\begin{split}
  m^2_h (h) =& - \mu^2 + 3 \lambda h^2 + \frac{m^2_h}{\xi}\, \sum_{n=1}^{\infty} \frac{c}{2^{n+2}} (n+2)(2n+3) \left ( \frac{h}{f} \right )^{2n+2}~
  \\
  =& 
  -\frac{m^2_h}{2} \left ( 1- \frac{3}{4}\, c\, \frac{\xi}{1-\xi/2} - \frac{5}{8}\, c\, \frac{\xi^2}{(1-\xi/2)^2}
 - \frac{1}{8}\, c\, \frac{\xi^3}{(1-\xi/2)^3} \right ) \\
  & + \frac{3}{2} m^2_h \left [ 1+ c \left ( 1- \frac{1}{(1-\xi/2)^3} \right ) \right ] \frac{h^2}{v^2}
  + c\, m^2_h\, \xi\, \frac{15}{8}\, \frac{h^4}{v^4}\frac{1-\frac{17}{30}\, \xi \frac{h^2}{v^2} + \frac{1}{10}\, \xi^2 \frac{h^4}{v^4}}{\left ( 1- \xi\frac{h^2}{2 v^2} \right )^3}~.
\end{split}
\end{equation}
We find that the Goldstone boson mass is given by
\begin{equation}
\begin{split}
  m^2_\chi (h) &= - \mu^2 + \lambda h^2 + \frac{m^2_h}{\xi}\, \sum_{n=1}^{\infty} \frac{c}{2^{n+2}} (n+2) \left ( \frac{h}{f} \right )^{2n+2}~.
\end{split}
\end{equation}
The thermal mass in the high-$T$ approximation can be easily obtained by using the above-mentioned field dependent masses into Eq.~(\ref{eq:VT:highTapprox}). The additional contribution to the thermal mass from the higher-dimensional operators with universal Wilson coefficients is given by
\begin{equation}
  \Delta \Pi_{h/\chi}(0) =  \frac{1}{ 2}\, c\, \left ( 1- \frac{1}{(1-\xi/2)^3} \right ) \frac{m^2_h}{2v^2} T^2~.
\end{equation}
The cubic and quartic couplings at tree-level are 
\begin{equation}\label{eq:34coupling:dimall}
\begin{split}
 \left . \lambda_3 = \frac{d^3V_{tree}(h)}{dh^3} \right |_{h=v} & = \frac{3m^2_h}{v} \left [ 1+ 16\, c\,  \frac{\xi}{(2-\xi)^4} \right ]~,\\[2.5pt]
 \left . \lambda_4 = \frac{d^4V_{tree}(h)}{dh^4} \right |_{h=v} & = \frac{3m^2_h}{v^2} \left [ 1+ 32\, c\,  \frac{(6+\xi) \xi}{(2-\xi)^5} \right ]~.
\end{split}
\end{equation}
It is interesting to note that the deviation of the quartic coupling is $2(6+\xi)/(2-\xi)$ times bigger than that of the cubic coupling, that is,
\begin{equation}
\frac{\lambda_3}{\lambda_{3\, SM}} -1 = 16\, c\,  \frac{\xi}{(2-\xi)^4}~,\quad  \frac{\lambda_4}{\lambda_{4\, SM}} -1 = 32\, c\,  \frac{(6+\xi) \xi}{(2-\xi)^5} = 2\, \frac{6+\xi}{2-\xi} \times 16\, c\,  \frac{\xi}{(2-\xi)^4}~.
\end{equation}
In the limit $f \rightarrow v$ (or $\xi \rightarrow 1$), the ratio $2(6+\xi)/(2-\xi)$ reaches a maximum value,
\begin{equation}
\frac{\lambda_3}{\lambda_{3\, SM}} -1 = 16\, c~,\quad  \frac{\lambda_4}{\lambda_{4\, SM}} -1 =  14 \times 16\, c~,
\end{equation}
where the deviation of the quartic coupling appears fourteen times bigger than the deviation of the cubic coupling.

While we highlighted the tree-level relations between cubic and quartic couplings in Eqs.~(\ref{eq:34coupling:dim6}) and (\ref{eq:34coupling:dimall}) which look very different from the Higgs portal case, throughout our simulation, the cubic and quartic couplings are numerically evaluated with the full effective potential (similarly for the Higgs portal).

\section{Strong First Order Electroweak Phase Transition and Higgs Self Coupling}
\label{sec:SFOPTnSelfcoupling}

\subsection{On the criteria of strong first order phase transition}

There have been some ambiguities in the literature pertaining to the exact criteria for SFOEPT. A quantity commonly used is the ratio of the critical Higgs VEV to the critical temperature, namely, $v_c/T_c$, which simply checks the existence of the degenerate vacua at $T_c$. The threshold value of the $v_c/T_c$ has been used in a certain range (for example, see~\cite{Patel:2011th}),
\beqna
\frac{v_c}{T_c} \gtrsim 0.6 - 1.4~.
\eeqna{eq:SFOEPT}
Once the degenerate vacua with the potential barrier is formed, eventually it should transit from one vacuum to the global vacua as the universe cools down. In a more sophisticated treatment, the Euclidean action, $\mathcal{S}_3$, is computed at a finite temperature, and we demand that the Euclidean action suppressing the tunneling rate is smaller than a certain value for the successful nucleation of the bubble~\cite{Dine:1992wr}. The corresponding temperature and the Higgs VEV in this approach are denoted by $T_N$ and $v_N$ whose values are also used in a certain range. For instance, $T_N$ is determined such that
\begin{equation}
 \frac{{\mathcal S}_3}{T_N} \sim 100 - 140~.
\end{equation}
Addressing the phenomenological impact on the Higgs self coupling of this criteria involves the calculation of the Euclidean action. The discrepancy between $(v_c, \, T_c)$ and $(v_N, \, T_N)$ can depend on the structure of the effective potential during the phase transition (see~\cite{Kurup:2017dzf} for a recent discussion on this aspect). While the overall ballpark of the parameter space does not seem to make a significant change under the variations mentioned above, there are several overlooked aspects which we will address below. The precision of the Higgs self couplings is subject to the $\mathcal{O}(1)$ uncertainty, which can make a dramatic impact on the prospect for future colliders. We will address these issues in Section~\ref{sec:validityEffpot} in detail.

\subsection{On the parameter space for strong first order phase transition and Higgs self couplings}

In the scenario of the SM extension with the scalar singlet as described in Section~\ref{sec:singlet}, the relevant parameters are the singlet mass, $m_S$, the coupling $\lambda_{HS}$ between the Higgs and the singlet, and the scalar singlet quartic coupling $\lambda_S$. For the one-step phase transition the quartic coupling $\lambda_S$ does not play much role directly in the phenomenology apart from ensuring the stability of the potential at a large field. In our simulation of the one-step phase transition, we simply fix $\lambda_S$ to zero and scan over the bare singlet mass $\mu_S$ and the quartic coupling $\lambda_{HS}$ in the intervals $\mu_S = [10, \, 1310]$ GeV (in steps of 10 GeV) and $\lambda_{HS} = [0,\, 5]$ (in steps of 0.05). On the contrary, when the phase transition proceeds via a two-step cascade, the $\lambda_S$ needs to stay above the minimum $\lambda_S^{min}$ in Eq.~(\ref{eq:twostep:lamSmin}) so that $(v,\, 0)$ remains the global minimum. We perform a scan only over $m_S$ and $\lambda_{HS}$ in the intervals $m_S = [65,\, 700]$ GeV (in steps of 5 GeV) and $\lambda_{HS} = [0,\, 5]$ (in steps of 0.05) for a few choices of $\lambda_S$, parameterized as $\lambda_S = \lambda_S^{min} + \delta_S$. We assume that the singlet mass is heavier than roughly $m_h/2$ to avoid the Higgs decays to the singlet scalar. We impose arbitrary hard cutoffs on the quartic couplings, namely $\lambda_{HS} < 5$ and $\lambda_S < 5$ (smaller than $4 \pi$ which is the typical unitarity bound), to avoid the strongly coupled regime. On the other hand, in the EFT approach with the $\mathcal{O}_6$ operator described in Section~\ref{sec:EFT}, we scan over $c_6$ in the interval $c_6 = [0,\, 4]$ (in steps of 0.025).  For the specific case where all higher-dimensional operators can be re-summed up to infinite order in the Higgs field, the universal Wilson coefficient $c$ is coarsely scanned over $c=[0,\, 5]$ (in steps of 0.05) along with $\xi\equiv (v/f)^2$ scanned over in the interval $\xi = [0,\, 1]$ (in steps of 0.02). For the special limit $f\rightarrow v$ (or $\xi \rightarrow 1$), we make a separate fine-grid scan over $c$ in the window $[0,\, 0.3]$ (in steps of 0.002). 
Throughout all our simulations, we do not include the Goldstone bosons (in the Landau gauge) in the effective potential~\footnote{However, we have checked, using the approximate prescription adopted in~\cite{Cline:2011mm} in our prescription A, that the contribution from the Goldstone bosons for the one-step phase transition in the Higgs portal scenario and for the EFT approach with higher-dimensional operators have only mild effect. The effect from the Goldstone bosons is found to be small in~\cite{Chiang:2018gsn} as well, using an alternative prescription for treating Goldstone bosons in the  Higgs portal scenario. Similarly, we find that the contribution from the Higgs is almost negligible for the one-step phase transition in the Higgs portal scenario and for the EFT approach with $\mathcal{O}_6$ operator.
}.
\\

\begin{figure}[!htb!] 
	\centering
	\includegraphics[width=0.3\linewidth]{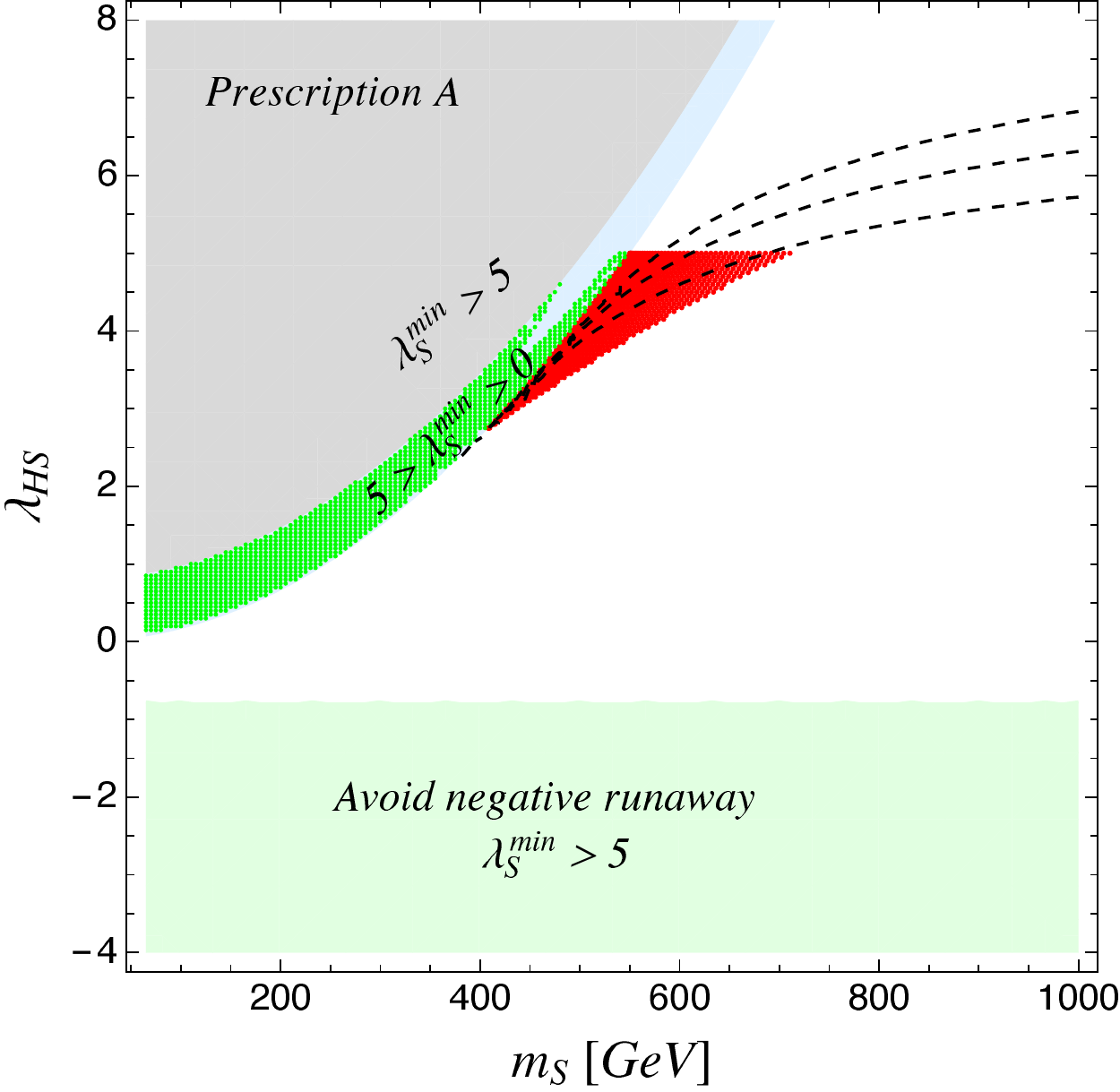}\quad
	\includegraphics[width=0.2985\linewidth]{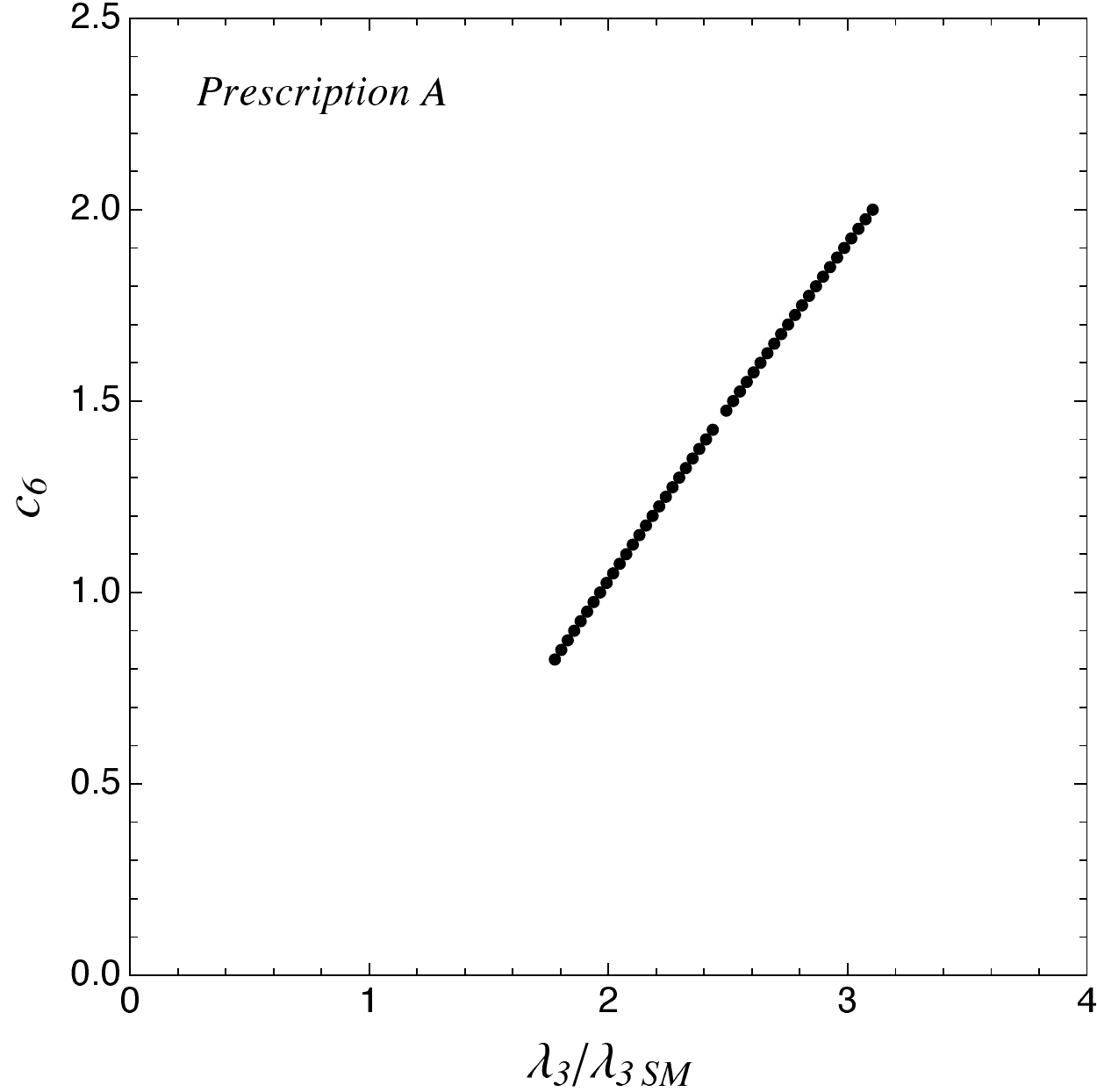}\quad
	\includegraphics[width=0.2985\linewidth]{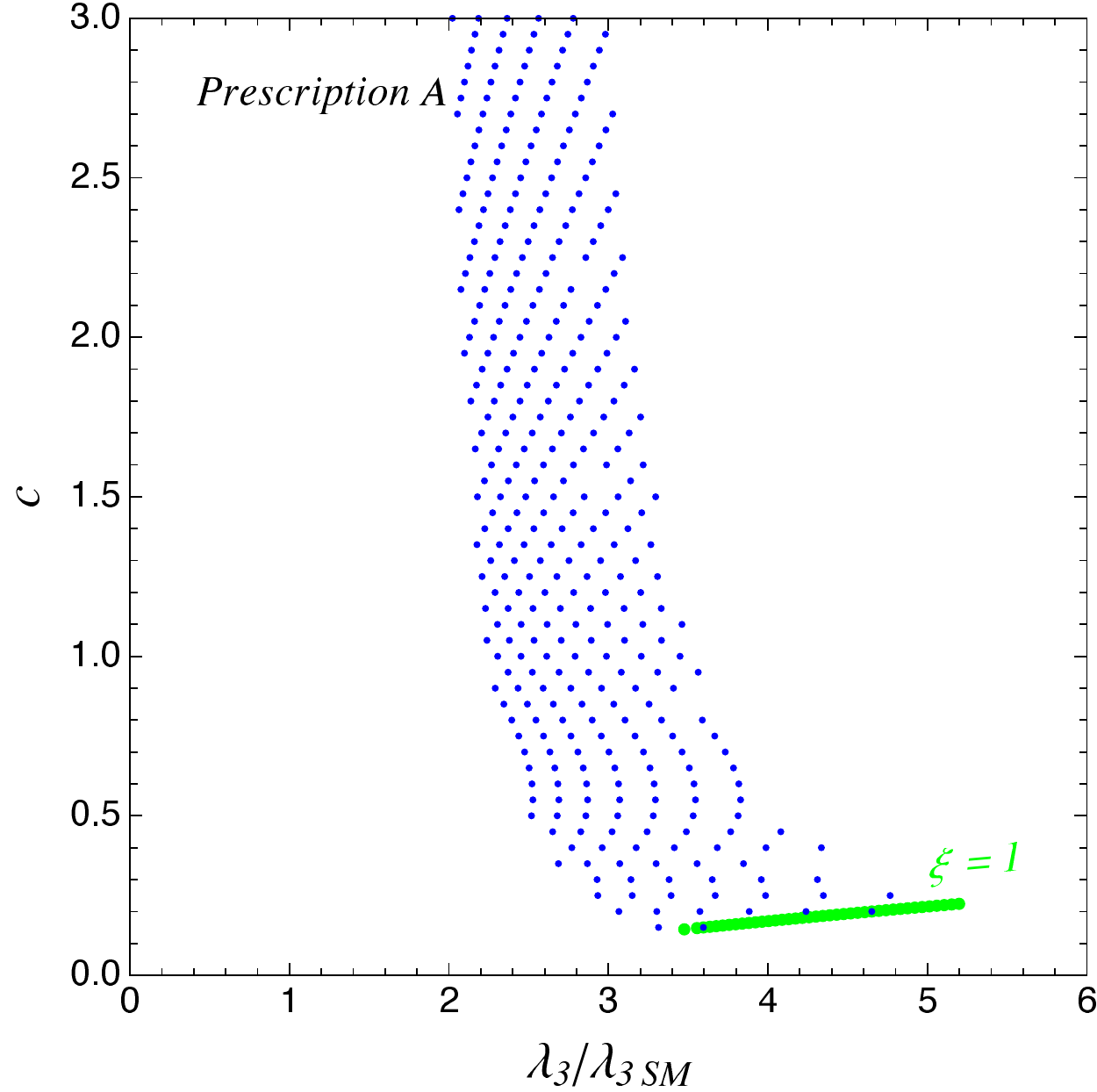}
	\caption{Left: The viable $(m_S,\, \lambda_{HS})$ region for the one-step SFOEPT (red) and for the two-step phase transition (green) in the  prescription A. The singlet quartic coupling is set to $\lambda_S = \lambda_S^{min}$ (or $\delta_S = 0$) for the two-step phase transition in the plot. In the left-panel, $(v,\, 0)$ is assumed to be the global minimum. Light-blue (part of them covered by green) corresponds to the strip with $0 < \lambda_S^{min} < 5$, and the grey region to $\lambda_S^{min} > 5$. 
The three dashed curves denote the lower bound of perturbative limits, where one-loop contribution to the Higgs quartic coupling becomes 0.4, 0.5, or 0.6 respectively from the bottom to the top. Middle: The viable ($c_6, \, \lambda_3/\lambda_{3\, SM}$) region for the SFOEPT in the EFT approach with $\mathcal{O}_6$ operator. Right: Similar region for the EFT approach with the resummed $|H|^{4+2n}$ ($n \geq 1$) operators with universal coefficients $c$ (blue) -- a particular limit $f\rightarrow v$ (or $\xi \rightarrow 1$) shown in green. In all plots, the parameter spaces are shown for $v_c/T_c > 1$.}
	\label{fig:lamHSvsMS}
\end{figure}
%
The viable parameter space for SFOEPT satisfying  $v_c/ T_c\geq 1$ in two classes of scenarios are shown in Fig.~\ref{fig:lamHSvsMS}. For the illustration in Fig.~\ref{fig:lamHSvsMS}, we chose the prescription A where the finite temperature potential is computed exactly, while the thermal mass entering into the potential is obtained using the high-$T$ approximation. In the plot on the left, we see that the Higgs portal with the scalar singlet becomes a plausible option for SFOEPT only for a strong coupling $\lambda_{HS}\sim \mathcal{O}(1)$ when it proceeds via one-step phase transition. This corresponds to a region of parameter space where a naive approach based on the one-loop effective potential requires a careful treatment -- the higher loop corrections to the one-loop Coleman-Weinberg potential may become large and the reliability of perturbative analysis may break down. Three dashed lines in the left panel of Fig.~\ref{fig:lamHSvsMS} represent a few rough estimates of the loop-contribution to the Higgs quartic coupling~\footnote{These three lines were drawn based on the formula in Eqs.~(3.2) and~(3.3) in~\cite{Curtin:2014jma}. We intend to use these contours as the guideline for the rough estimate of higher-loop contributions to the effective potential at zero temperature, and we do not impose this on our parameter space.}.  The issue of the strong coupling is a bit ameliorated for the case of two-step phase transition as the SFOEPT can be realized for a wide-range of coupling sizes including the small coupling region. However, the perturbative region is still limited due to the constraint on the minimum quartic coupling of the singlet, $\lambda_S \geq \lambda_S^{min}$. In the left panel of Fig.~\ref{fig:lamHSvsMS}, the corresponding region to $5 > \lambda_S^{min} > 0$ appears as a narrow strip. In the presence of a strong coupling, especially for the finite-temperature quantum field theory, a consistent treatment of the effective potential involves a thermal resummation of various types of diagrams which is beyond the scope of this work (see~\cite{Curtin:2016urg} for a recent discussion). 
A systematic approach for the power counting and thermal resummation remains to be developed to correctly address the plausibility of the SFOEPT. 

\begin{figure}[!htb!] 
	\centering
	\includegraphics[width=0.4\linewidth]{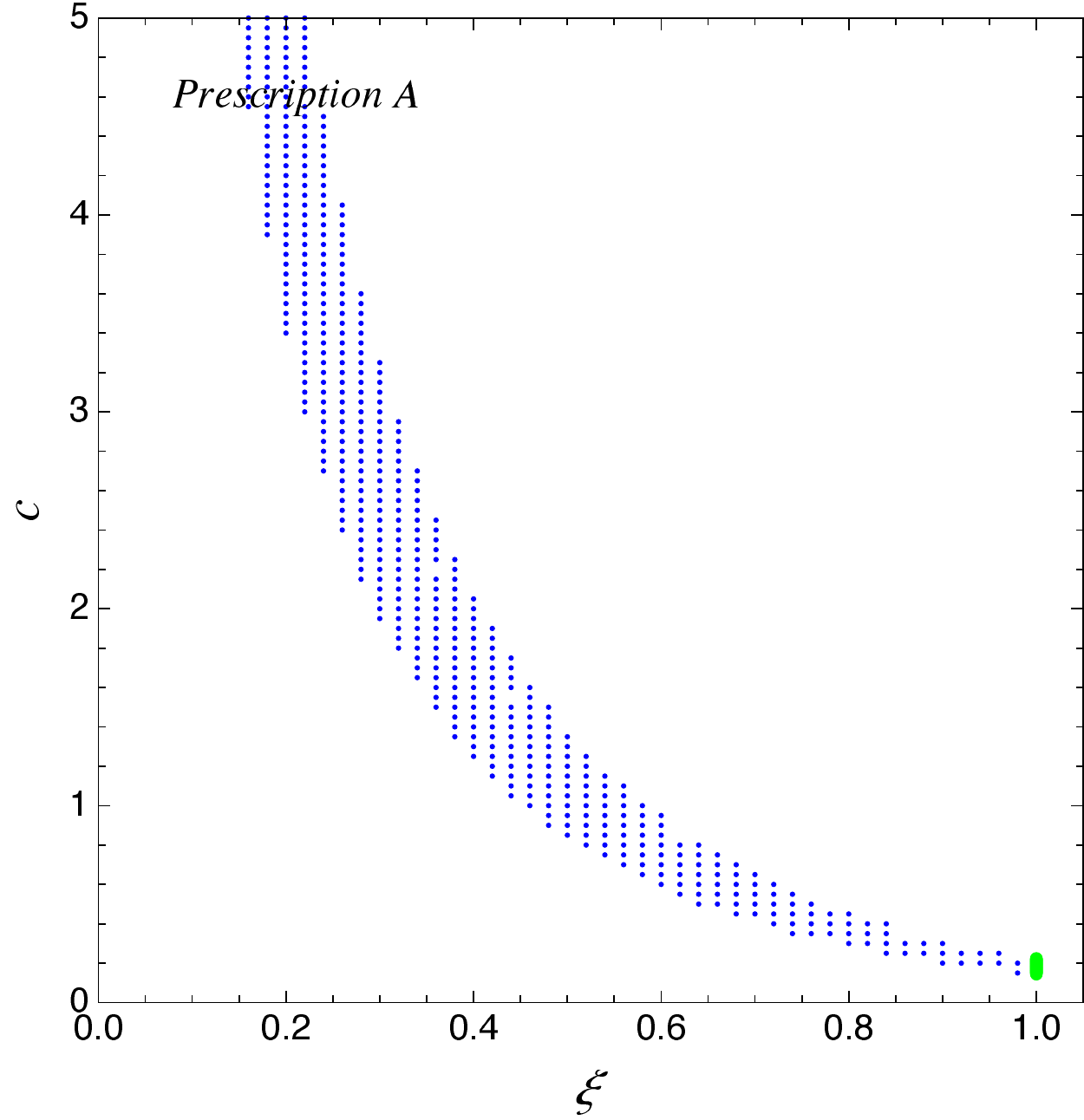}
	\caption{The viable ($\xi,\, c$) region for the EFT approach where all higher-dimensional operators, $|H|^{4+2n}$ ($n \geq 1$) are resummed to all orders in the Higgs field for universal Wilson coefficients (blue) -- a particular limit $f\rightarrow v$ (or $\xi \rightarrow 1$) shown in green.}
	\label{fig:lam3VSc}
\end{figure}

\begin{figure}[!htb!] 
	\centering
	\includegraphics[width=0.301\linewidth]{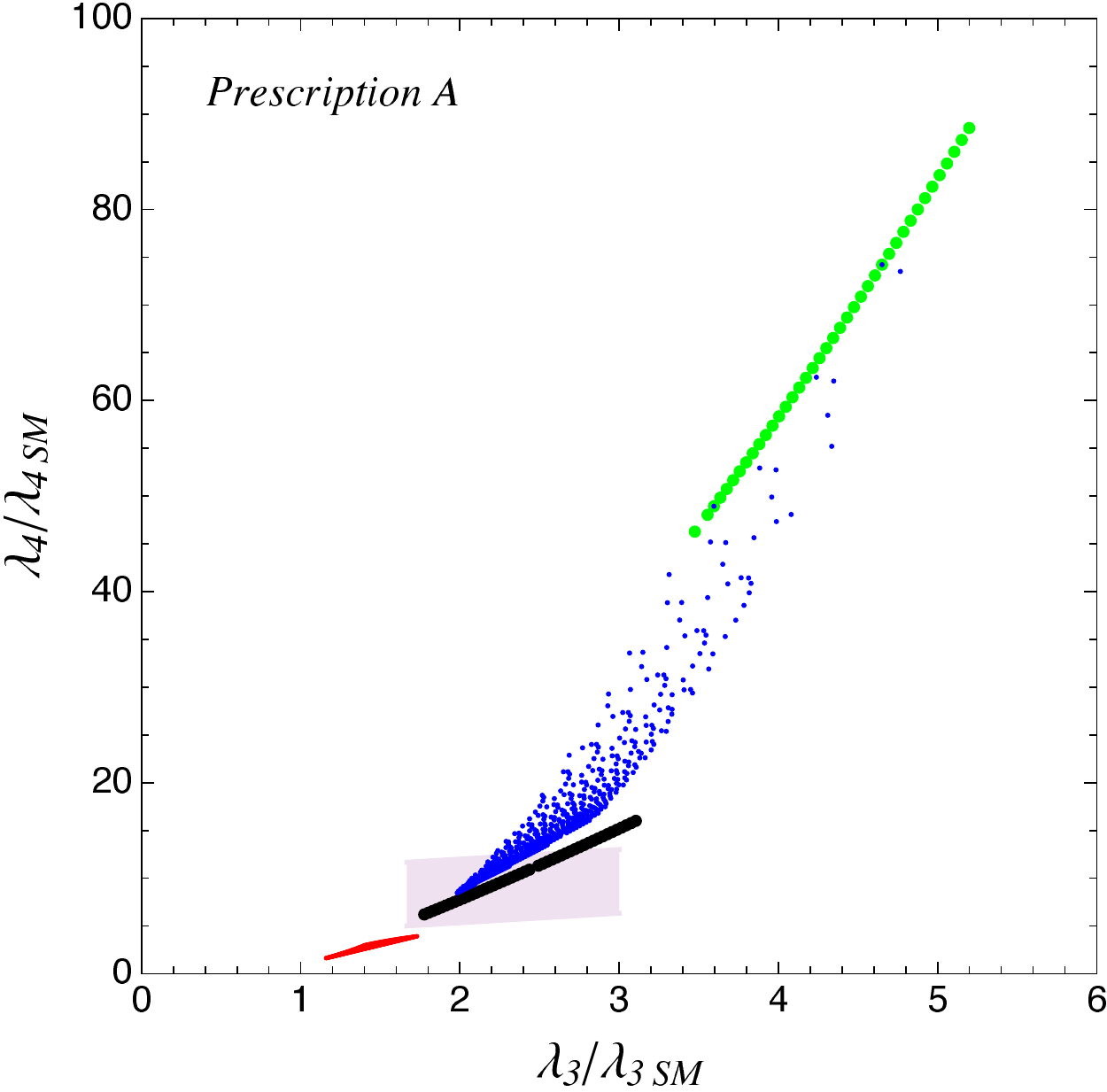}\quad
	\includegraphics[width=0.295\linewidth]{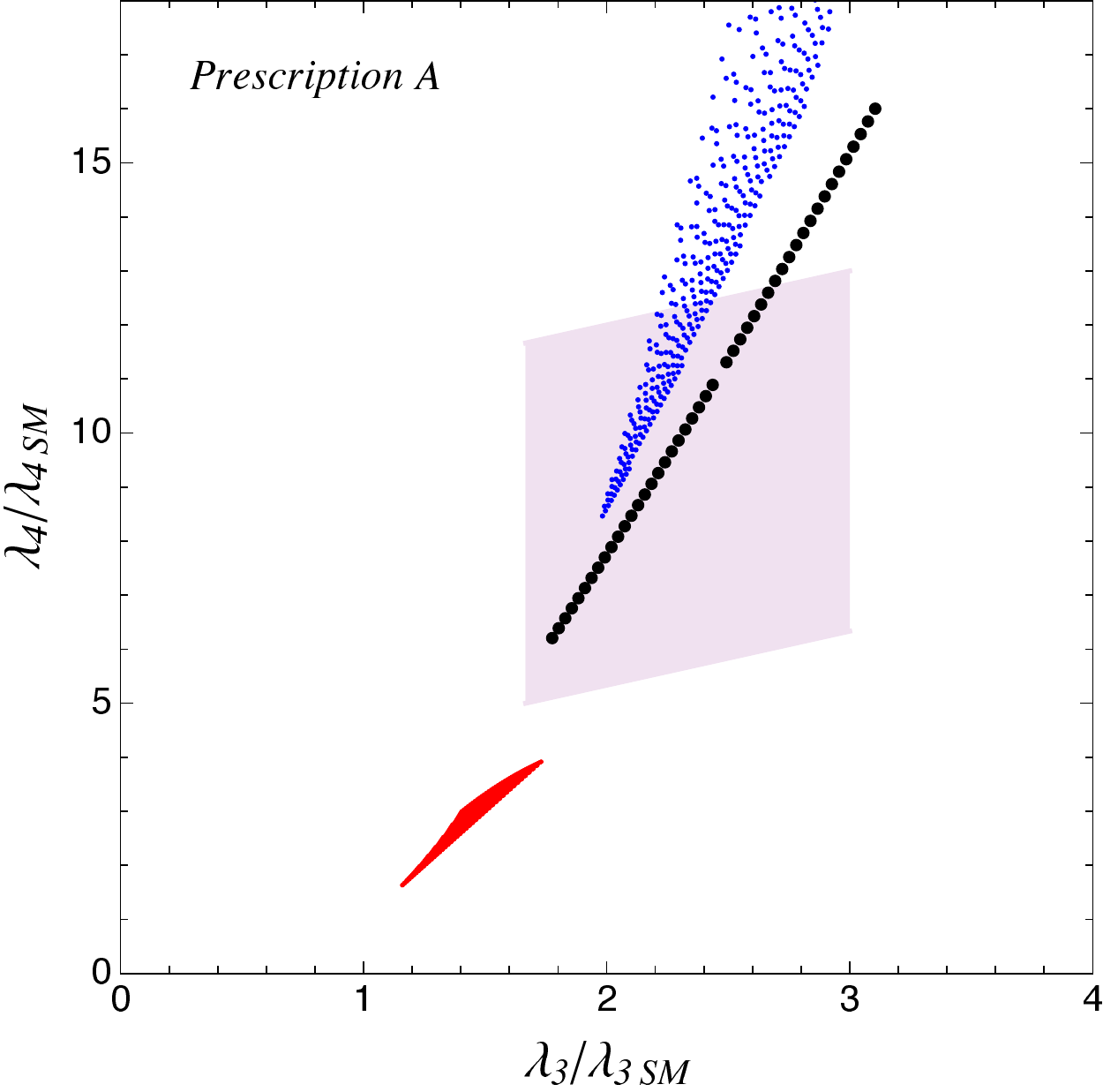}\quad
	\includegraphics[width=0.303\linewidth]{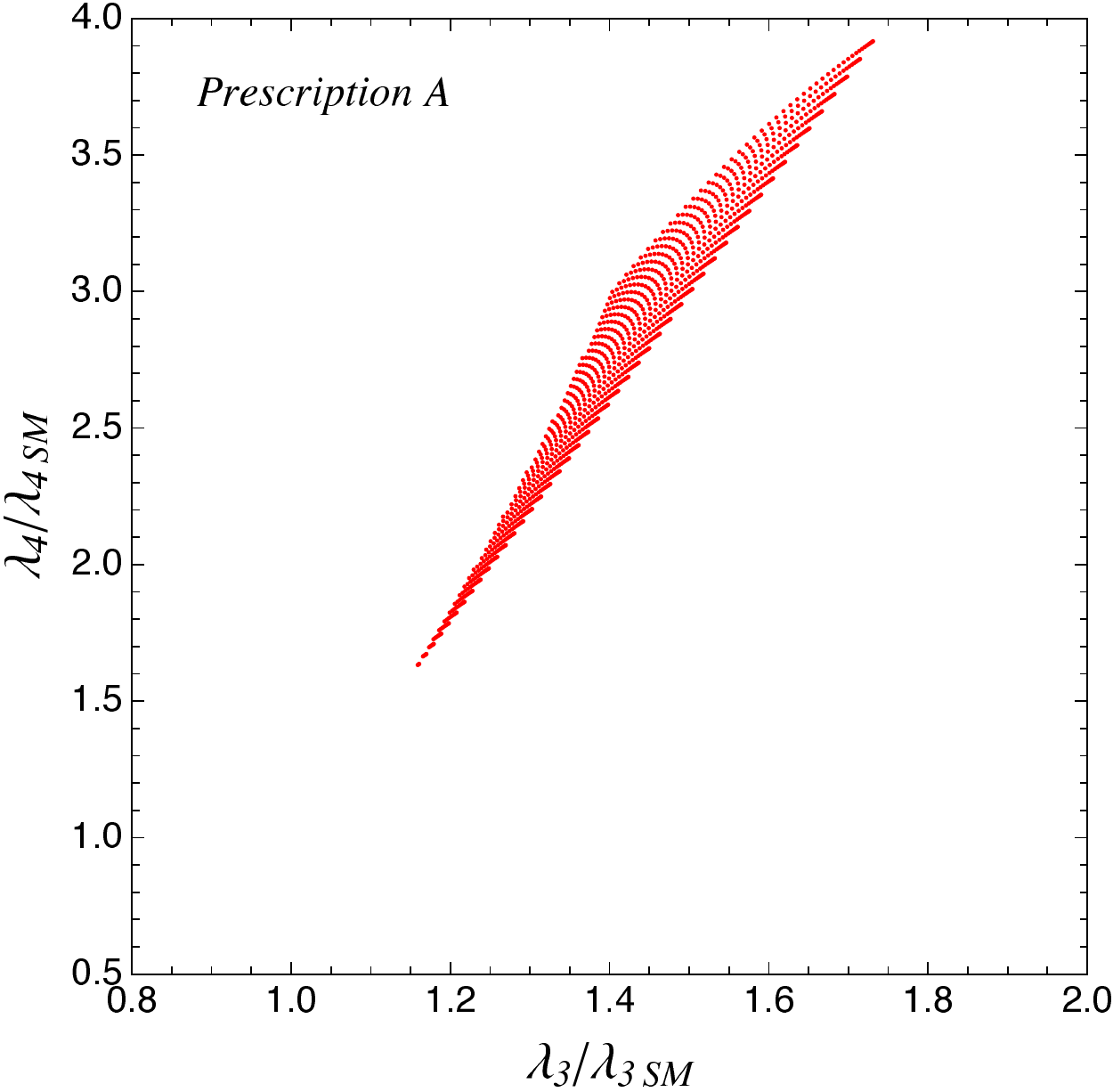}
	\caption{The correlation between the cubic and quartic Higgs self couplings in two classes of BSM scenarios: the one-step SFOEPT in the BSM with a singlet scalar (red) and the phase transition in the EFT approach with only $\mathcal{O}_6$ operator (black) and with all the higher-dimensional operators, $|H|^{4+2n}$ ($n \geq 1$), with universal coefficients (blue) -- a particular limit $f\rightarrow v$ (or $\xi \rightarrow 1$) shown in green.  The parameter spaces (except light-purple region) are shown for $v_c/T_c > 1$. Light-purple is the corresponding region to Eq.~(\ref{eq:c6:crudeRange}) using a crude high-$T$ approximation. The middle (last) plot is the zoomed-in version of the left (middle) plot. }
	\label{fig:lam3VSlam4}
\end{figure}

In the middle panel of Fig.~\ref{fig:lamHSvsMS}, we find that the coefficient of dimension six operator for SFOEPT with $v_c/ T_c\geq 1$ criteria appears to have a large deviation, $c_6\sim \mathcal{O}(1)$. As aforementioned, a large coefficient is alarming from the EFT perspective, and the truncation at the level of the dimension-six operators should be taken with grain of salt. Given a large coefficient of the dimension-six operator, dimension-eight operators or higher may not be ignored, and the presence of higher-dimensional operators with non-negligible coefficients may change the details of the physics relevant for the SFOEPT.  A large value of $c_6$ may indicate a strongly-coupled New Physics not far away from the TeV scale although the exact translation depends on the assumption of the UV completion. In the right panel of Fig.~\ref{fig:lamHSvsMS} and Fig.~\ref{fig:lam3VSc}, we show the result of the EFT approach when all higher-dimensional operators with universal coefficients are resummed to all orders in the Higgs field. In the region of $c\sim \mathcal{O}(1)$ in Fig.~\ref{fig:lamHSvsMS}, the range of $\lambda_3/\lambda_{3\, SM}$ values satisfying $v_c/T_c > 1$ is similar to the case only with the dimension-six operator. The right panel of Fig.~\ref{fig:lamHSvsMS} shows that the cubic coupling can deviate by larger amount with increasing $\xi$ (equivalently to decreasing $c$ as is evident in Fig.~\ref{fig:lam3VSc}). In the special limit $f\rightarrow v$ (or $\xi \rightarrow 1$), the coefficient $c$ is well below one, or $c \sim [0.14,\, 0.23]$, and the overall deviation of the cubic coupling reaches the maximum, $\lambda_3/\lambda_{3\, SM} \sim [3.45, \, 5.35]$ shown in the plot\footnote{Note that blue points on top of the green line, in the right panel of Fig.~\ref{fig:lamHSvsMS} (similarly for the left panel of Fig.~\ref{fig:lam3VSlam4}), which corresponds to the special limit $\xi \rightarrow 1$ do not cover the entire green line simply because of the coarse two-dimensional scan over $c$ and $\xi$.}. The deviation of the cubic coupling is much larger than the case with only the $\mathcal{O}_6$ operator.
%
In a situation where we observe $\lambda_3/\lambda_{3\, SM} \sim \mathcal{O}(1)$ deviation of the cubic coupling from the SM value, it could be induced by various types of New Physics models.  When the phenomenological disentanglement of various New Physics scenarios becomes challenging, the measurement of the Higgs quartic coupling can be beneficial.  As is illustrated in Fig.~\ref{fig:lam3VSlam4}, two different classes of scenarios based on either the Higgs portal with a singlet scalar or the EFT approach are widely separated in $(\lambda_3/\lambda_{3\, SM},\,  \lambda_4/\lambda_{4\, SM})$ plane~\footnote{One should not take the Higgs self couplings at face values in Fig.~\ref{fig:lam3VSlam4} as they could be out of EFT validity region in which case their values are subject to $\mathcal{O}(1)$ fluctuation.}. Within the EFT approach, as was discussed in~\ref{sec:EFT} and illustrated in Fig.~\ref{fig:lam3VSlam4}, the relation between $\lambda_3/\lambda_{3\, SM}$ and $\lambda_4/\lambda_{4\, SM}$ widely varies depending on the details of the underlying models. While the resulting quartic coupling can still be in the perturbative regime for the singlet-assisted BSM and the EFT only with $\mathcal{O}_6$ operator due to the small $\lambda_{4\, SM}$, the situation is likely pointing toward a strongly coupled dynamics for the EFT with the resummed higher-dimensional operators. The latter case with a very large quartic coupling might be constrained by other means. For instance, the unitarity bound on the cubic and quartic couplings were discussed in~\cite{DiLuzio:2017tfn} and a possibility of constraining models via the $S$ and $T$ electroweak precision parameters induced by large Higgs self couplings was considered in~\cite{Kribs:2017znd}. On the other hand, the High-luminosity LHC (HL LHC) may have a sensitivity on the order one deviation of the Higgs self coupling as will be discussed in detail in Section~\ref{sec:ProsFutureCollider}. It implies that a large fraction of the parameter space of the cubic coupling in Fig.~\ref{fig:lam3VSlam4} can be tested at the HL LHC.

In two classes of benchmark scenarios that we considered in our study, the deviation of the Higgs self couplings compatible with the SFOEPT are positive. It will be interesting to explore the BSM models which predict the large negative deviation of the Higgs self coupling as the negative deviation has better sensitivity at the colliders.

\subsection{Prospect for Future Collider}
\label{sec:ProsFutureCollider}

The measurement of the Higgs self coupling at the LHC is very challenging. The currently available projections of the Higgs cubic coupling on the HL LHC assuming 3 ab$^{-1}$ indicates too poor sensitivity at 95\%CL~\cite{ATL-PHYS-PUB-2014-019, ATL-PHYS-PUB-2017-001,CMS-PAS-FTR-15-002} to test any part of the parameter space shown in Fig.~\ref{fig:lam3VSlam4}. However, demanding the sensitivity at 68\%CL may have a chance to access a chunk of the parameter space in Fig.~\ref{fig:lam3VSlam4}. It has been shown in~\cite{Azatov:2015oxa}, using the same luminosity at the HL LHC, that the sensitivity of $\lambda_3/\lambda_{3\, SM}-1$ at 68\%CL has two intervals, $[-1.0, 1.8]\ \cup \ [3.5, 5.1]$ (see~\cite{DiVita:2017eyz} for the related discussion). The first interval around SM value can test the Higgs cubic coupling with order one deviation, and this will exclude most EFT cases considered in this work. Any improvement of the Higgs self coupling at the HL LHC will be beneficial in testing the BSM scenarios for the EWBG based on the SFOEPT. The sensitivity of the cubic coupling gets significantly improved at 100 TeV $pp$ collider due to the increased signal rate, and its sensitivity can reach up to $\sim 3$\% level~\cite{Contino:2016spe}. The cubic coupling can also be accessed at the ILC, and its sensitivity dominantly comes from the Vector Boson Fusion (VBF) process at the high center of mass energy. The VBF process at 1 TeV, assuming that the integrated luminosity can reach 5 ab$^{-1}$, can measure the cubic coupling up to $\sim 10$\%~\cite{Baer:2013cma}. 

One might naively think that the measurement of the quartic coupling is extremely difficult as the SM cross section is tiny even at 100 TeV $pp$ collider. The cross section $\sigma_{SM}(pp\rightarrow hhh)$ (before folding in Higgs decays) at 100 TeV is a few fb. 
However, unlike a common lore that there is no meaningful sensitivity for the Higgs quartic coupling at 100 TeV $pp$-collider unless the deviation of the quartic coupling from the SM is very large, the results in~\cite{Papaefstathiou:2015paa,Fuks:2017zkg} suggest that the 100 TeV $pp$ collider would have a meaningful sensitivity to the quartic coupling in a situation that the deviation of the cubic coupling is as big as (or bigger than) $\sim 40$\% (see Fig. 6 of~\cite{Papaefstathiou:2015paa} for the $2\sigma$ sensitivity on the cubic and quartic couplings). As is evident in Fig.~\ref{fig:lam3VSlam4}, all EFT scenarios for the SFOEPT in this study can be well differentiated by the quartic coupling at 100 TeV $pp$ collider. Even for the Higgs portal with a real singlet scalar which predicts rather a small size of the quartic coupling, almost half of the quartic coupling compatible with the SFOEPT can have a 2$\sigma$ sensitivity at 100 TeV $pp$ collider. Our novel observation highlights the utility of the quartic coupling as a way to disentangle various BSM scenarios for the SFOEPT. The analyses of $hhh\rightarrow b\bar{b}b\bar{b}\gamma\gamma$~\cite{Papaefstathiou:2015paa} and $hhh\rightarrow b\bar{b}b\bar{b}\tau^+\tau^-$~\cite{Fuks:2017zkg} also show that two different channels are sensitive to the different regions in $(\lambda_3, \, \lambda_4)$ space in such a way that they are complementary~\footnote{The exclusion plots in~\cite{Papaefstathiou:2015paa,Fuks:2017zkg} are the 2$\sigma$ sensitivity contours -- excluding at 68\%CL will be much stronger.}. While the $hhh\rightarrow b\bar{b}b\bar{b}\gamma\gamma$ channel has a better sensitivity on the positive deviation of the quartic coupling for the case with a positive large deviation of the cubic coupling, the $hhh\rightarrow b\bar{b}b\bar{b}\tau^+\tau^-$ channel has a better sensitivity on the negative deviation of the quartic coupling for the same deviation of the cubic coupling. 

\section{Validity of Effective Potential}
\label{sec:validityEffpot}
\begin{figure}[!htb!] 
	\centering
	\includegraphics[width=0.225\linewidth]{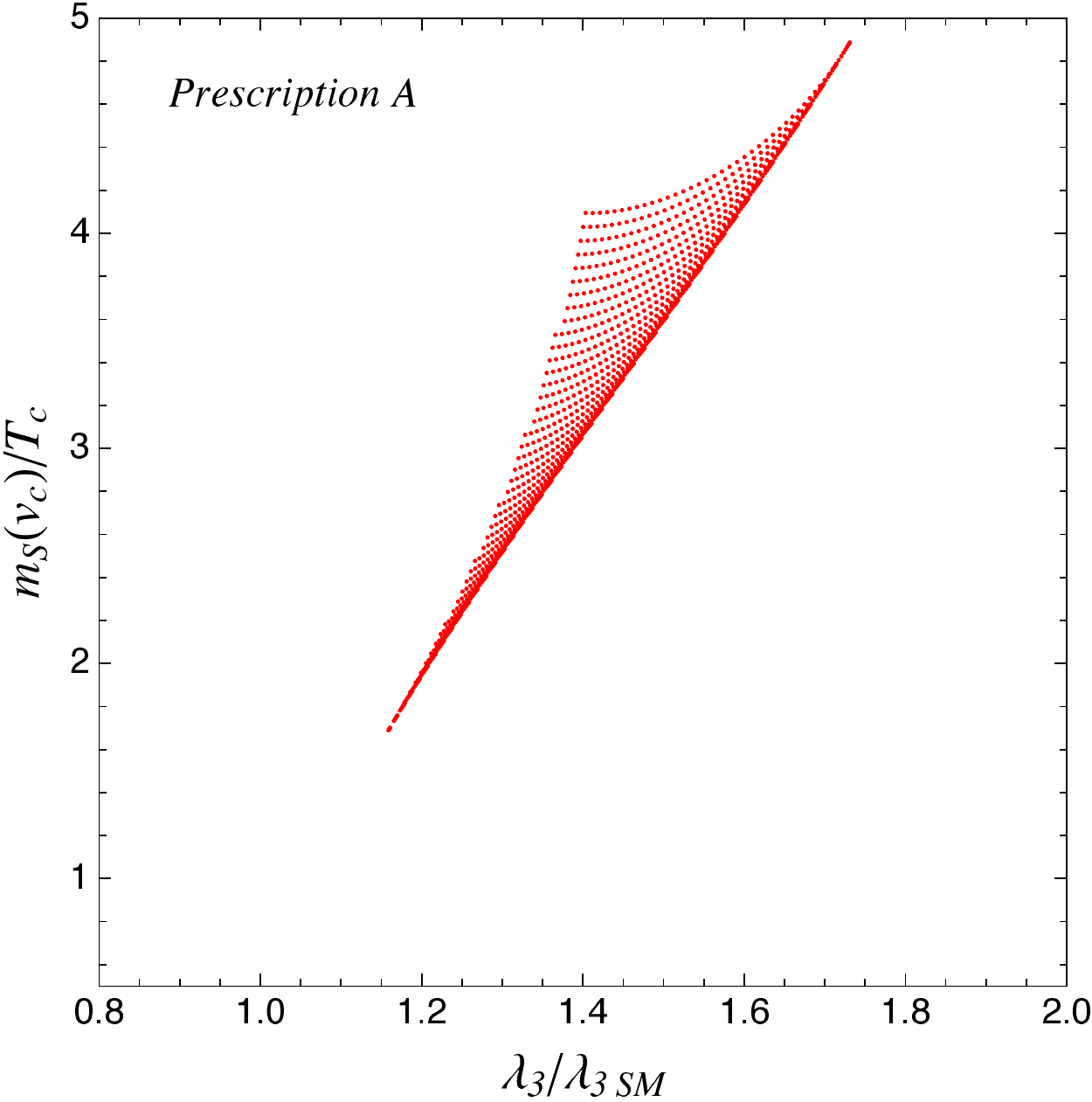}
	\includegraphics[width=0.225\linewidth]{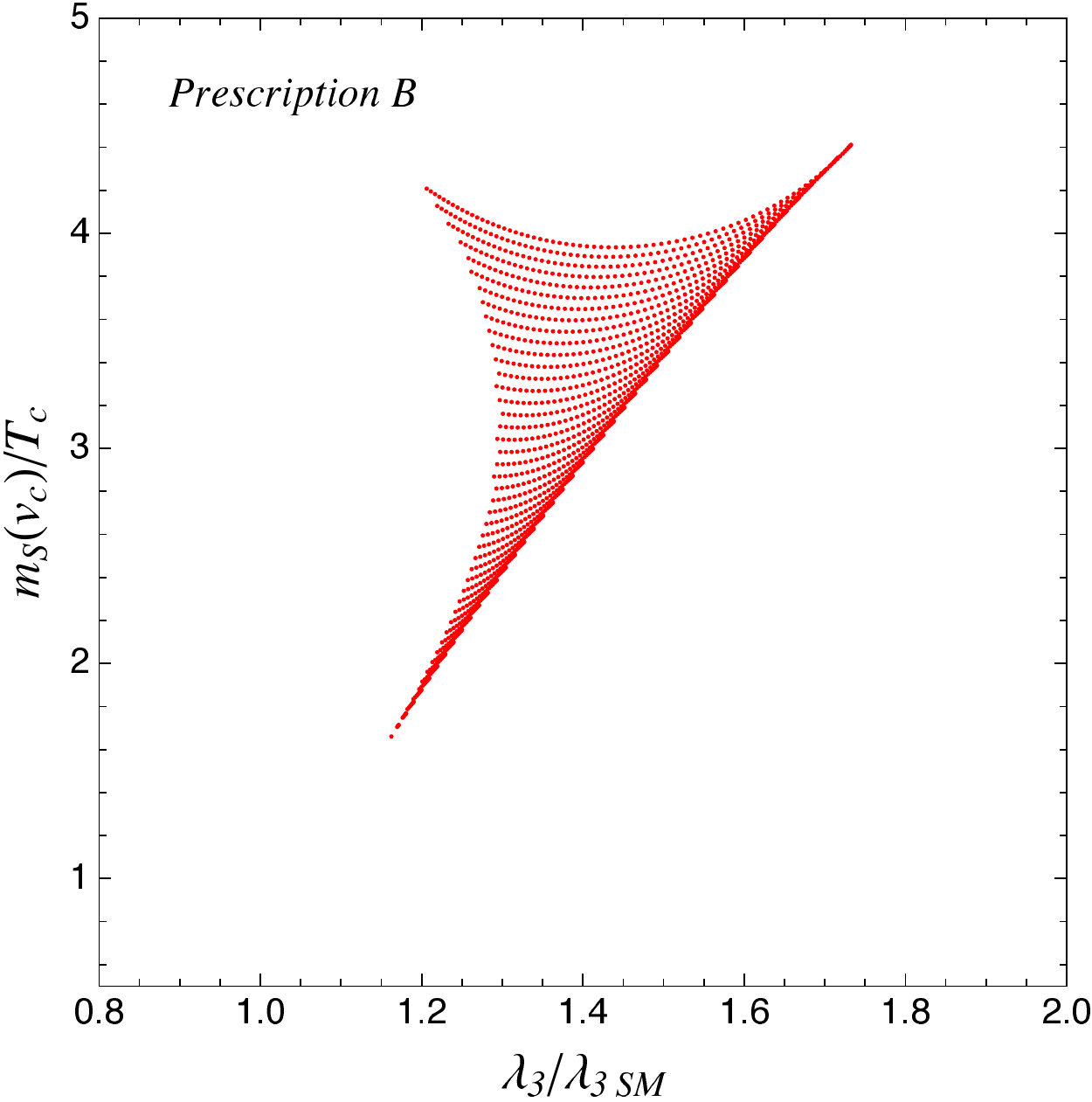}
	\includegraphics[width=0.230\linewidth]{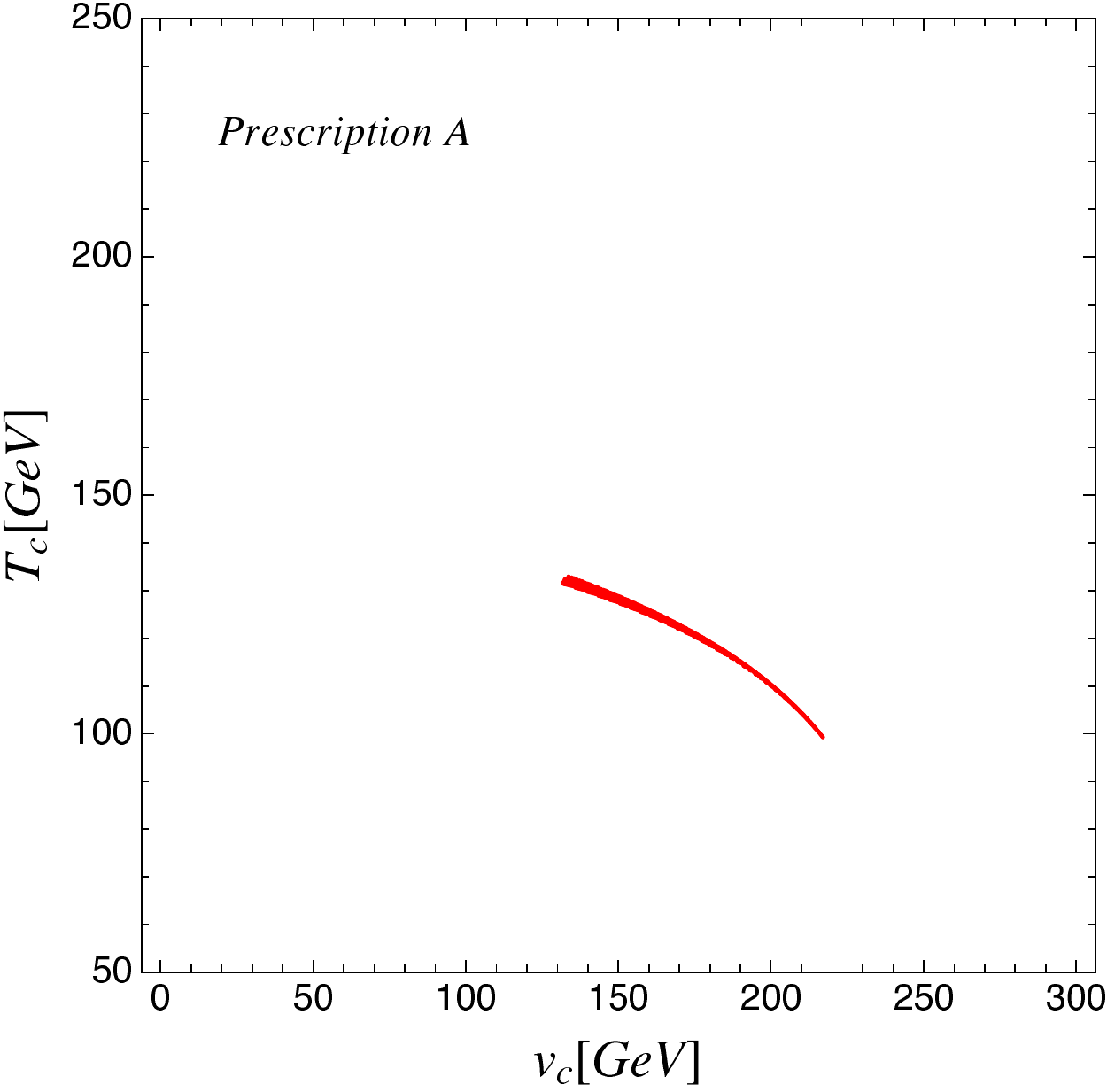}
	\includegraphics[width=0.230\linewidth]{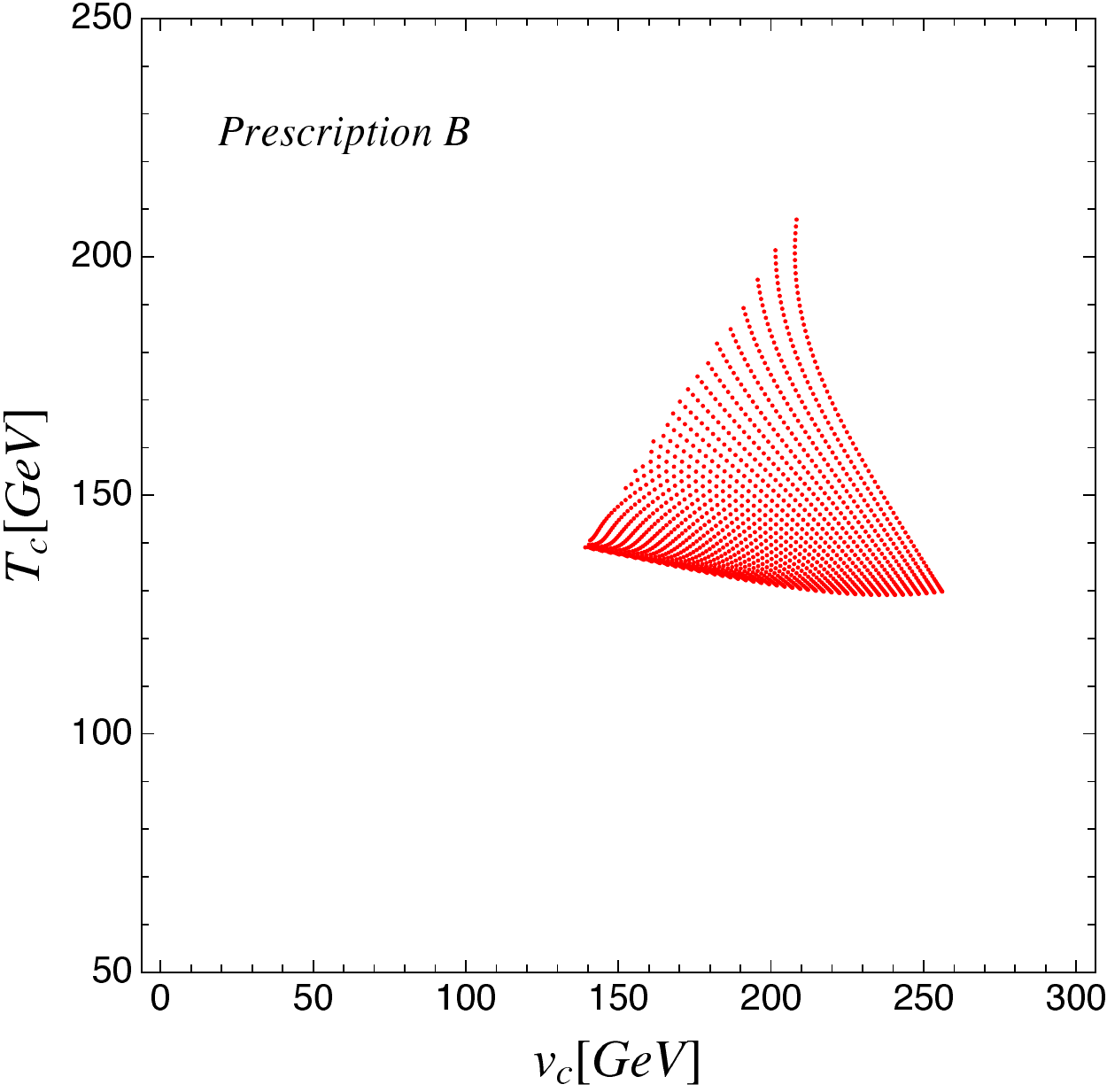}	
	\caption{The illustration of the discrepancy between prescriptions A and B for the one-step SFOEPT of the Higgs portal with a singlet scalar. The $m_S(v_c)/ T_c$ vs $\lambda_3/\lambda_{3\, SM}$ (left two plots), and $T_c$ vs $v_c$ (right two plots) for the SFOEPT using two prescriptions.}
	\label{fig:AorB}
\end{figure}

Among several issues regarding the validity of the effective potential, in this section, we will focus on the issue caused by the break-down of the high-temperature approximation of the thermal potential and its impact or uncertainty on the precision of the Higgs self couplings. As is evident in Fig.~\ref{fig:lamHSvsMS}, the coupling $\lambda_{HS} \sim \mathcal{O}(1)$ for the one-step SFOEPT. For the two-step case, $\lambda_{HS}$ can be a bit relaxed, but still the region for perturbative $\lambda_S$ is limited. When the naive criteria for the SFOEPT is satisfied, or $v_c \gtrsim T_c$, with $\mathcal{O}(1)$ coupling, the field dependent mass parameter in the potential is not small compared to the critical temperature, or
\begin{equation}\label{eq:mT:NDA}
  {m^2(v_c)\over T_c^2} \sim \mathcal{O}(1)\times {v^2_c \over T^2_c} \gtrsim 1~,
\end{equation}
which invalidates the high-$T$ approximation. The exact amount of the uncertainty depends on the degree of the violation of the high-$T$ approximation -- the value of $m/T$ (e.g. $m/T \sim 2-5$) shown in the first and second panels in Fig.~\ref{fig:AorB} indicates that the high-$T$ approximation is not appropriate for the one-step SFOEPT. Its violation is more pronounced for the two-step SFOEPT. On the other hand, a coupling with $\mathcal{O}(1)$ strength can be the signal of the breakdown of the perturbation theory. 

The discrepancy between two prescriptions is very pronounced in the values of the critical VEV and critical temperature, as is seen in the third and fourth panels in Fig.~\ref{fig:AorB}. The prescription A leads to more focused region where the VEV at the critical temperature is always smaller than zero-temperature VEV, or $v_c \lesssim v=246$ GeV. The same plots show that the critical temperatures in the prescription A are densely populated in the vicinity of $T_c \sim \mathcal{O}(100)$ GeV. 

\begin{figure}[!htb!] 
	\centering
	\includegraphics[width=0.32\linewidth]{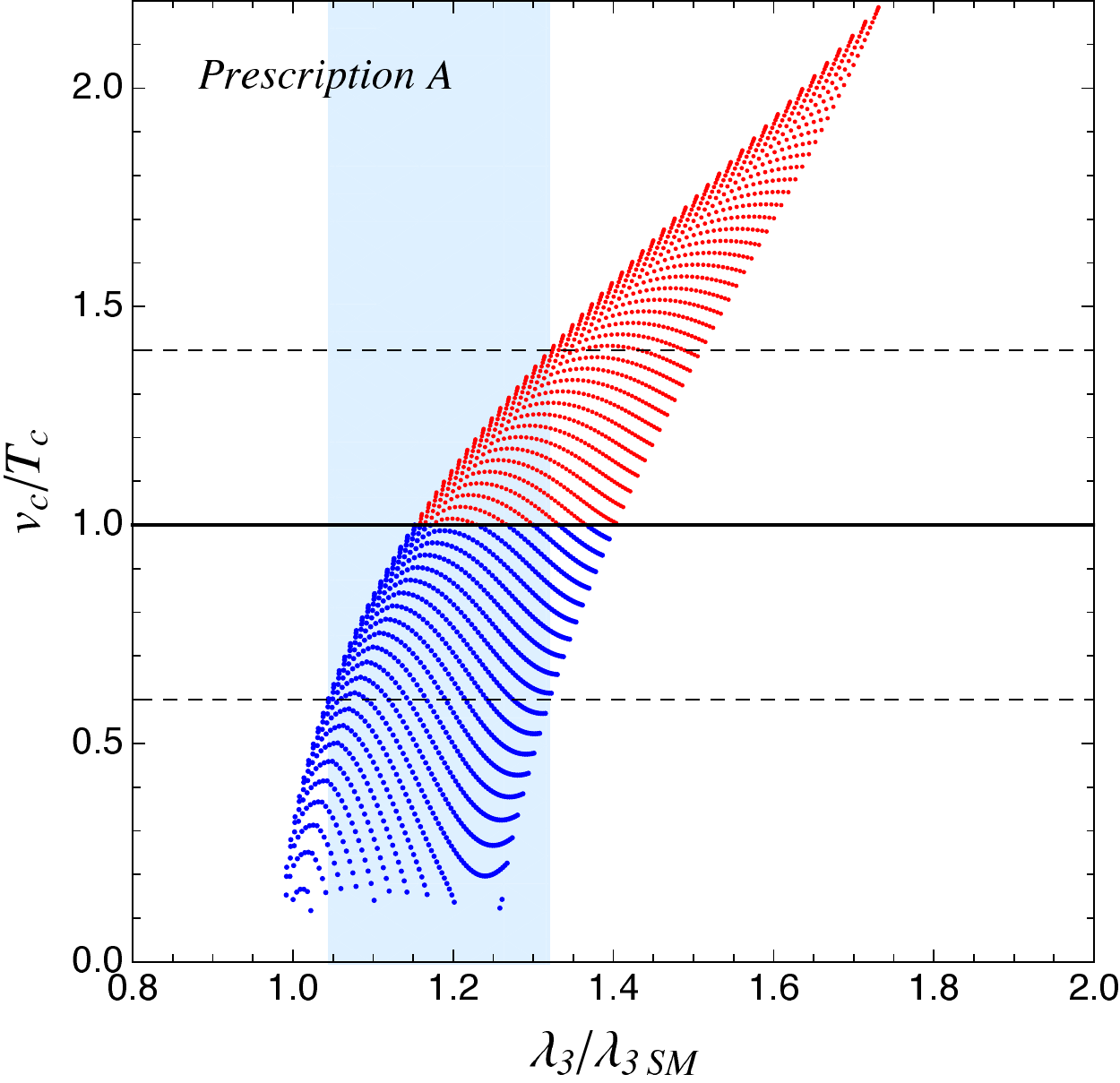} 
	\includegraphics[width=0.32\linewidth]{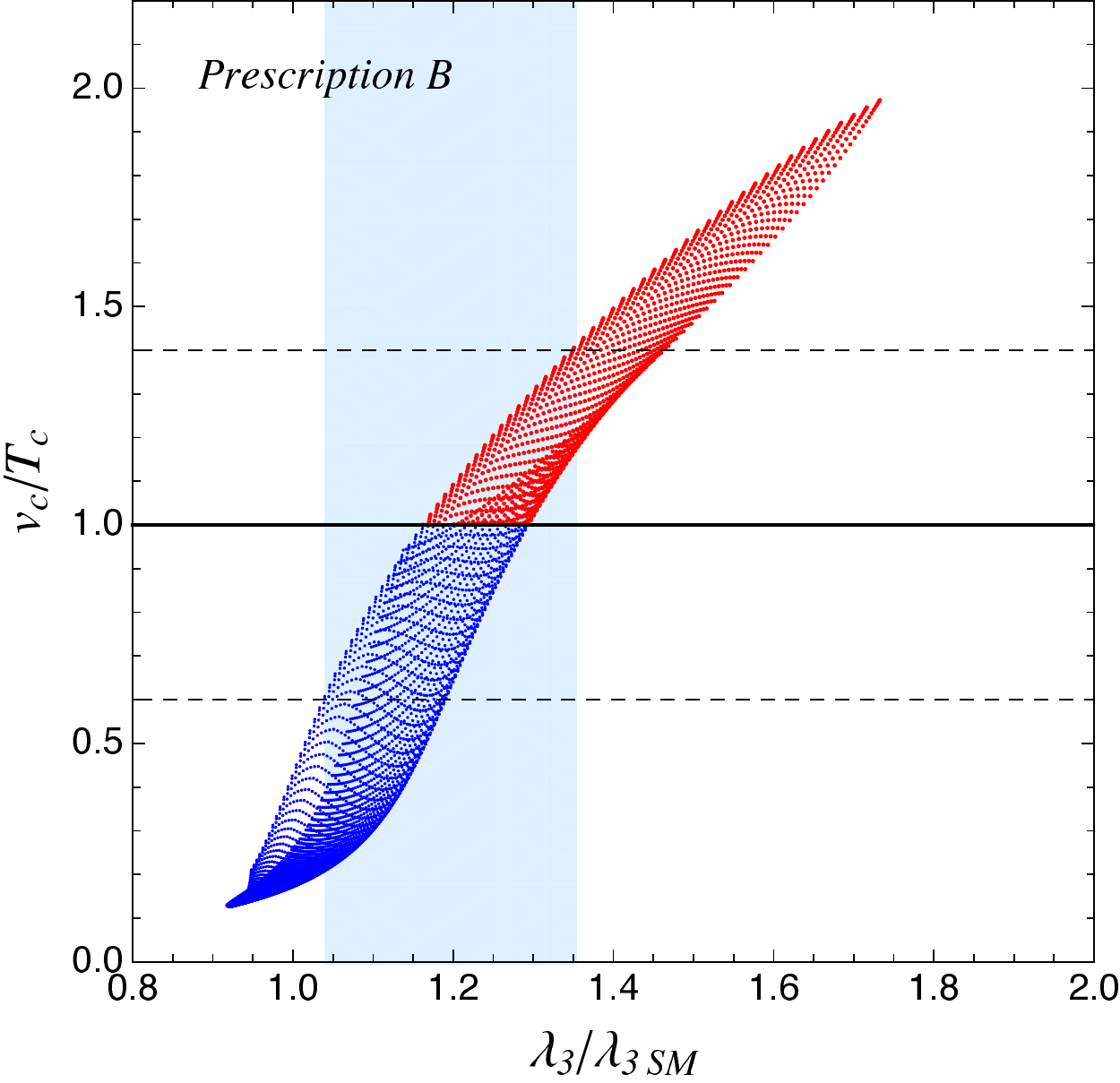} 
	\includegraphics[width=0.32\linewidth]{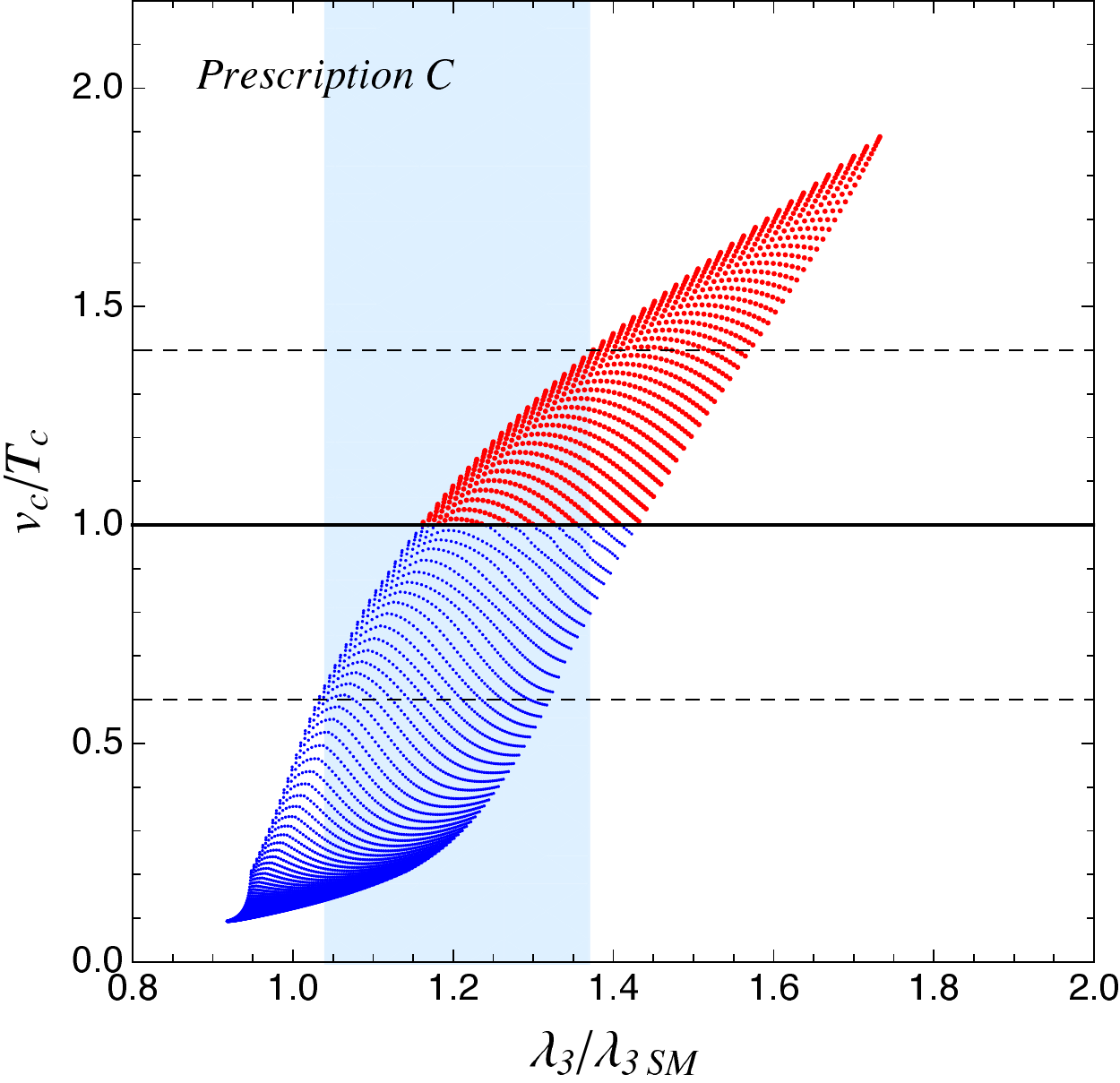}
	\caption{The correlation between $v_c/T_c$ and $\lambda_3/\lambda_{3\, SM}$ for one-step phase transition in three different prescriptions. The color is divided by $v_c/T_c = 1$. In all plots, the bare mass was scanned over the window $\mu_S = [10, 900]$ GeV in steps of 10 GeV. The light-blue band represents the variation of the highest precision of $\lambda_3/\lambda_{3\, SM}$ corresponding to the variation of the criteria on $v_c/T_c$ in the interval $v_c/T_c \gtrsim [0.6\, 1.4]$.}
	\label{fig:vTratioHiggsSelf}
\end{figure}

We estimate the highest precision of the Higgs self couplings that future colliders need to achieve to rule out the considered scenario. The situation is illustrated in Fig.~\ref{fig:vTratioHiggsSelf}. The highest precision corresponds to the left boundaries of both plots in Fig.~\ref{fig:vTratioHiggsSelf} whose contributions are due to the lower singlet masses
\footnote{In the first two plots of Fig.~\ref{fig:vTratioHiggsSelf:msrange} in Appendix~\ref{app:nondecoupling}, we present similar plots to Fig.~\ref{fig:vTratioHiggsSelf} but including the contributions only from the low mass range, $\mu_S = [10, \, 90]$ GeV, in both prescriptions. They clearly show that the lower masses are responsible for the vicinity of the smallest deviation of the Higgs cubic coupling.} -- exactly where the high-temperature approximation relatively works the best amongst viable parameter space in Fig.~\ref{fig:vTratioHiggsSelf} and three prescriptions agree well
\footnote{The typical $m_s(v_c)/T_c$ is roughly 2 along the left boundaries which indicates relatively good agreement between the high-$T$ approximation and the exact evaluation according to Fig.~\ref{fig:VTExactVSapprox}.}. 
Demanding $v_c/T_c \gtrsim 1$ translates to $\sim 15$\% deviation of the Higgs cubic coupling as the smallest in all prescriptions as is seen in Fig.~\ref{fig:vTratioHiggsSelf}. As was mentioned in Section~\ref{sec:ProsFutureCollider}, the 15\% precision can be achieved in both 100 TeV $pp$ collider and ILC (via variety of the processes).
On demanding $v_c/T_c \gtrsim 0.6$, ruling out the scenario requires $\sim$ 5\% precision of the Higgs cubic coupling and this precision can be achieved only at the 100 $pp$ collider~\cite{Contino:2016spe}. In other words, the exact numerical criteria on the $v_c/T_c$ has a drastic impact on the prospect for future colliders. An interesting question regarding this observation will be to know if the $v_c/T_c$ has a preferred value in this specific nightmare scenario instead of equally probable values in the finite range.

\begin{figure}[!htb!] 
	\centering
	\includegraphics[width=0.32\linewidth]{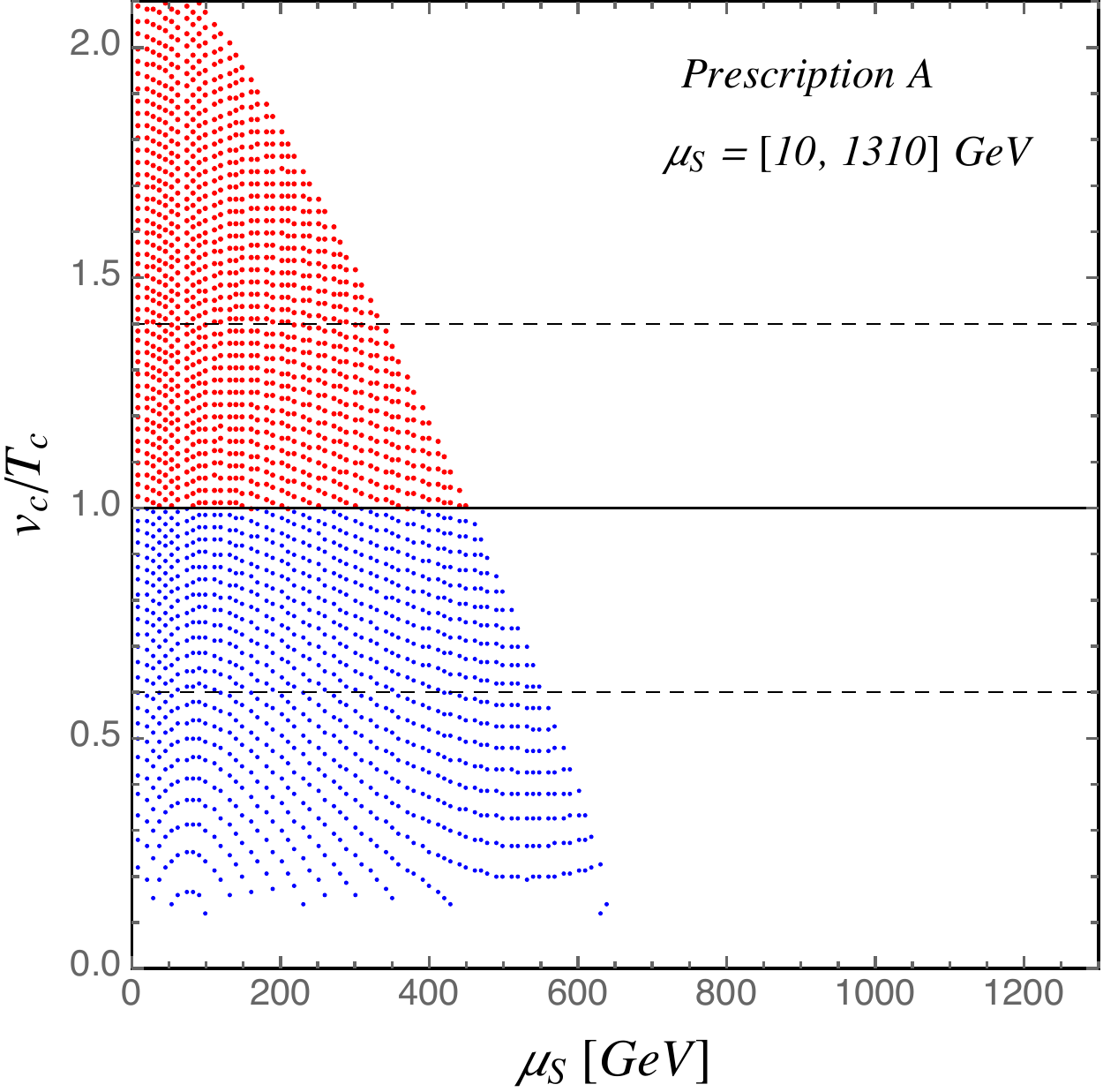}
	\includegraphics[width=0.32\linewidth]{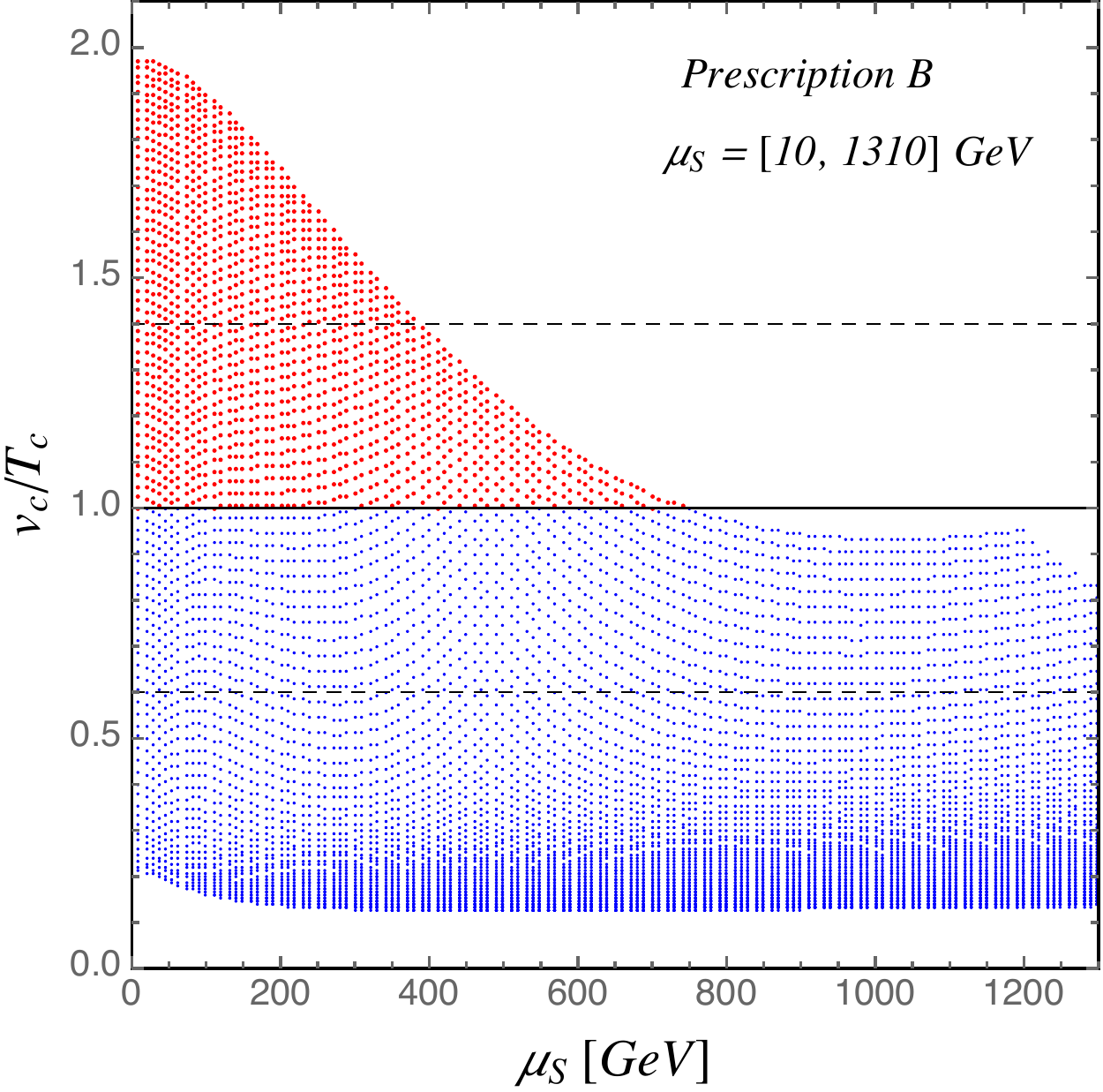}
	\includegraphics[width=0.32\linewidth]{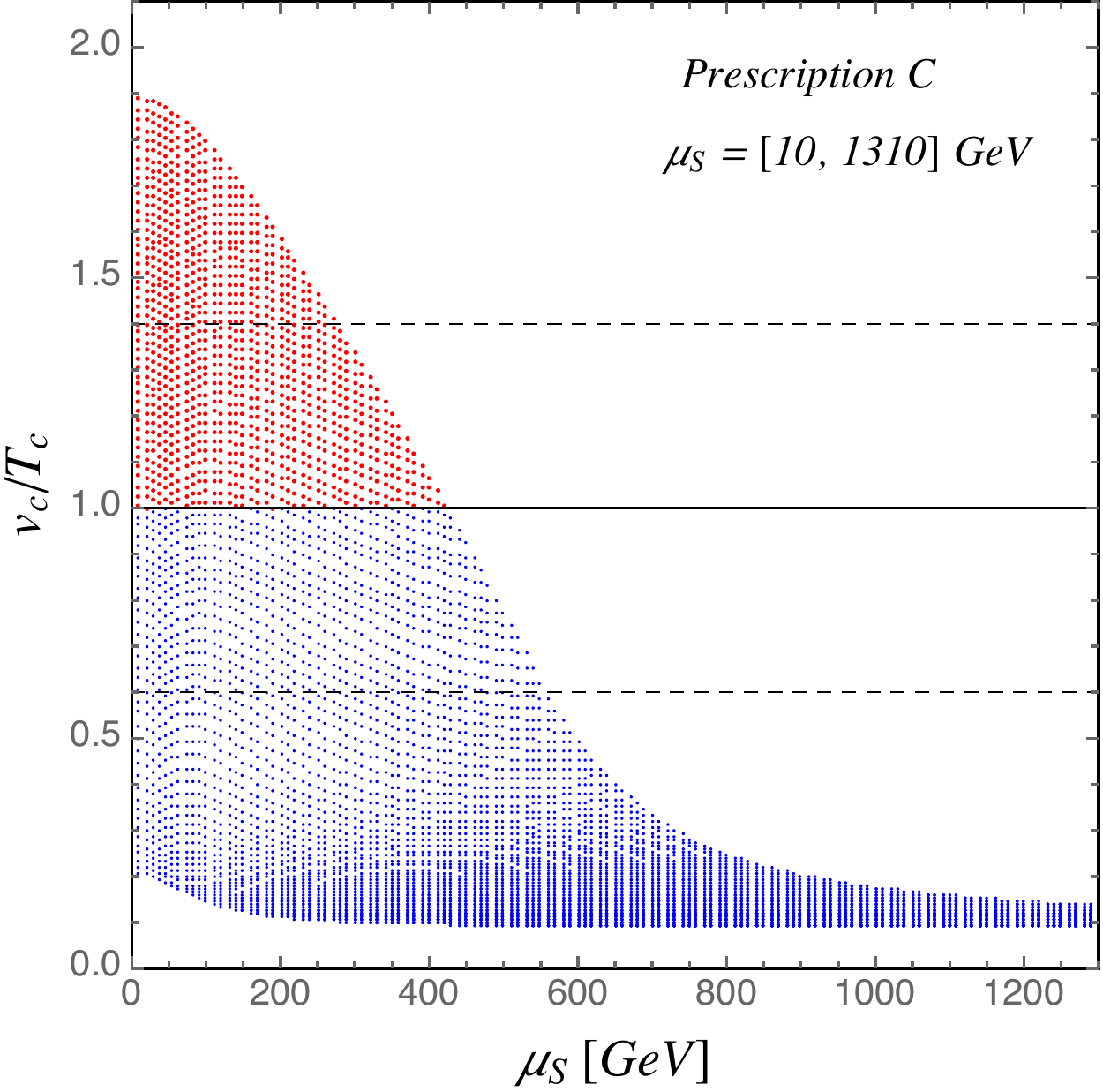}
	\caption{The illustration of the decoupling (non-decoupling) of the high singlet masses for one-step phase transition in prescription A (prescription B). In all plots, the bare mass was scanned over the window $\mu_S = [10, 1310]$ GeV in steps of 10 GeV and $\lambda_{HS} = [0,\, 5]$ (in steps of 0.05).}
	\label{fig:highmass:decoupling}
\end{figure}

\begin{figure}[!htb!] 
	\centering
	\includegraphics[width=0.4\linewidth]{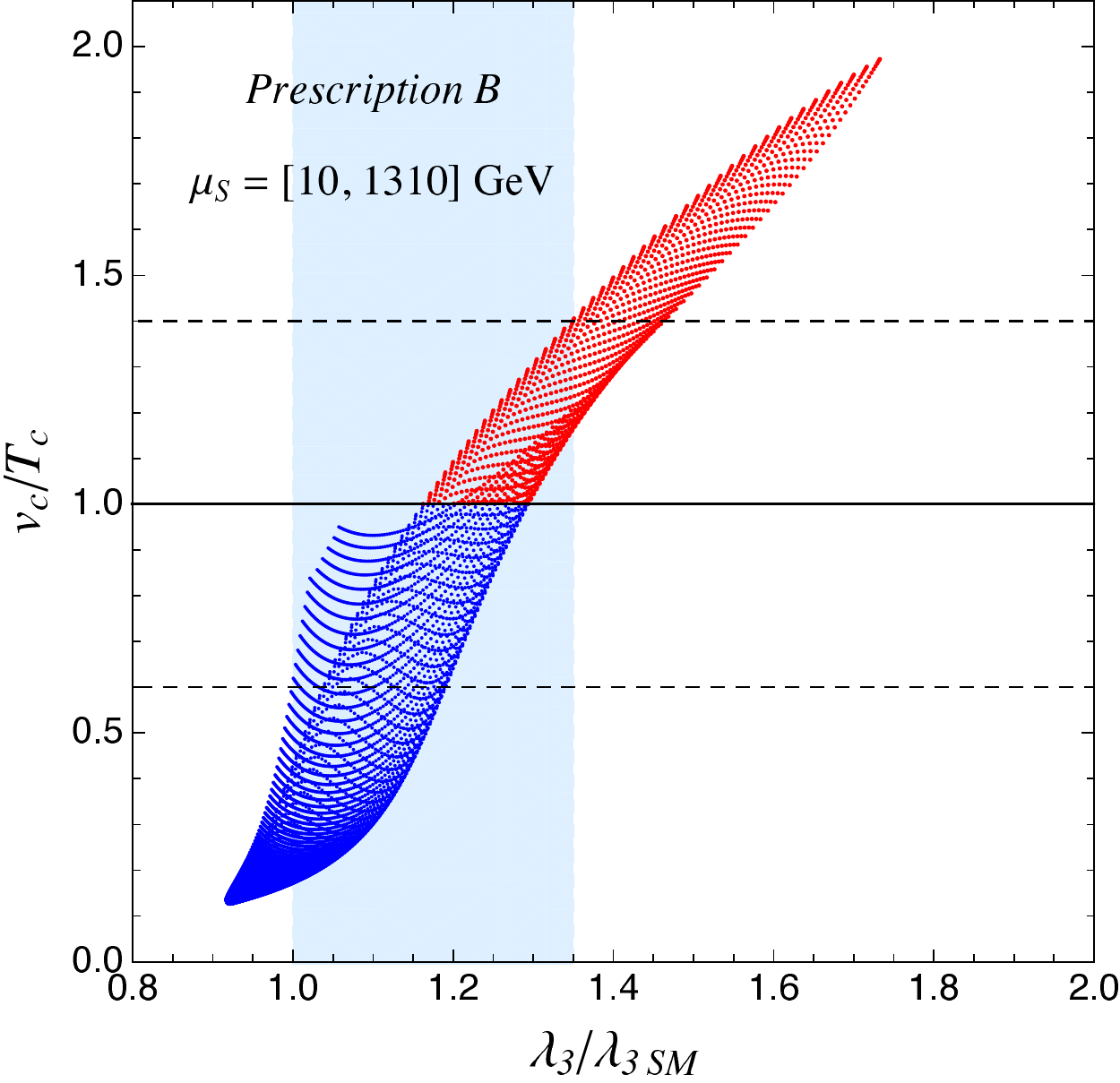}
	\caption{A similar plot to Fig.~\ref{fig:vTratioHiggsSelf} but with $\mu_S = [10,\, 1310]$ GeV.}
	\label{fig:highT:bigbaremass}
\end{figure}

In making plots in Fig.~\ref{fig:vTratioHiggsSelf}, we have included only the bare singlet masses in the window $\mu_S = [10, \, 900]$ GeV. While the contribution from higher singlet masses to the $v_c/T_c \gtrsim [0.6,\, 1.4]$ region naturally decouples around $\mu_S \sim 550$ GeV in prescription A and C, a similar decoupling does not occur in the prescription B using the high-$T$ approximation of the thermal potential~\footnote{The non-decoupling behavior we found is not the same non-decoupling issue addressed in the literature (see for example~\cite{Curtin:2016urg}, where different type of non-decoupling issue is mentioned): since the high-temperature approximation enters into the effective potential either indirectly via the truncated thermal mass or directly via the approximated thermal potential itself, it reveals different form of non-decouplings. The non-decoupling of the heavy singlet mentioned is about the truncated thermal mass at leading order in high-temperature approximation. Since the leading order thermal mass is proportional to the temperature without depending on the singlet mass, the thermal mass does not show the decoupling behavior even when the singlet mass approaches to the infinity. This type of non-decoupling universally exists in all prescriptions as long as one uses the leading order thermal mass (as commonly done in most literature). On the other hand, the non-decoupling of the heavy singlet that we newly addressed in our work is apparent only in {\bf prescription B}, and it is very different aspect of the high-temperature approximation of the thermal potential.} . Instead, the higher masses in the prescription B continue to contribute to the SFOEPT parameter space, as is seen in Figs.~\ref{fig:highmass:decoupling} and~\ref{fig:highT:bigbaremass}, and severely affect the precision of $\lambda_3$ when using rather conservative criterion of $v_c/T_c > 0.6 - 0.9$.
If more commonly used $v_c/T_c > 1$ is adopted as the criterion for the SFOEPT, this problem will not affect the observable such as the Higgs self coupling. While the newly added region, [900, 1310] GeV, corresponds to bigger $m_s(v_c)/T_c$ where the high-$T$ approximation fails more badly (thus one should not trust the result), it is interesting to observe that their contribution severely distorts the left boundary of the plot where the precision was previously set by the contribution from lower singlet masses. For clarification, a separate plot showing only the contribution from $\mu_S = [910,\, 1310]$ GeV using the prescription B is presented in the last panel of Fig.~\ref{fig:vTratioHiggsSelf:msrange} in Appendix~\ref{app:nondecoupling}.

\begin{figure}[!htb!] 
	\centering
	\includegraphics[width=0.40\linewidth]{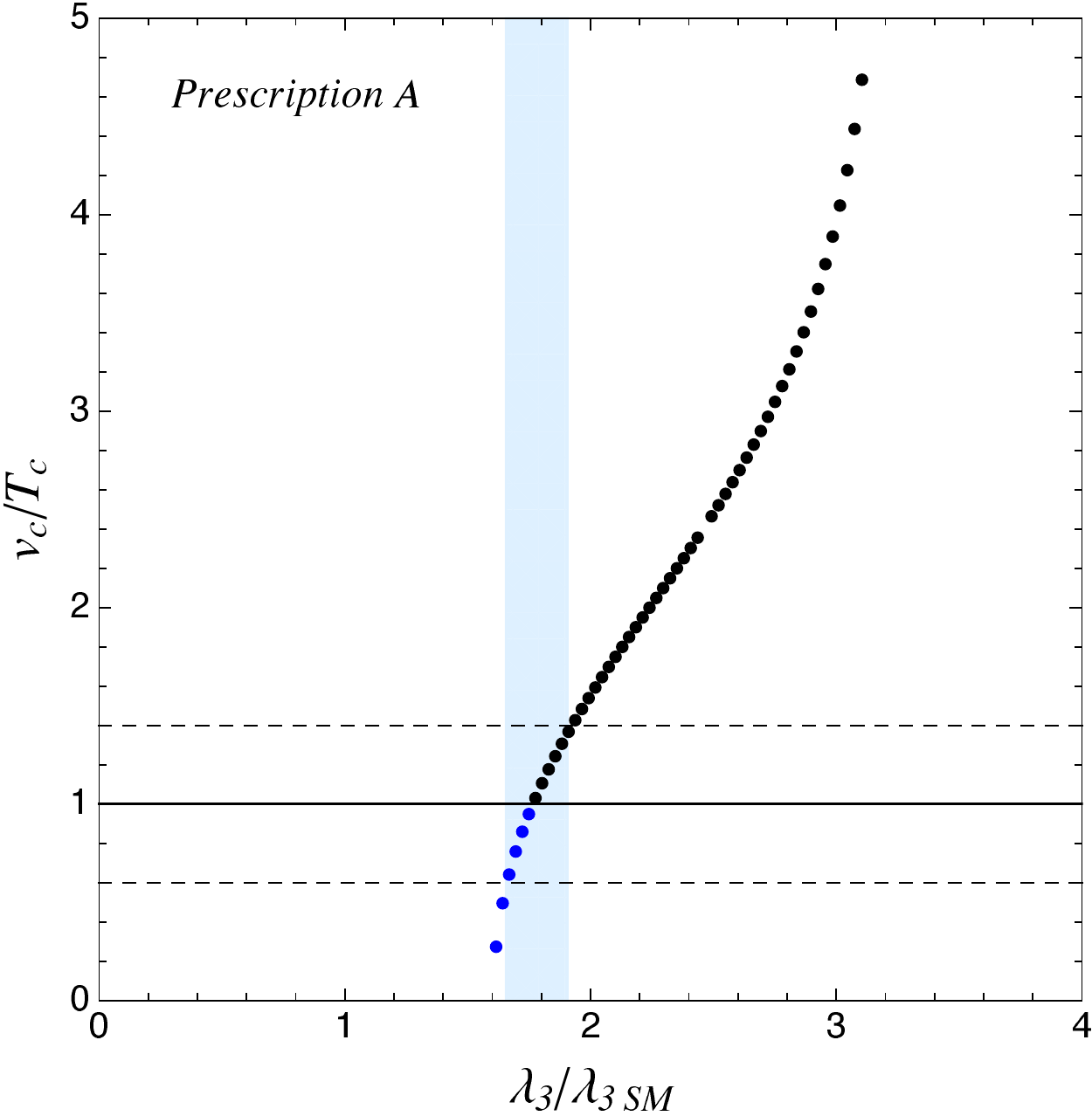} \quad
	\includegraphics[width=0.40\linewidth]{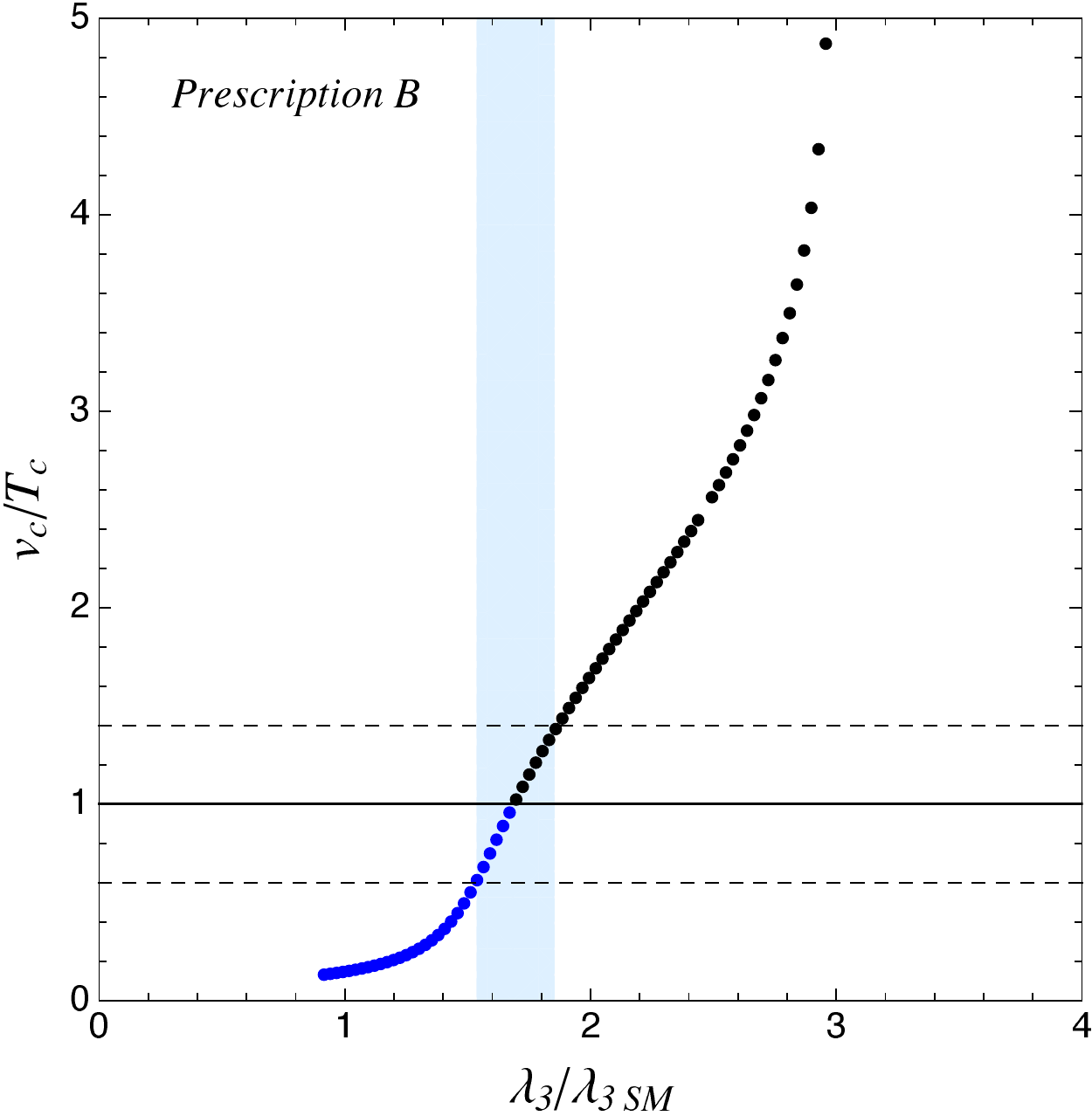}
	\caption{The correlation between $v_c/T_c$ and $\lambda_3/\lambda_{3\, SM}$ for the EFT approach only with $\mathcal{O}_6$ operator in two different prescriptions. The color is divided by $v_c/T_c = 1$. The light-blue band represents the variation of the highest precision of $\lambda_3/\lambda_{3\, SM}$ corresponding to the variation of $v_c/T_c$ in the interval $v_c/T_c \gtrsim [0.6,\, 1.4]$.}
	\label{fig:vTratioHiggsSelf:EFT}
\end{figure}
\begin{figure}[!htb!] 
	\centering
	\includegraphics[width=0.40\linewidth]{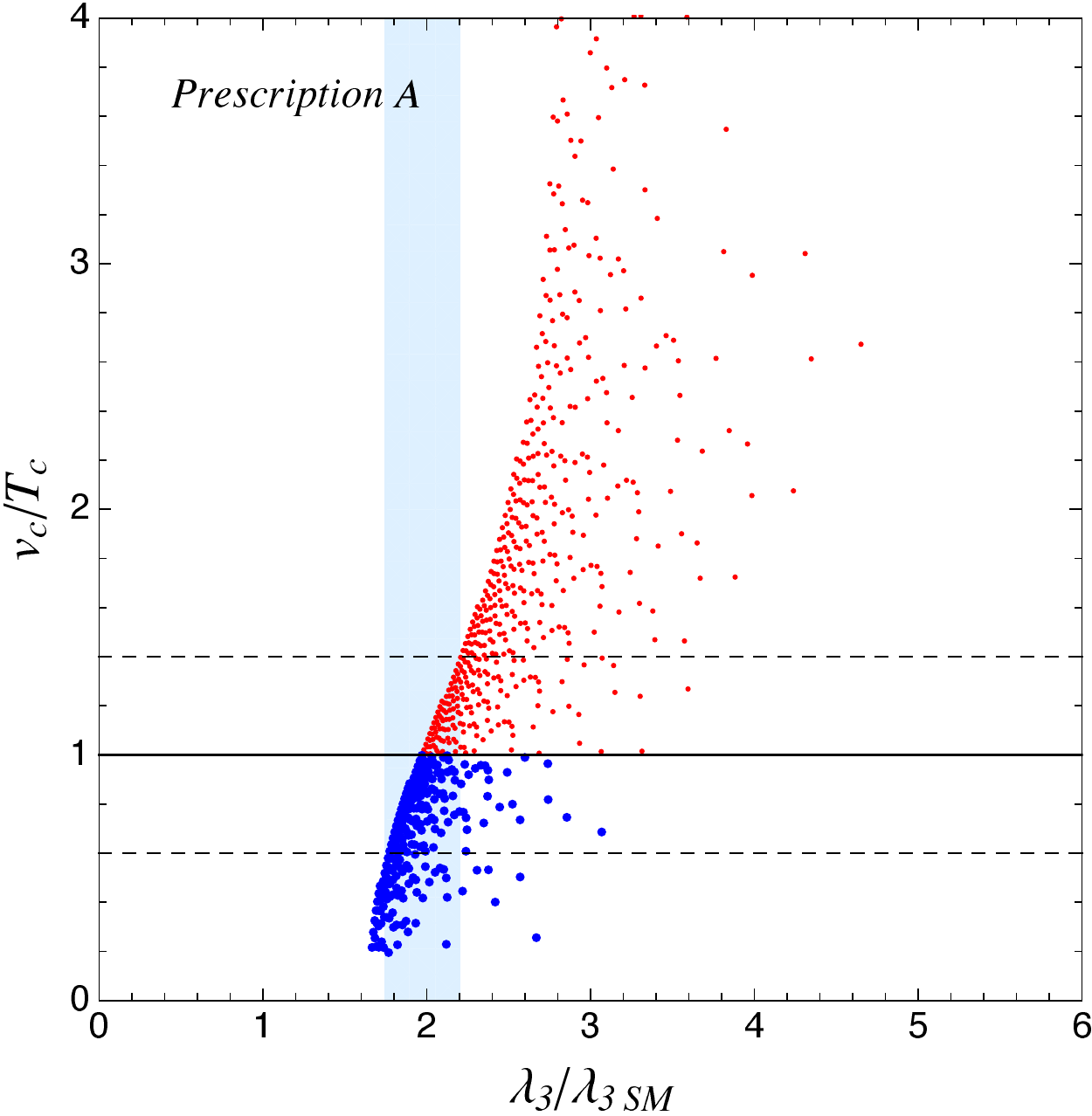} \quad
	\includegraphics[width=0.40\linewidth]{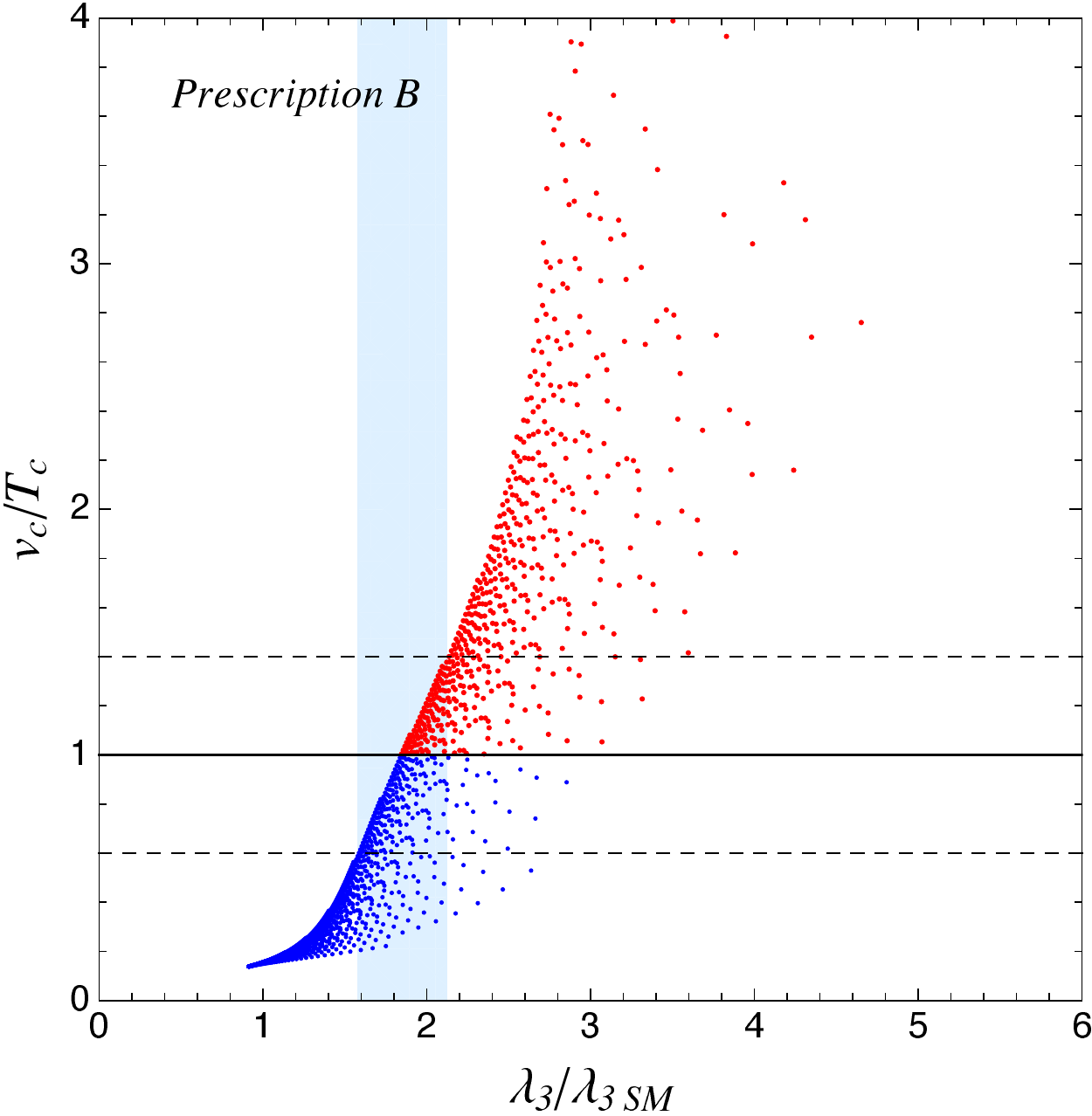}
	\caption{The correlation between $v_c/T_c$ and $\lambda_3/\lambda_{3\, SM}$ for the EFT approach with the re-summed higher-dimensional operators with universal coefficients in two different prescriptions. The color is divided by $v_c/T_c = 1$. The light-blue band represents the variation of the highest precision of $\lambda_3/\lambda_{3\, SM}$ corresponding to the variation of $v_c/T_c$ in the interval $v_c/T_c \gtrsim [0.6,\, 1.4]$ -- they are $[1.77,\, 2.2]$ for the prescription A and $[1.58,\, 2.12]$ for the prescription B.}
	\label{fig:vTratioHiggsSelf:EFTallOrder}
\end{figure}

We show the correlation between $v_c/T_c$ and $\lambda_3/\lambda_{3\, SM}$ for the EFT approach with the $\mathcal{O}_6$ operator in Fig.~\ref{fig:vTratioHiggsSelf:EFT}. Similar results for the EFT approach with re-summed higher-dimensional operators are presented in Fig.~\ref{fig:vTratioHiggsSelf:EFTallOrder}. As is evident in Fig.~\ref{fig:vTratioHiggsSelf:EFT} and Fig.~\ref{fig:vTratioHiggsSelf:EFTallOrder}, the viable parameter spaces in two prescriptions look similar in the region of interest although the prescription A favors slightly larger values. For instance, while we read off $\lambda_3/\lambda_{3\, SM} \gtrsim [1.66,\, 1.91]$ in Fig.~\ref{fig:vTratioHiggsSelf:EFT} for $v_c/T_c \gtrsim [0.6,\, 1.4]$ in the prescription A, it becomes $\lambda_3/\lambda_{3\, SM} \gtrsim [1.54,\, 1.85]$ in the prescription B for the same $v_c/T_c$.  For the EFT approach with the re-summed higher-dimensional operators, the corresponding values are extracted from Fig.~\ref{fig:vTratioHiggsSelf:EFTallOrder} and they are $\lambda_3/\lambda_{3\, SM} \gtrsim [1.77,\, 2.2]$ in the prescription A and $\lambda_3/\lambda_{3\, SM} \gtrsim [1.58,\, 2.12]$ in the prescription B. The similar results between two prescriptions are easily understood since there is no large $m/T$ parameter involved in this scenario unlike the $m_S(v_c)/T_c$ that became large in the Higgs portal with a singlet scalar. The smallest deviation of $\lambda_3$ in the EFT approach which reads $\sim 60\%$ can be easily accessed by any future colliders unlike the situation of the scenario with the singlet scalar which predicts $\sim 5\%$.\\

\section{Conclusion}
\label{sec:conclusion}
In this paper, we have numerically examined a few issues regarding the validity of the effective potential and its implication on the Higgs self couplings. Amongst several issues that can cause non-negligible uncertainties in the Higgs self couplings, we have focused on the validity of the high-temperature approximation and on the impact of the threshold $v_c/T_c$ value in the range of $[0.6,\, 1.4]$ on the precision of the Higgs self couplings. To this end, we have adopted three prescriptions of the effective potential which differ by the treatment of the thermal potential. We have also explored the correlation between the Higgs cubic and the quartic couplings in the scenarios where electroweak phase transition is strongly first order. We have addressed these subjects in the context of two types of BSM scenarios, namely: (i) the Higgs portal with a singlet scalar with the $\mathds{Z}_2$ symmetry and (ii) the EFT approach with higher-dimensional operators.

 We have shown that the precision of the Higgs self couplings behaves very differently under two prescriptions for the case of the Higgs portal with the singlet scalar. While the singlet mass contributing to the viable region with the one-step SFOEPT naturally decouples at high masses in the prescription using the exact evaluation of the thermal potential, a similar decoupling does not happen when using high-$T$ approximation of the thermal potential with a rather conservative criterion of $v_c/T_c \gtrsim 0.6-0.9$. 
We demonstrate that the precision of the cubic Higgs self coupling that has to be achieved to rule out a minimal singlet scalar case with the $\mathds{Z}_2$ symmetry that proceeds via one-step phase transition significantly varies depending on the $v_c/T_c$ values. While demanding $v_c/T_c > 1\, (1.4)$ requires the measurement of the coupling at $\sim 15\%$ ($35\%$) precision which is achievable at various future colliders such as ILC (via VBF process at higher c.o.m energy) and 100 TeV $pp$ collider, more conservative criteria, $v_c/T_c > 0.6$ requires $\sim 5\%$ precision of the cubic Higgs self coupling which is likely plausible only at 100 TeV $pp$ collider. 

We repeated similar exercises for the EFT approach. We found that, unlike the Higgs portal scenario, the EFT approach shows a similar pattern of the Higgs self couplings under all prescriptions in the region of interest, namely, $v_c/T_c \geq 0.6$, except overall shift of the deviation of the Higgs self couplings. We observe that the prescription using the exact thermal potential favors slightly higher deviation.  The smallest deviation of the trilinear Higgs self coupling compatible with the SFOEPT is found to be higher than about $60\%$, typically $\delta (\lambda_3/\lambda_{3\, SM}) \sim \mathcal{O}(1)$. Therefore, the EFT approach explored in this paper will be well tested in various future colliders. We have pointed out that a large fraction of the parameter space of the EFT approach can already be tested at the HL LHC at 68\% CL.

We examined the correlations between $\lambda_3/\lambda_{3\, SM}$ and $\lambda_4/\lambda_{4\, SM}$ in our benchmark scenarios when they are compatible with the SFOEPT. Interestingly, we found that the various New Physics scenarios for the SFOEPT not only appear widely separated in the two-dimensional Higgs self coupling space, but also the actual coupling sizes can be quite large enough for them to be tested via either direct or indirect measurements. For instance, we show that the quartic coupling, $\lambda_4/\lambda_{4\, SM}$, can reach very large value as big as $\mathcal{O}(1-10)$ for the EFT approach where all higher-dimensional operators are resummed assuming universal coefficients. The deviations of both Higgs cubic and quartic self couplings in the Higgs portal scenario with a singlet scalar is less pronounced. However, even in that case, we found that about a half of parameter space allowed for the SFOEPT has a meaningful sensitivity for the Higgs quartic coupling at the 100 TeV $pp$ collider. And therefore, Higgs quartic coupling measurement will be relevant for the study of the SFOEPT at the 100 TeV $pp$ collider.

In this work, we have not considered the impact on the Higgs self couplings caused by the discrepancy between the critical temperature, $T_c$, and the nucleation temperature, $T_N$, as well as the effect due to the finite range of $\mathcal{S}_3/T_N$ which is used to determine $T_N$. We have also not studied an issue caused by the coupling $\lambda_{HS}$ with order one strength at the finite temperature in the context of our specific BSM scenarios. Addressing the above issues requires a reliable computation of the tunneling rate from the electroweak symmetric vacua to the broken vacua in the first order phase transition\footnote{See~\cite{Spannowsky:2016ile,Gan:2017mcv} for the recent computations of the sphaleron in the context of the composite (and non-standard) Higgs models and the EFT with dimension-six operators.}, and the systematic classification of all non-negligible thermal diagrams and the re-summation of those diagrams~\cite{Curtin:2016urg}. Including all sources of uncertainties (as well as what we have considered in this work) at the same time to accurately estimate the impact of them on the precision of the Higgs self couplings will be an important exercise to know the plausibility of testing BSM scenarios in various future colliders.


\section*{Acknowledgements}
MS thanks Roberto Contino for useful suggestions. The authors are grateful to the CERN-CKC for its hospitality where part of this project was done. This work was supported by the National Research Foundation of Korea (NRF) grant funded by the Korea government (MEST) (No. NRF-2015R1A2A1A15052408). MS was supported by Samsung Science and Technology Foundation under Project Number SSTF-BA1602-04. SL was also supported in part by the Korean Research Foundation (KRF) through the Korea-CERN collaboration program (NRF-2016R1D1A3B01010529). The work of BJ was partially supported by the  S\~ao Paulo
Research Foundation (FAPESP) under Grants No. 2016/01343-7  and No. 2017/05770-0.

\begin{appendix}

\section{One-step SFOEPT in Higgs portal with a singlet scalar}
\label{app:onestep}
\begin{figure}[!htb!] 
	\centering
	\includegraphics[width=0.30\linewidth]{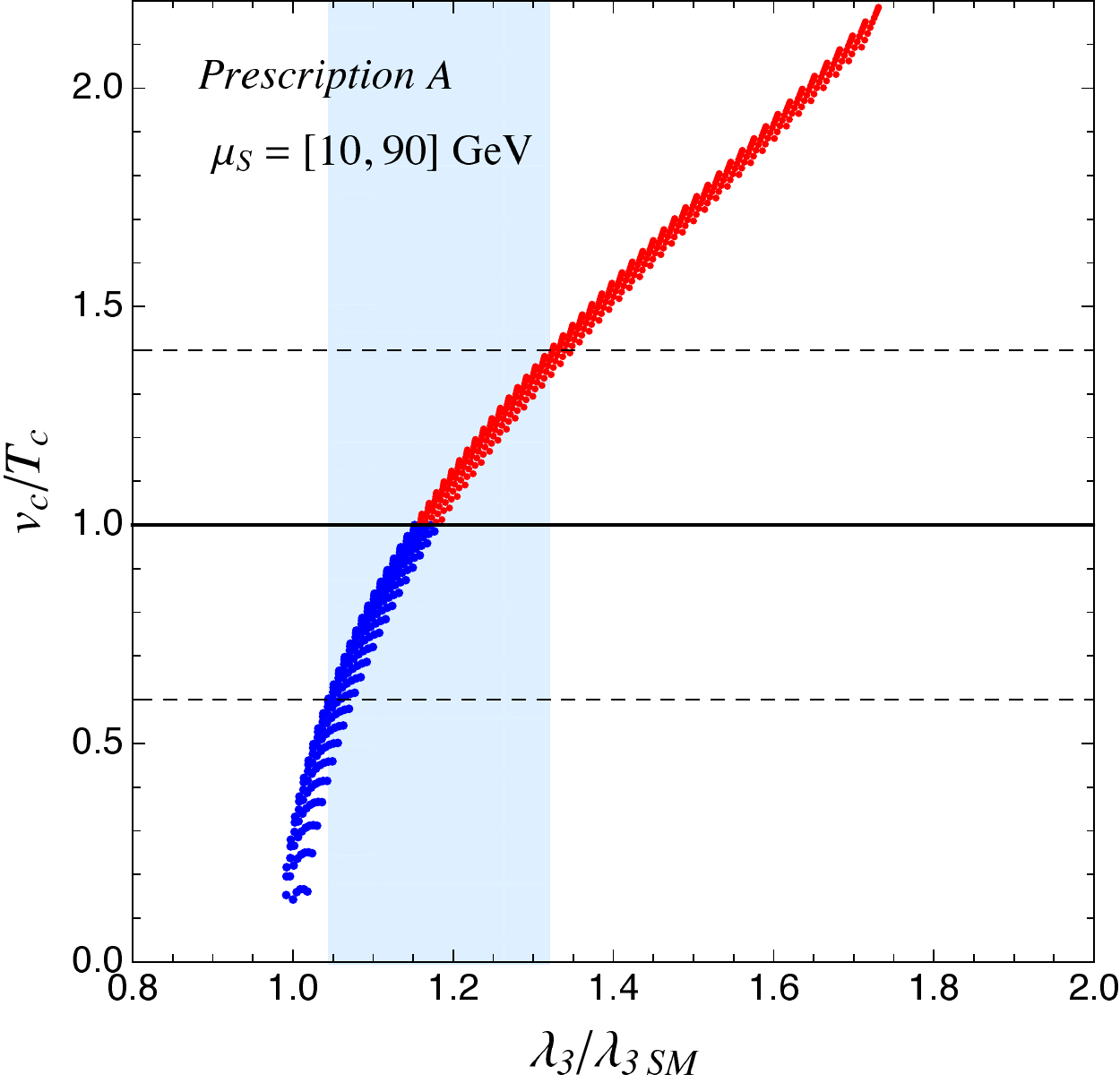} \quad
	\includegraphics[width=0.30\linewidth]{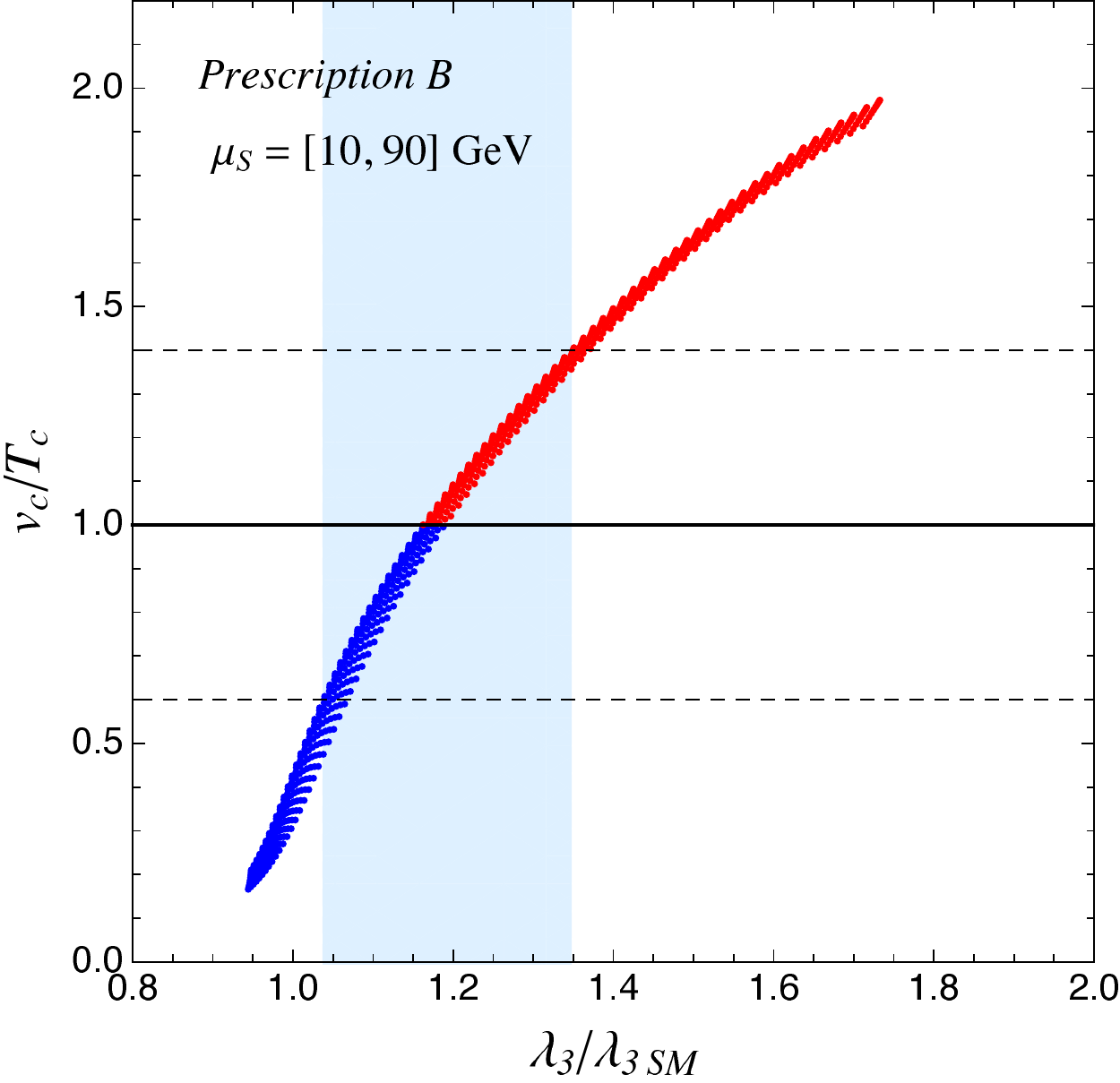} \quad
	\includegraphics[width=0.30\linewidth]{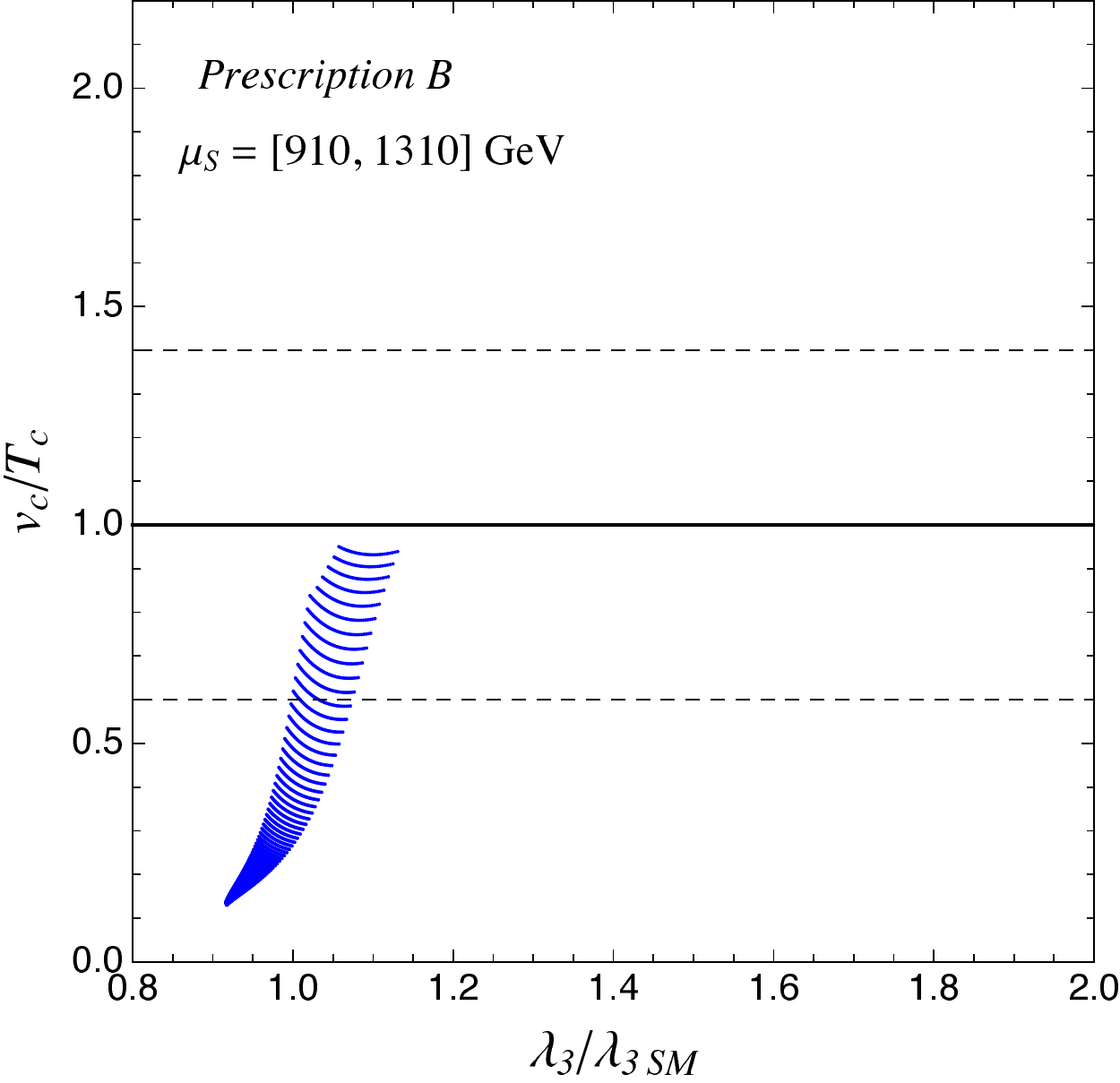}
	\caption{The correlation between $v_c/T_c$ and $\lambda_3/\lambda_{3\, SM}$ for one-step SFOEPT in two different prescriptions. The color is divided by $v_c/T_c = 1$. In first two plots (last plot), the bare mass was scanned over the window $\mu_S = [10, 90]$ GeV ($[910,\, 1310]$ GeV). The light-blue band represents the variation of the highest precision of $\lambda_3/\lambda_{3\, SM}$ corresponding to the variation of the criteria on $v_c/T_c$ in the interval $v_c/T_c \gtrsim [0.6\, 1.4]$.}
	\label{fig:vTratioHiggsSelf:msrange}
\end{figure}

In Fig.~\ref{fig:vTratioHiggsSelf:msrange}, we present the similar plots to Fig.~\ref{fig:vTratioHiggsSelf} but with the contributions from the mass region, $\mu_S = [10,\, 90]$ GeV in the first two plots and with the contribution from the mass region, $\mu_S =[910,\, 1310]$ GeV, in the last plot.
\label{app:nondecoupling}

\section{Higgs portal with a singlet scalar in the prescription B}
\label{app:precB}
\begin{figure}[!htb!] 
	\centering
	\includegraphics[width=0.30\linewidth]{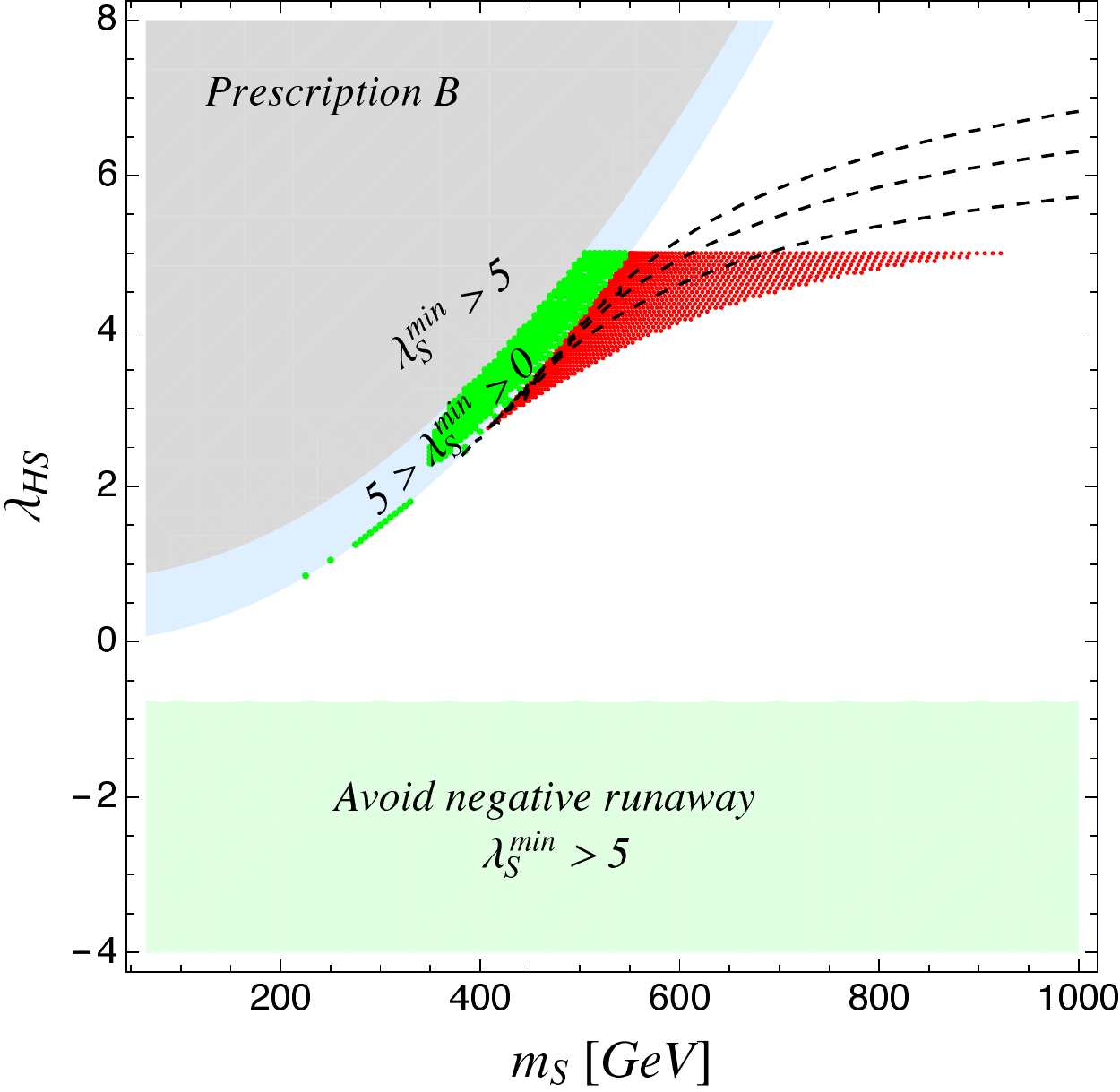}\quad 
	\includegraphics[width=0.2980\linewidth]{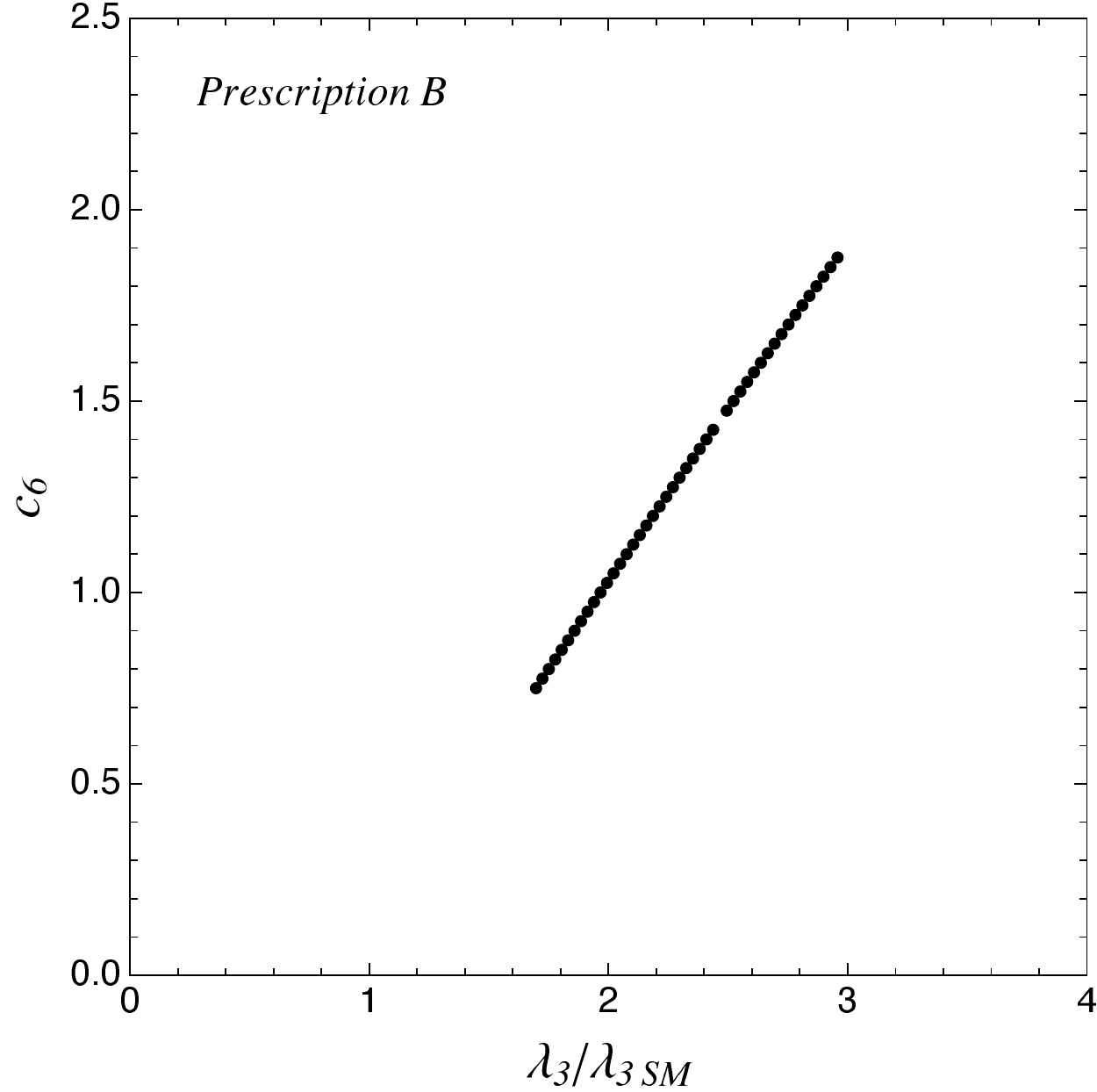}\quad
	\includegraphics[width=0.2868\linewidth]{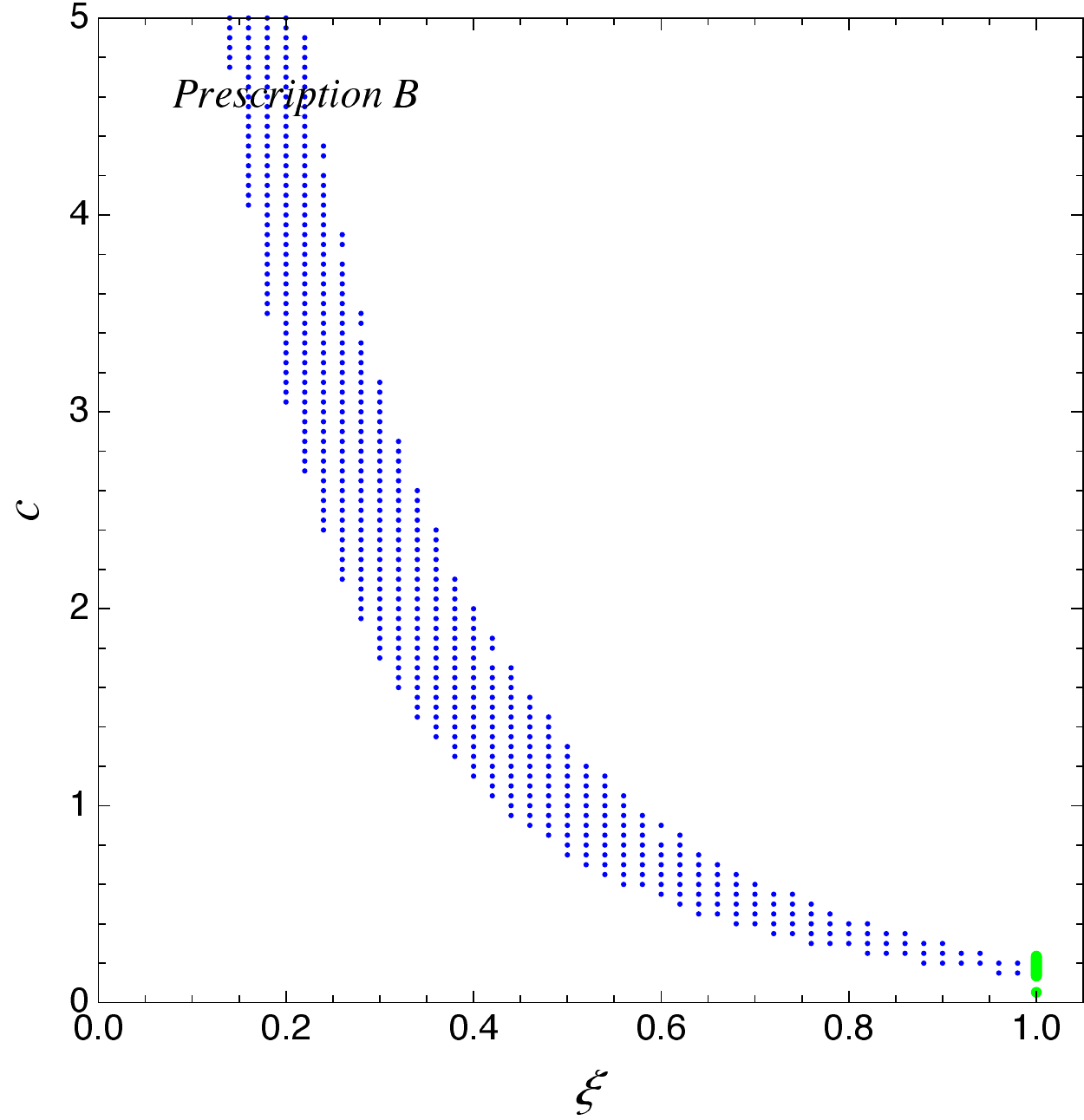}
	\caption{Similar plots to Fig.~\ref{fig:lamHSvsMS} with the prescription B.}
	\label{fig:lamHSvsMS:presB}
\end{figure}

\begin{figure}[!htb!] 
	\centering
	\includegraphics[width=0.301\linewidth]{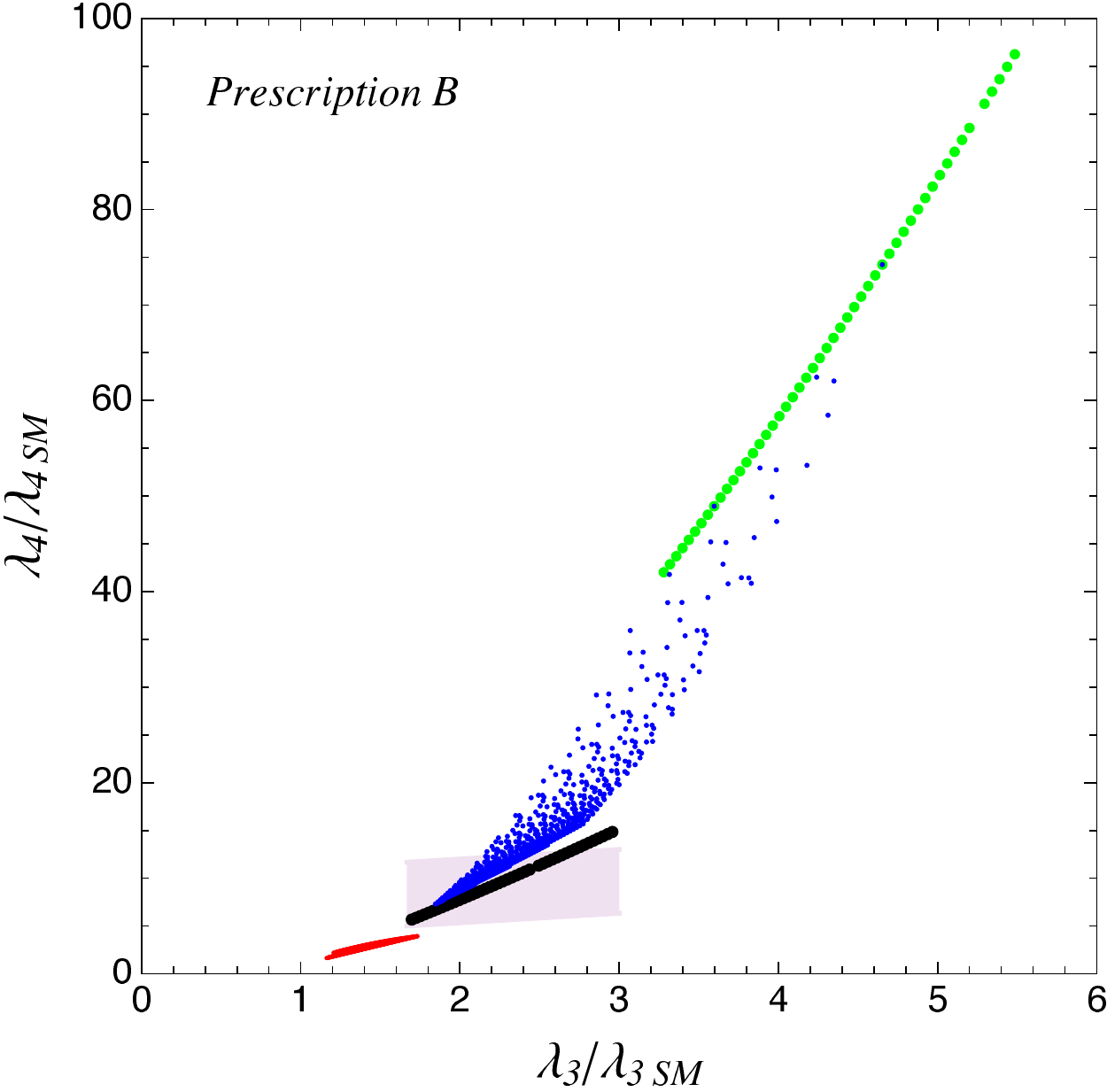}\quad
	\includegraphics[width=0.295\linewidth]{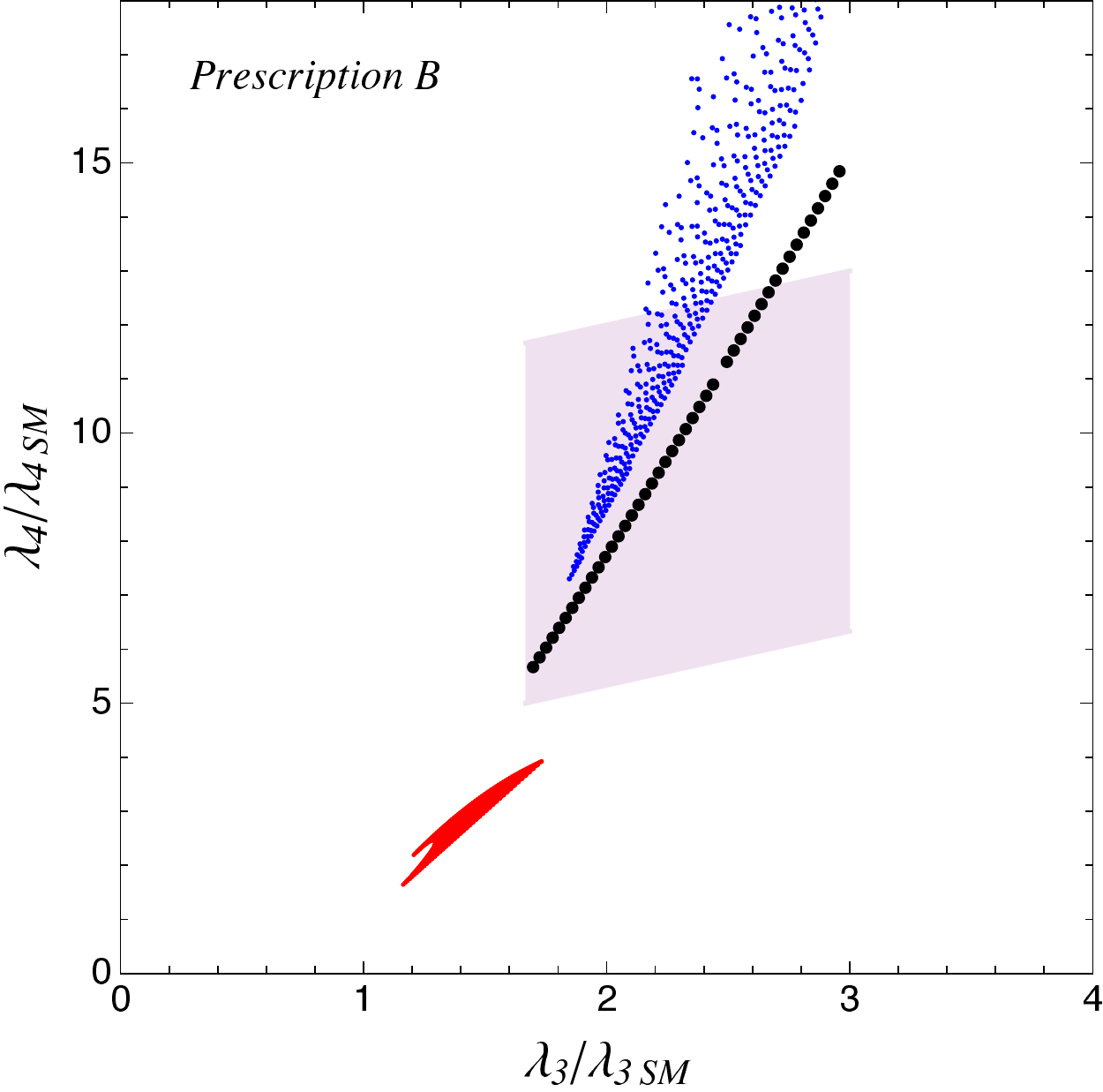}\quad
	\includegraphics[width=0.303\linewidth]{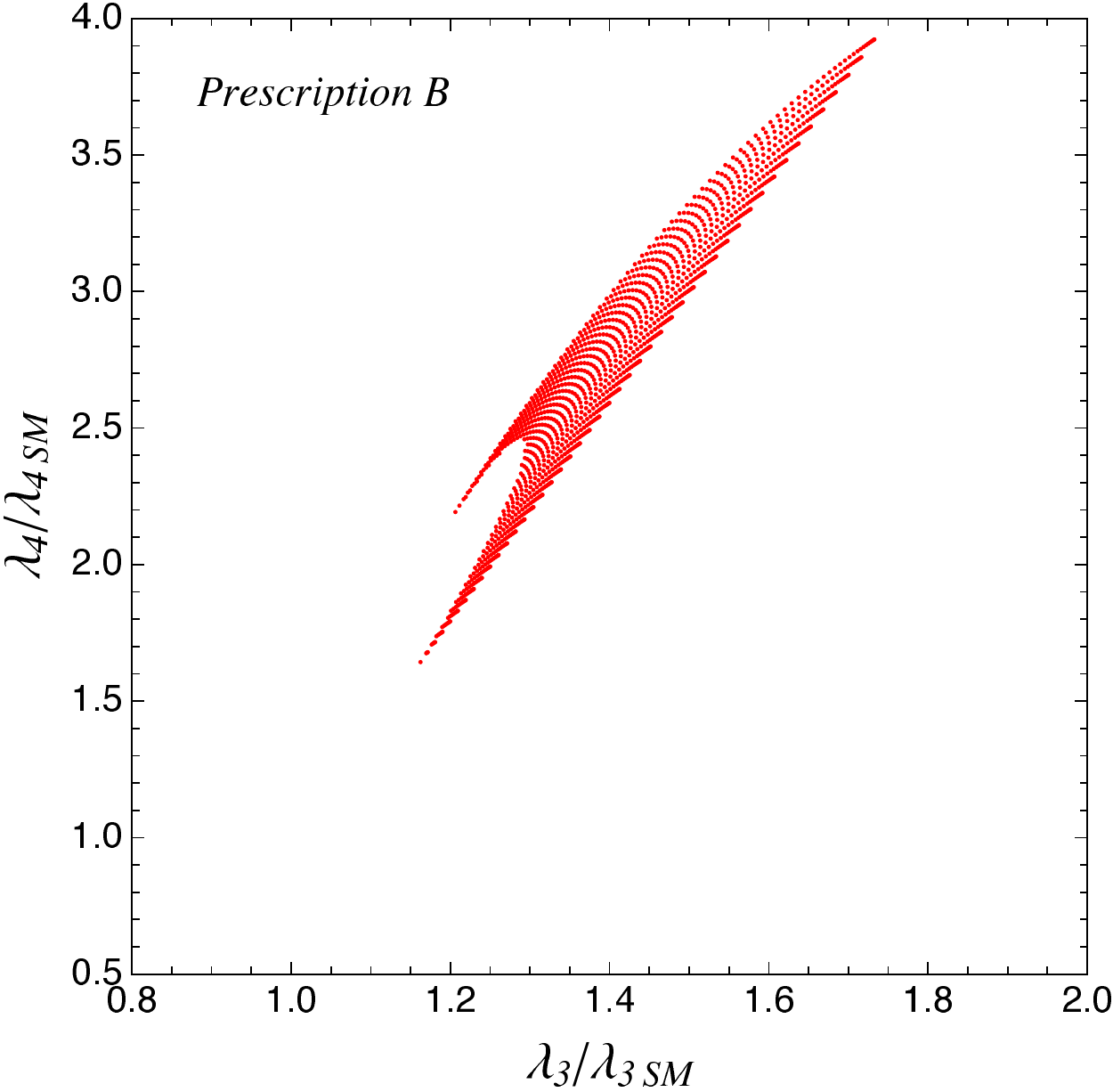}
	\caption{Similar plots to Fig.~\ref{fig:lam3VSlam4} with the prescription B.}
	\label{fig:lam3VSlam4:presB}
\end{figure}

In Fig.~\ref{fig:lamHSvsMS:presB} (Fig.~\ref{fig:lam3VSlam4:presB}), we present similar plots to Fig.~\ref{fig:lamHSvsMS} (Fig.~\ref{fig:lam3VSlam4}) but made using the prescription B. In the prescription B, the thermal potential in the high-$T$ approximation is used as in Eq.~(\ref{eq:VT:highTapprox}).
\begin{figure}[!htb!] 
	\centering
	\includegraphics[width=0.240\linewidth]{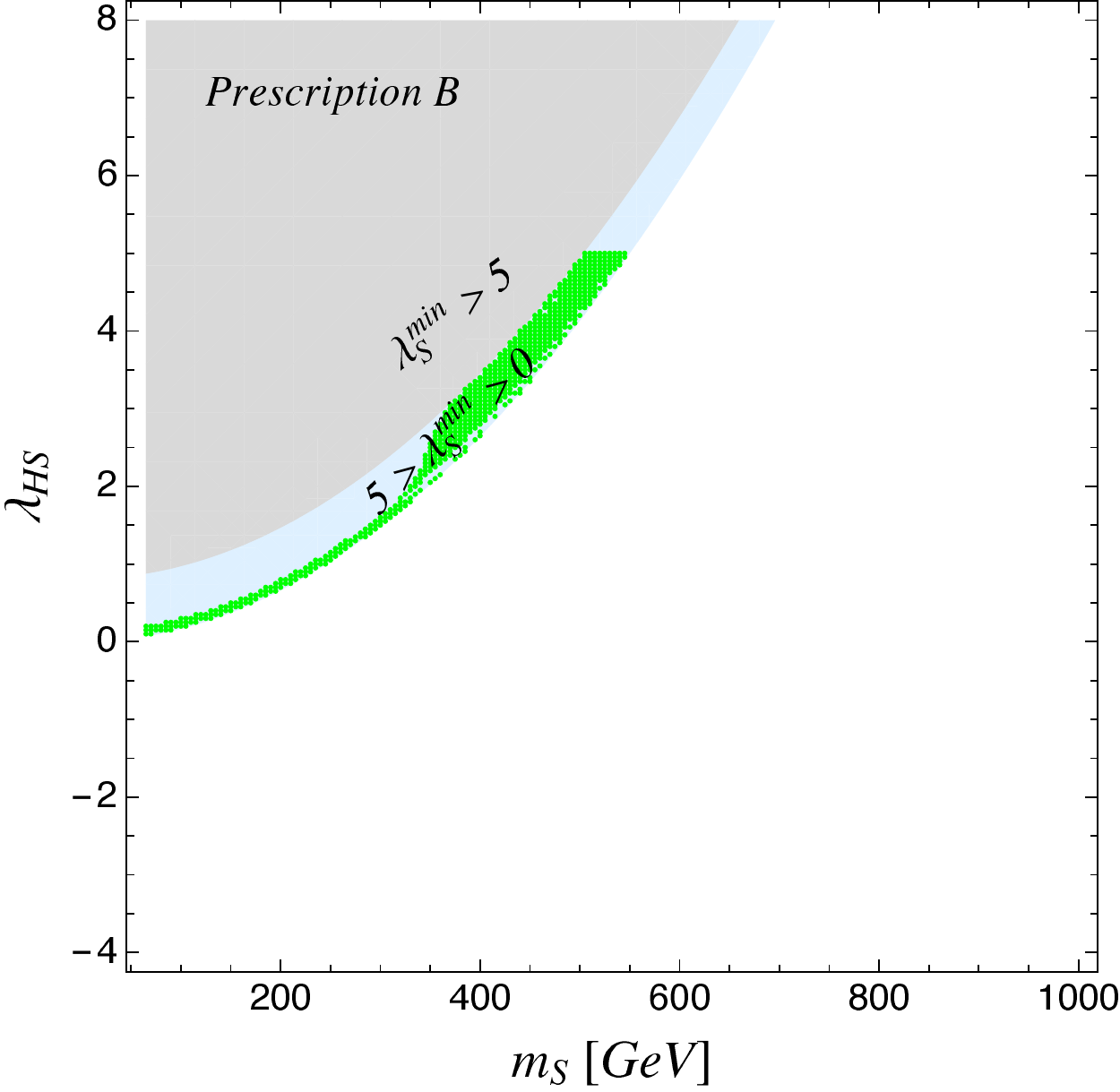}
	\includegraphics[width=0.240\linewidth]{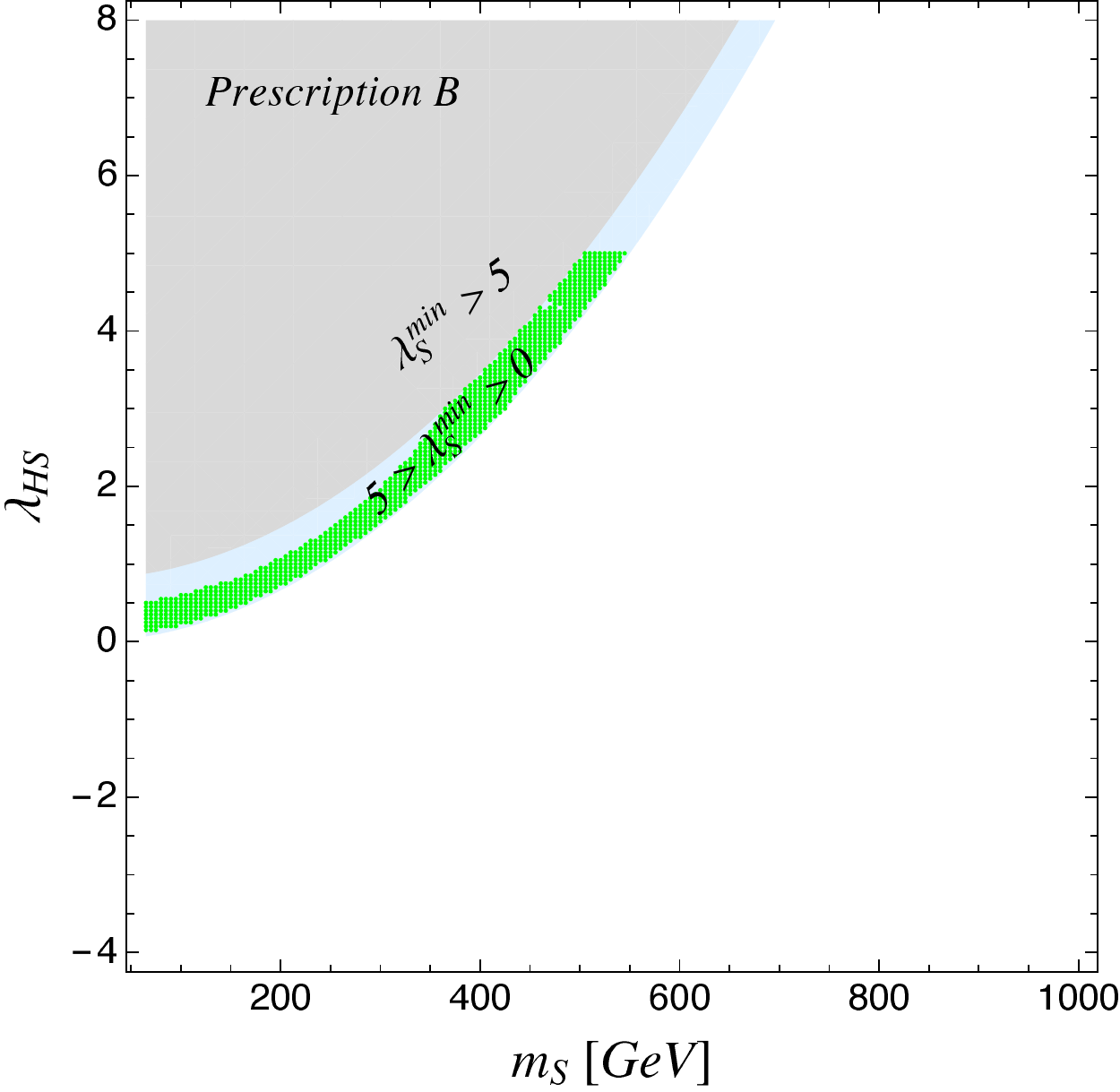}
	\includegraphics[width=0.240\linewidth]{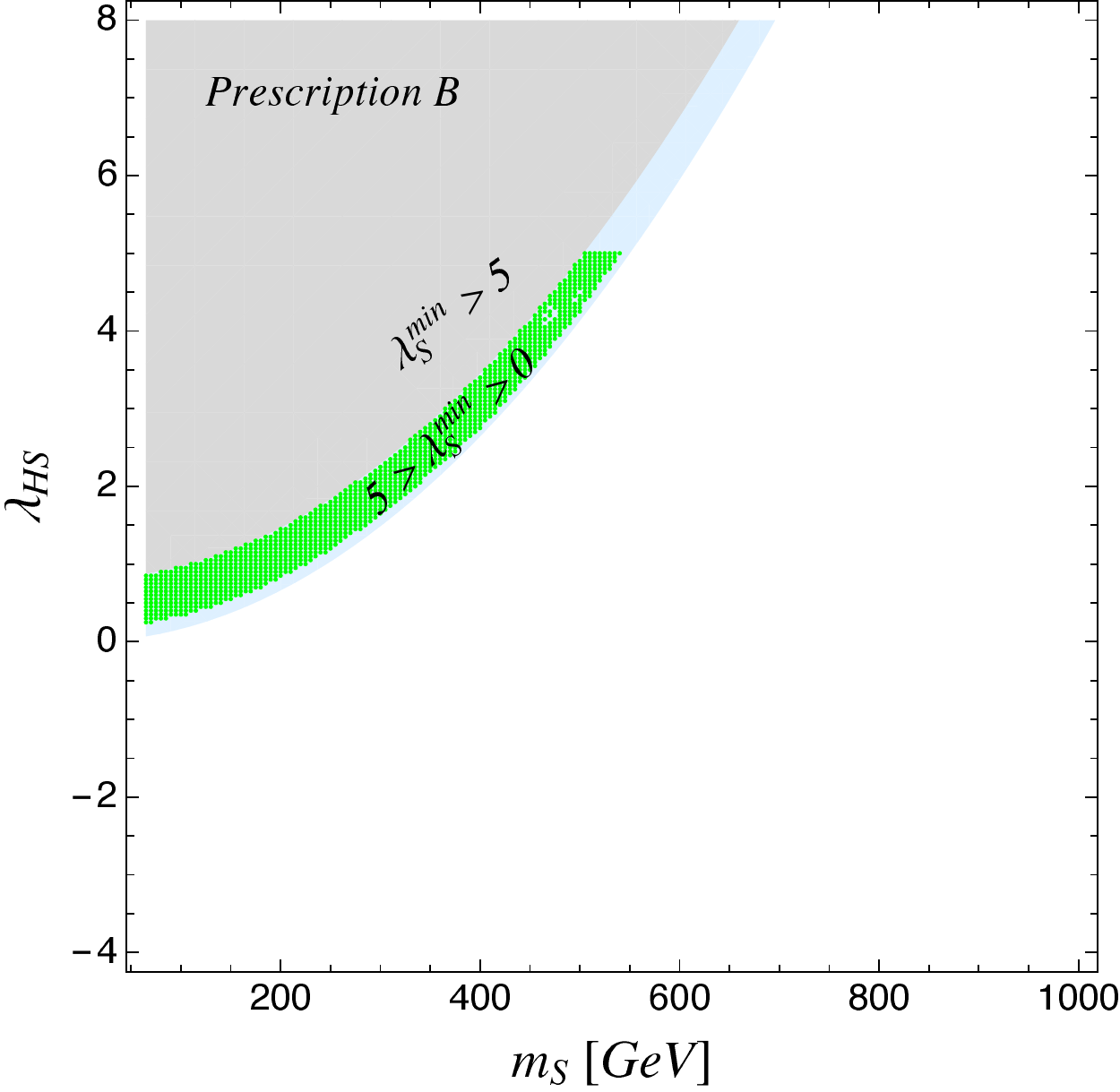}
        \includegraphics[width=0.240\linewidth]{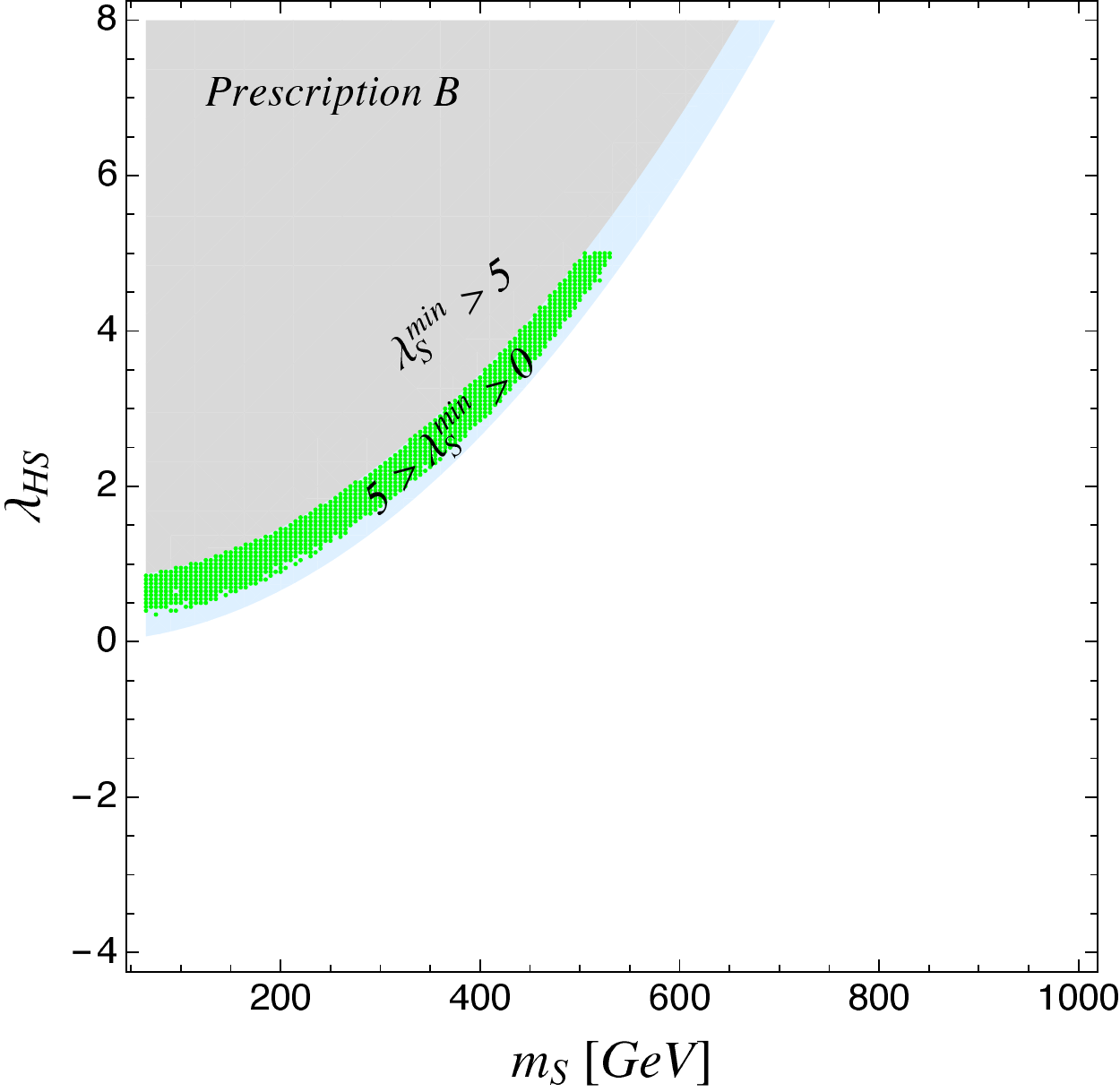}
	\caption{Similar plot to the left panel of Fig.~\ref{fig:lamHSvsMS} with the prescription B for different values of $\delta_S$: $\delta_S = 0.01$ (first), $\delta_S = 0.1$ (second), $\delta_S=1.0$ (third), and $\delta_S =2.5$ (last)}
	\label{fig:lamHSvsMS:presB:quarticS}
\end{figure}
 At a glance, the ballpark of the viable parameter space for the SFOEPT looks similar to those obtained from the prescription A. Looking at them closely we observe a few obvious discrepancies between two prescriptions.  Firstly, the viable singlet scalar masses for the SFOEPT in the prescription B extend to higher values with the increasing $\lambda_{HS}$ value as is seen in the left panel of Fig.~\ref{fig:lamHSvsMS:presB}. We suspect that the region with larger values of the $\lambda_{HS}$ and $m_S$ is where the high-temperature approximation more badly fails as was indicated in Eq.~(\ref{eq:mT:NDA}). Also, comparing right panels of Figs.~\ref{fig:lam3VSlam4} and~\ref{fig:lam3VSlam4:presB} shows the different shapes of the compatible region with the SFOEPT.

As was briefly mentioned in Section~\ref{sec:validityEffpot}, we have observed that the discrepancy between two prescriptions is more pronounced in the case of the two-step SFOEPT. In Fig.~\ref{fig:lamHSvsMS:presB:quarticS}, we show the viable parameter $(\lambda_{HS},\, m_S)$ space for the two-step SFOEPT. Four plots in Fig.~\ref{fig:lamHSvsMS:presB:quarticS} (along with the left panel of Fig.~\ref{fig:lamHSvsMS:presB}) differ by the quartic coupling, $\lambda_S \equiv \lambda_{S}^{\rm min} + \delta_S$ where $\lambda_S^{\rm min}$ was defined in Eq.~(\ref{eq:twostep:lamSmin}), and they illustrate how sensitively the parameter space depends on the quartic coupling, $\lambda_S$.

\end{appendix}

\bibliography{EWPT}

\begin{thebibliography}{70}
\expandafter\ifx\csname natexlab\endcsname\relax\def\natexlab#1{#1}\fi
\expandafter\ifx\csname bibnamefont\endcsname\relax
  \def\bibnamefont#1{#1}\fi
\expandafter\ifx\csname bibfnamefont\endcsname\relax
  \def\bibfnamefont#1{#1}\fi
\expandafter\ifx\csname citenamefont\endcsname\relax
  \def\citenamefont#1{#1}\fi
\expandafter\ifx\csname url\endcsname\relax
  \def\url#1{\texttt{#1}}\fi
\expandafter\ifx\csname urlprefix\endcsname\relax\def\urlprefix{URL }\fi
\providecommand{\bibinfo}[2]{#2}
\providecommand{\eprint}[2][]{\url{#2}}

\bibitem[{\citenamefont{Aad et~al.}(2015{\natexlab{a}})}]{Aad:2015xja}
\bibinfo{author}{\bibfnamefont{G.}~\bibnamefont{Aad}} \bibnamefont{et~al.}
  (\bibinfo{collaboration}{ATLAS}), \bibinfo{journal}{Phys. Rev.}
  \textbf{\bibinfo{volume}{D92}}, \bibinfo{pages}{092004}
  (\bibinfo{year}{2015}{\natexlab{a}}), \eprint{1509.04670}.

\bibitem[{\citenamefont{Aad et~al.}(2015{\natexlab{b}})}]{Aad:2015uka}
\bibinfo{author}{\bibfnamefont{G.}~\bibnamefont{Aad}} \bibnamefont{et~al.}
  (\bibinfo{collaboration}{ATLAS}), \bibinfo{journal}{Eur. Phys. J.}
  \textbf{\bibinfo{volume}{C75}}, \bibinfo{pages}{412}
  (\bibinfo{year}{2015}{\natexlab{b}}), \eprint{1506.00285}.

\bibitem[{\citenamefont{Khachatryan et~al.}(2016)}]{Khachatryan:2016sey}
\bibinfo{author}{\bibfnamefont{V.}~\bibnamefont{Khachatryan}}
  \bibnamefont{et~al.} (\bibinfo{collaboration}{CMS}), \bibinfo{journal}{Phys.
  Rev.} \textbf{\bibinfo{volume}{D94}}, \bibinfo{pages}{052012}
  (\bibinfo{year}{2016}), \eprint{1603.06896}.

\bibitem[{\citenamefont{Kuzmin et~al.}(1985)\citenamefont{Kuzmin, Rubakov, and
  Shaposhnikov}}]{Kuzmin:1985mm}
\bibinfo{author}{\bibfnamefont{V.~A.} \bibnamefont{Kuzmin}},
  \bibinfo{author}{\bibfnamefont{V.~A.} \bibnamefont{Rubakov}},
  \bibnamefont{and} \bibinfo{author}{\bibfnamefont{M.~E.}
  \bibnamefont{Shaposhnikov}}, \bibinfo{journal}{Phys. Lett.}
  \textbf{\bibinfo{volume}{155B}}, \bibinfo{pages}{36} (\bibinfo{year}{1985}).

\bibitem[{\citenamefont{Kirzhnits}(1972)}]{Kirzhnits:1972iw}
\bibinfo{author}{\bibfnamefont{D.~A.} \bibnamefont{Kirzhnits}},
  \bibinfo{journal}{JETP Lett.} \textbf{\bibinfo{volume}{15}},
  \bibinfo{pages}{529} (\bibinfo{year}{1972}), \bibinfo{note}{[Pisma Zh. Eksp.
  Teor. Fiz.15,745(1972)]}.

\bibitem[{\citenamefont{Kirzhnits and Linde}(1972)}]{Kirzhnits:1972ut}
\bibinfo{author}{\bibfnamefont{D.~A.} \bibnamefont{Kirzhnits}}
  \bibnamefont{and} \bibinfo{author}{\bibfnamefont{A.~D.} \bibnamefont{Linde}},
  \bibinfo{journal}{Phys. Lett.} \textbf{\bibinfo{volume}{42B}},
  \bibinfo{pages}{471} (\bibinfo{year}{1972}).

\bibitem[{\citenamefont{Dolan and Jackiw}(1974)}]{Dolan:1973qd}
\bibinfo{author}{\bibfnamefont{L.}~\bibnamefont{Dolan}} \bibnamefont{and}
  \bibinfo{author}{\bibfnamefont{R.}~\bibnamefont{Jackiw}},
  \bibinfo{journal}{Phys. Rev.} \textbf{\bibinfo{volume}{D9}},
  \bibinfo{pages}{3320} (\bibinfo{year}{1974}).

\bibitem[{\citenamefont{Weinberg}(1974)}]{Weinberg:1974hy}
\bibinfo{author}{\bibfnamefont{S.}~\bibnamefont{Weinberg}},
  \bibinfo{journal}{Phys. Rev.} \textbf{\bibinfo{volume}{D9}},
  \bibinfo{pages}{3357} (\bibinfo{year}{1974}).

\bibitem[{\citenamefont{Sakharov}(1967)}]{Sakharov:1967dj}
\bibinfo{author}{\bibfnamefont{A.~D.} \bibnamefont{Sakharov}},
  \bibinfo{journal}{Pisma Zh. Eksp. Teor. Fiz.} \textbf{\bibinfo{volume}{5}},
  \bibinfo{pages}{32} (\bibinfo{year}{1967}), \bibinfo{note}{[Usp. Fiz.
  Nauk161,61(1991)]}.

\bibitem[{\citenamefont{Shaposhnikov}(1987)}]{Shaposhnikov:1987tw}
\bibinfo{author}{\bibfnamefont{M.~E.} \bibnamefont{Shaposhnikov}},
  \bibinfo{journal}{Nucl. Phys.} \textbf{\bibinfo{volume}{B287}},
  \bibinfo{pages}{757} (\bibinfo{year}{1987}).

\bibitem[{\citenamefont{Shaposhnikov}(1988)}]{Shaposhnikov:1987pf}
\bibinfo{author}{\bibfnamefont{M.~E.} \bibnamefont{Shaposhnikov}},
  \bibinfo{journal}{Nucl. Phys.} \textbf{\bibinfo{volume}{B299}},
  \bibinfo{pages}{797} (\bibinfo{year}{1988}).

\bibitem[{\citenamefont{Shaposhnikov}(1992)}]{Shaposhnikov:1991cu}
\bibinfo{author}{\bibfnamefont{M.~E.} \bibnamefont{Shaposhnikov}},
  \bibinfo{journal}{Phys. Lett.} \textbf{\bibinfo{volume}{B277}},
  \bibinfo{pages}{324} (\bibinfo{year}{1992}), \bibinfo{note}{[Erratum: Phys.
  Lett.B282,483(1992)]}.

\bibitem[{\citenamefont{Morrissey and Ramsey-Musolf}(2012)}]{Morrissey:2012db}
\bibinfo{author}{\bibfnamefont{D.~E.} \bibnamefont{Morrissey}}
  \bibnamefont{and} \bibinfo{author}{\bibfnamefont{M.~J.}
  \bibnamefont{Ramsey-Musolf}}, \bibinfo{journal}{New J. Phys.}
  \textbf{\bibinfo{volume}{14}}, \bibinfo{pages}{125003}
  (\bibinfo{year}{2012}), \eprint{1206.2942}.

\bibitem[{\citenamefont{Noble and Perelstein}(2008)}]{Noble:2007kk}
\bibinfo{author}{\bibfnamefont{A.}~\bibnamefont{Noble}} \bibnamefont{and}
  \bibinfo{author}{\bibfnamefont{M.}~\bibnamefont{Perelstein}},
  \bibinfo{journal}{Phys. Rev.} \textbf{\bibinfo{volume}{D78}},
  \bibinfo{pages}{063518} (\bibinfo{year}{2008}), \eprint{0711.3018}.

\bibitem[{\citenamefont{Huang et~al.}(2016{\natexlab{a}})\citenamefont{Huang,
  Joglekar, Li, and Wagner}}]{Huang:2015tdv}
\bibinfo{author}{\bibfnamefont{P.}~\bibnamefont{Huang}},
  \bibinfo{author}{\bibfnamefont{A.}~\bibnamefont{Joglekar}},
  \bibinfo{author}{\bibfnamefont{B.}~\bibnamefont{Li}}, \bibnamefont{and}
  \bibinfo{author}{\bibfnamefont{C.~E.~M.} \bibnamefont{Wagner}},
  \bibinfo{journal}{Phys. Rev.} \textbf{\bibinfo{volume}{D93}},
  \bibinfo{pages}{055049} (\bibinfo{year}{2016}{\natexlab{a}}),
  \eprint{1512.00068}.

\bibitem[{\citenamefont{Katz and Perelstein}(2014)}]{Katz:2014bha}
\bibinfo{author}{\bibfnamefont{A.}~\bibnamefont{Katz}} \bibnamefont{and}
  \bibinfo{author}{\bibfnamefont{M.}~\bibnamefont{Perelstein}},
  \bibinfo{journal}{JHEP} \textbf{\bibinfo{volume}{07}}, \bibinfo{pages}{108}
  (\bibinfo{year}{2014}), \eprint{1401.1827}.

\bibitem[{\citenamefont{Curtin et~al.}(2014)\citenamefont{Curtin, Meade, and
  Yu}}]{Curtin:2014jma}
\bibinfo{author}{\bibfnamefont{D.}~\bibnamefont{Curtin}},
  \bibinfo{author}{\bibfnamefont{P.}~\bibnamefont{Meade}}, \bibnamefont{and}
  \bibinfo{author}{\bibfnamefont{C.-T.} \bibnamefont{Yu}},
  \bibinfo{journal}{JHEP} \textbf{\bibinfo{volume}{11}}, \bibinfo{pages}{127}
  (\bibinfo{year}{2014}), \eprint{1409.0005}.

\bibitem[{\citenamefont{Patel and Ramsey-Musolf}(2011)}]{Patel:2011th}
\bibinfo{author}{\bibfnamefont{H.~H.} \bibnamefont{Patel}} \bibnamefont{and}
  \bibinfo{author}{\bibfnamefont{M.~J.} \bibnamefont{Ramsey-Musolf}},
  \bibinfo{journal}{JHEP} \textbf{\bibinfo{volume}{07}}, \bibinfo{pages}{029}
  (\bibinfo{year}{2011}), \eprint{1101.4665}.

\bibitem[{\citenamefont{Kurup and Perelstein}(2017)}]{Kurup:2017dzf}
\bibinfo{author}{\bibfnamefont{G.}~\bibnamefont{Kurup}} \bibnamefont{and}
  \bibinfo{author}{\bibfnamefont{M.}~\bibnamefont{Perelstein}},
  \bibinfo{journal}{Phys. Rev.} \textbf{\bibinfo{volume}{D96}},
  \bibinfo{pages}{015036} (\bibinfo{year}{2017}), \eprint{1704.03381}.

\bibitem[{\citenamefont{Espinosa and Quiros}(2007)}]{Espinosa:2007qk}
\bibinfo{author}{\bibfnamefont{J.~R.} \bibnamefont{Espinosa}} \bibnamefont{and}
  \bibinfo{author}{\bibfnamefont{M.}~\bibnamefont{Quiros}},
  \bibinfo{journal}{Phys. Rev.} \textbf{\bibinfo{volume}{D76}},
  \bibinfo{pages}{076004} (\bibinfo{year}{2007}), \eprint{hep-ph/0701145}.

\bibitem[{\citenamefont{Barger et~al.}(2008)\citenamefont{Barger, Langacker,
  McCaskey, Ramsey-Musolf, and Shaughnessy}}]{Barger:2007im}
\bibinfo{author}{\bibfnamefont{V.}~\bibnamefont{Barger}},
  \bibinfo{author}{\bibfnamefont{P.}~\bibnamefont{Langacker}},
  \bibinfo{author}{\bibfnamefont{M.}~\bibnamefont{McCaskey}},
  \bibinfo{author}{\bibfnamefont{M.~J.} \bibnamefont{Ramsey-Musolf}},
  \bibnamefont{and}
  \bibinfo{author}{\bibfnamefont{G.}~\bibnamefont{Shaughnessy}},
  \bibinfo{journal}{Phys. Rev.} \textbf{\bibinfo{volume}{D77}},
  \bibinfo{pages}{035005} (\bibinfo{year}{2008}), \eprint{0706.4311}.

\bibitem[{\citenamefont{Espinosa et~al.}(2008)\citenamefont{Espinosa,
  Konstandin, No, and Quiros}}]{Espinosa:2008kw}
\bibinfo{author}{\bibfnamefont{J.~R.} \bibnamefont{Espinosa}},
  \bibinfo{author}{\bibfnamefont{T.}~\bibnamefont{Konstandin}},
  \bibinfo{author}{\bibfnamefont{J.~M.} \bibnamefont{No}}, \bibnamefont{and}
  \bibinfo{author}{\bibfnamefont{M.}~\bibnamefont{Quiros}},
  \bibinfo{journal}{Phys. Rev.} \textbf{\bibinfo{volume}{D78}},
  \bibinfo{pages}{123528} (\bibinfo{year}{2008}), \eprint{0809.3215}.

\bibitem[{\citenamefont{Espinosa et~al.}(2012)\citenamefont{Espinosa,
  Konstandin, and Riva}}]{Espinosa:2011ax}
\bibinfo{author}{\bibfnamefont{J.~R.} \bibnamefont{Espinosa}},
  \bibinfo{author}{\bibfnamefont{T.}~\bibnamefont{Konstandin}},
  \bibnamefont{and} \bibinfo{author}{\bibfnamefont{F.}~\bibnamefont{Riva}},
  \bibinfo{journal}{Nucl. Phys.} \textbf{\bibinfo{volume}{B854}},
  \bibinfo{pages}{592} (\bibinfo{year}{2012}), \eprint{1107.5441}.

\bibitem[{\citenamefont{Cline and Kainulainen}(2013)}]{Cline:2012hg}
\bibinfo{author}{\bibfnamefont{J.~M.} \bibnamefont{Cline}} \bibnamefont{and}
  \bibinfo{author}{\bibfnamefont{K.}~\bibnamefont{Kainulainen}},
  \bibinfo{journal}{JCAP} \textbf{\bibinfo{volume}{1301}}, \bibinfo{pages}{012}
  (\bibinfo{year}{2013}), \eprint{1210.4196}.

\bibitem[{\citenamefont{Cline et~al.}(2013)\citenamefont{Cline, Kainulainen,
  Scott, and Weniger}}]{Cline:2013gha}
\bibinfo{author}{\bibfnamefont{J.~M.} \bibnamefont{Cline}},
  \bibinfo{author}{\bibfnamefont{K.}~\bibnamefont{Kainulainen}},
  \bibinfo{author}{\bibfnamefont{P.}~\bibnamefont{Scott}}, \bibnamefont{and}
  \bibinfo{author}{\bibfnamefont{C.}~\bibnamefont{Weniger}},
  \bibinfo{journal}{Phys. Rev.} \textbf{\bibinfo{volume}{D88}},
  \bibinfo{pages}{055025} (\bibinfo{year}{2013}), \bibinfo{note}{[Erratum:
  Phys. Rev.D92,no.3,039906(2015)]}, \eprint{1306.4710}.

\bibitem[{\citenamefont{Alanne et~al.}(2014)\citenamefont{Alanne, Tuominen, and
  Vaskonen}}]{Alanne:2014bra}
\bibinfo{author}{\bibfnamefont{T.}~\bibnamefont{Alanne}},
  \bibinfo{author}{\bibfnamefont{K.}~\bibnamefont{Tuominen}}, \bibnamefont{and}
  \bibinfo{author}{\bibfnamefont{V.}~\bibnamefont{Vaskonen}},
  \bibinfo{journal}{Nucl. Phys.} \textbf{\bibinfo{volume}{B889}},
  \bibinfo{pages}{692} (\bibinfo{year}{2014}), \eprint{1407.0688}.

\bibitem[{\citenamefont{Vaskonen}(2017)}]{Vaskonen:2016yiu}
\bibinfo{author}{\bibfnamefont{V.}~\bibnamefont{Vaskonen}},
  \bibinfo{journal}{Phys. Rev.} \textbf{\bibinfo{volume}{D95}},
  \bibinfo{pages}{123515} (\bibinfo{year}{2017}), \eprint{1611.02073}.

\bibitem[{\citenamefont{Marzola et~al.}(2017)\citenamefont{Marzola, Racioppi,
  and Vaskonen}}]{Marzola:2017jzl}
\bibinfo{author}{\bibfnamefont{L.}~\bibnamefont{Marzola}},
  \bibinfo{author}{\bibfnamefont{A.}~\bibnamefont{Racioppi}}, \bibnamefont{and}
  \bibinfo{author}{\bibfnamefont{V.}~\bibnamefont{Vaskonen}},
  \bibinfo{journal}{Eur. Phys. J.} \textbf{\bibinfo{volume}{C77}},
  \bibinfo{pages}{484} (\bibinfo{year}{2017}), \eprint{1704.01034}.

\bibitem[{\citenamefont{Grojean et~al.}(2005)\citenamefont{Grojean, Servant,
  and Wells}}]{Grojean:2004xa}
\bibinfo{author}{\bibfnamefont{C.}~\bibnamefont{Grojean}},
  \bibinfo{author}{\bibfnamefont{G.}~\bibnamefont{Servant}}, \bibnamefont{and}
  \bibinfo{author}{\bibfnamefont{J.~D.} \bibnamefont{Wells}},
  \bibinfo{journal}{Phys. Rev.} \textbf{\bibinfo{volume}{D71}},
  \bibinfo{pages}{036001} (\bibinfo{year}{2005}), \eprint{hep-ph/0407019}.

\bibitem[{\citenamefont{Bodeker et~al.}(2005)\citenamefont{Bodeker, Fromme,
  Huber, and Seniuch}}]{Bodeker:2004ws}
\bibinfo{author}{\bibfnamefont{D.}~\bibnamefont{Bodeker}},
  \bibinfo{author}{\bibfnamefont{L.}~\bibnamefont{Fromme}},
  \bibinfo{author}{\bibfnamefont{S.~J.} \bibnamefont{Huber}}, \bibnamefont{and}
  \bibinfo{author}{\bibfnamefont{M.}~\bibnamefont{Seniuch}},
  \bibinfo{journal}{JHEP} \textbf{\bibinfo{volume}{02}}, \bibinfo{pages}{026}
  (\bibinfo{year}{2005}), \eprint{hep-ph/0412366}.

\bibitem[{\citenamefont{Delaunay et~al.}(2008)\citenamefont{Delaunay, Grojean,
  and Wells}}]{Delaunay:2007wb}
\bibinfo{author}{\bibfnamefont{C.}~\bibnamefont{Delaunay}},
  \bibinfo{author}{\bibfnamefont{C.}~\bibnamefont{Grojean}}, \bibnamefont{and}
  \bibinfo{author}{\bibfnamefont{J.~D.} \bibnamefont{Wells}},
  \bibinfo{journal}{JHEP} \textbf{\bibinfo{volume}{04}}, \bibinfo{pages}{029}
  (\bibinfo{year}{2008}), \eprint{0711.2511}.

\bibitem[{\citenamefont{Gan et~al.}(2017)\citenamefont{Gan, Long, and
  Wang}}]{Gan:2017mcv}
\bibinfo{author}{\bibfnamefont{X.}~\bibnamefont{Gan}},
  \bibinfo{author}{\bibfnamefont{A.~J.} \bibnamefont{Long}}, \bibnamefont{and}
  \bibinfo{author}{\bibfnamefont{L.-T.} \bibnamefont{Wang}}
  (\bibinfo{year}{2017}), \eprint{1708.03061}.

\bibitem[{\citenamefont{Giudice et~al.}(2007)\citenamefont{Giudice, Grojean,
  Pomarol, and Rattazzi}}]{Giudice:2007fh}
\bibinfo{author}{\bibfnamefont{G.~F.} \bibnamefont{Giudice}},
  \bibinfo{author}{\bibfnamefont{C.}~\bibnamefont{Grojean}},
  \bibinfo{author}{\bibfnamefont{A.}~\bibnamefont{Pomarol}}, \bibnamefont{and}
  \bibinfo{author}{\bibfnamefont{R.}~\bibnamefont{Rattazzi}},
  \bibinfo{journal}{JHEP} \textbf{\bibinfo{volume}{06}}, \bibinfo{pages}{045}
  (\bibinfo{year}{2007}), \eprint{hep-ph/0703164}.

\bibitem[{\citenamefont{Maltoni et~al.}(2014)\citenamefont{Maltoni, Vryonidou,
  and Zaro}}]{Maltoni:2014eza}
\bibinfo{author}{\bibfnamefont{F.}~\bibnamefont{Maltoni}},
  \bibinfo{author}{\bibfnamefont{E.}~\bibnamefont{Vryonidou}},
  \bibnamefont{and} \bibinfo{author}{\bibfnamefont{M.}~\bibnamefont{Zaro}},
  \bibinfo{journal}{JHEP} \textbf{\bibinfo{volume}{11}}, \bibinfo{pages}{079}
  (\bibinfo{year}{2014}), \eprint{1408.6542}.

\bibitem[{\citenamefont{Plehn and Rauch}(2005)}]{Plehn:2005nk}
\bibinfo{author}{\bibfnamefont{T.}~\bibnamefont{Plehn}} \bibnamefont{and}
  \bibinfo{author}{\bibfnamefont{M.}~\bibnamefont{Rauch}},
  \bibinfo{journal}{Phys. Rev.} \textbf{\bibinfo{volume}{D72}},
  \bibinfo{pages}{053008} (\bibinfo{year}{2005}), \eprint{hep-ph/0507321}.

\bibitem[{\citenamefont{Binoth et~al.}(2006)\citenamefont{Binoth, Karg, Kauer,
  and Ruckl}}]{Binoth:2006ym}
\bibinfo{author}{\bibfnamefont{T.}~\bibnamefont{Binoth}},
  \bibinfo{author}{\bibfnamefont{S.}~\bibnamefont{Karg}},
  \bibinfo{author}{\bibfnamefont{N.}~\bibnamefont{Kauer}}, \bibnamefont{and}
  \bibinfo{author}{\bibfnamefont{R.}~\bibnamefont{Ruckl}},
  \bibinfo{journal}{Phys. Rev.} \textbf{\bibinfo{volume}{D74}},
  \bibinfo{pages}{113008} (\bibinfo{year}{2006}), \eprint{hep-ph/0608057}.

\bibitem[{\citenamefont{Papaefstathiou and
  Sakurai}(2016)}]{Papaefstathiou:2015paa}
\bibinfo{author}{\bibfnamefont{A.}~\bibnamefont{Papaefstathiou}}
  \bibnamefont{and} \bibinfo{author}{\bibfnamefont{K.}~\bibnamefont{Sakurai}},
  \bibinfo{journal}{JHEP} \textbf{\bibinfo{volume}{02}}, \bibinfo{pages}{006}
  (\bibinfo{year}{2016}), \eprint{1508.06524}.

\bibitem[{\citenamefont{Chen et~al.}(2016)\citenamefont{Chen, Yan, Zhao, Zhong,
  and Zhao}}]{Chen:2015gva}
\bibinfo{author}{\bibfnamefont{C.-Y.} \bibnamefont{Chen}},
  \bibinfo{author}{\bibfnamefont{Q.-S.} \bibnamefont{Yan}},
  \bibinfo{author}{\bibfnamefont{X.}~\bibnamefont{Zhao}},
  \bibinfo{author}{\bibfnamefont{Y.-M.} \bibnamefont{Zhong}}, \bibnamefont{and}
  \bibinfo{author}{\bibfnamefont{Z.}~\bibnamefont{Zhao}},
  \bibinfo{journal}{Phys. Rev.} \textbf{\bibinfo{volume}{D93}},
  \bibinfo{pages}{013007} (\bibinfo{year}{2016}), \eprint{1510.04013}.

\bibitem[{\citenamefont{Fuks et~al.}(2016)\citenamefont{Fuks, Kim, and
  Lee}}]{Fuks:2015hna}
\bibinfo{author}{\bibfnamefont{B.}~\bibnamefont{Fuks}},
  \bibinfo{author}{\bibfnamefont{J.~H.} \bibnamefont{Kim}}, \bibnamefont{and}
  \bibinfo{author}{\bibfnamefont{S.~J.} \bibnamefont{Lee}},
  \bibinfo{journal}{Phys. Rev.} \textbf{\bibinfo{volume}{D93}},
  \bibinfo{pages}{035026} (\bibinfo{year}{2016}), \eprint{1510.07697}.

\bibitem[{\citenamefont{Contino et~al.}(2017)}]{Contino:2016spe}
\bibinfo{author}{\bibfnamefont{R.}~\bibnamefont{Contino}} \bibnamefont{et~al.},
  \bibinfo{journal}{CERN Yellow Report} pp. \bibinfo{pages}{255--440}
  (\bibinfo{year}{2017}), \eprint{1606.09408}.

\bibitem[{\citenamefont{Fuks et~al.}(2017)\citenamefont{Fuks, Kim, and
  Lee}}]{Fuks:2017zkg}
\bibinfo{author}{\bibfnamefont{B.}~\bibnamefont{Fuks}},
  \bibinfo{author}{\bibfnamefont{J.~H.} \bibnamefont{Kim}}, \bibnamefont{and}
  \bibinfo{author}{\bibfnamefont{S.~J.} \bibnamefont{Lee}},
  \bibinfo{journal}{Phys. Lett.} \textbf{\bibinfo{volume}{B771}},
  \bibinfo{pages}{354} (\bibinfo{year}{2017}), \eprint{1704.04298}.

\bibitem[{\citenamefont{Gross et~al.}(1981)\citenamefont{Gross, Pisarski, and
  Yaffe}}]{Gross:1980br}
\bibinfo{author}{\bibfnamefont{D.~J.} \bibnamefont{Gross}},
  \bibinfo{author}{\bibfnamefont{R.~D.} \bibnamefont{Pisarski}},
  \bibnamefont{and} \bibinfo{author}{\bibfnamefont{L.~G.} \bibnamefont{Yaffe}},
  \bibinfo{journal}{Rev. Mod. Phys.} \textbf{\bibinfo{volume}{53}},
  \bibinfo{pages}{43} (\bibinfo{year}{1981}).

\bibitem[{\citenamefont{Parwani}(1992)}]{Parwani:1991gq}
\bibinfo{author}{\bibfnamefont{R.~R.} \bibnamefont{Parwani}},
  \bibinfo{journal}{Phys. Rev.} \textbf{\bibinfo{volume}{D45}},
  \bibinfo{pages}{4695} (\bibinfo{year}{1992}), \bibinfo{note}{[Erratum: Phys.
  Rev.D48,5965(1993)]}, \eprint{hep-ph/9204216}.

\bibitem[{\citenamefont{Arnold and Espinosa}(1993)}]{Arnold:1992rz}
\bibinfo{author}{\bibfnamefont{P.~B.} \bibnamefont{Arnold}} \bibnamefont{and}
  \bibinfo{author}{\bibfnamefont{O.}~\bibnamefont{Espinosa}},
  \bibinfo{journal}{Phys. Rev.} \textbf{\bibinfo{volume}{D47}},
  \bibinfo{pages}{3546} (\bibinfo{year}{1993}), \bibinfo{note}{[Erratum: Phys.
  Rev.D50,6662(1994)]}, \eprint{hep-ph/9212235}.

\bibitem[{\citenamefont{Carrington}(1992)}]{Carrington:1991hz}
\bibinfo{author}{\bibfnamefont{M.~E.} \bibnamefont{Carrington}},
  \bibinfo{journal}{Phys. Rev.} \textbf{\bibinfo{volume}{D45}},
  \bibinfo{pages}{2933} (\bibinfo{year}{1992}).

\bibitem[{\citenamefont{Curtin et~al.}(2016)\citenamefont{Curtin, Meade, and
  Ramani}}]{Curtin:2016urg}
\bibinfo{author}{\bibfnamefont{D.}~\bibnamefont{Curtin}},
  \bibinfo{author}{\bibfnamefont{P.}~\bibnamefont{Meade}}, \bibnamefont{and}
  \bibinfo{author}{\bibfnamefont{H.}~\bibnamefont{Ramani}}
  (\bibinfo{year}{2016}), \eprint{1612.00466}.

\bibitem[{\citenamefont{Boyd et~al.}(1993)\citenamefont{Boyd, Brahm, and
  Hsu}}]{Boyd:1992xn}
\bibinfo{author}{\bibfnamefont{C.~G.} \bibnamefont{Boyd}},
  \bibinfo{author}{\bibfnamefont{D.~E.} \bibnamefont{Brahm}}, \bibnamefont{and}
  \bibinfo{author}{\bibfnamefont{S.~D.~H.} \bibnamefont{Hsu}},
  \bibinfo{journal}{Phys. Rev.} \textbf{\bibinfo{volume}{D48}},
  \bibinfo{pages}{4952} (\bibinfo{year}{1993}), \eprint{hep-ph/9206235}.

\bibitem[{\citenamefont{Dine et~al.}(1992)\citenamefont{Dine, Leigh, Huet,
  Linde, and Linde}}]{Dine:1992wr}
\bibinfo{author}{\bibfnamefont{M.}~\bibnamefont{Dine}},
  \bibinfo{author}{\bibfnamefont{R.~G.} \bibnamefont{Leigh}},
  \bibinfo{author}{\bibfnamefont{P.~Y.} \bibnamefont{Huet}},
  \bibinfo{author}{\bibfnamefont{A.~D.} \bibnamefont{Linde}}, \bibnamefont{and}
  \bibinfo{author}{\bibfnamefont{D.~A.} \bibnamefont{Linde}},
  \bibinfo{journal}{Phys. Rev.} \textbf{\bibinfo{volume}{D46}},
  \bibinfo{pages}{550} (\bibinfo{year}{1992}), \eprint{hep-ph/9203203}.

\bibitem[{\citenamefont{Coleman and Weinberg}(1973)}]{Coleman:1973jx}
\bibinfo{author}{\bibfnamefont{S.~R.} \bibnamefont{Coleman}} \bibnamefont{and}
  \bibinfo{author}{\bibfnamefont{E.~J.} \bibnamefont{Weinberg}},
  \bibinfo{journal}{Phys. Rev.} \textbf{\bibinfo{volume}{D7}},
  \bibinfo{pages}{1888} (\bibinfo{year}{1973}).

\bibitem[{\citenamefont{Fendley}(1987)}]{Fendley:1987ef}
\bibinfo{author}{\bibfnamefont{P.}~\bibnamefont{Fendley}},
  \bibinfo{journal}{Phys. Lett.} \textbf{\bibinfo{volume}{B196}},
  \bibinfo{pages}{175} (\bibinfo{year}{1987}).

\bibitem[{\citenamefont{Anderson and Hall}(1992)}]{Anderson:1991zb}
\bibinfo{author}{\bibfnamefont{G.~W.} \bibnamefont{Anderson}} \bibnamefont{and}
  \bibinfo{author}{\bibfnamefont{L.~J.} \bibnamefont{Hall}},
  \bibinfo{journal}{Phys. Rev.} \textbf{\bibinfo{volume}{D45}},
  \bibinfo{pages}{2685} (\bibinfo{year}{1992}).

\bibitem[{\citenamefont{Garny and Konstandin}(2012)}]{Garny:2012cg}
\bibinfo{author}{\bibfnamefont{M.}~\bibnamefont{Garny}} \bibnamefont{and}
  \bibinfo{author}{\bibfnamefont{T.}~\bibnamefont{Konstandin}},
  \bibinfo{journal}{JHEP} \textbf{\bibinfo{volume}{07}}, \bibinfo{pages}{189}
  (\bibinfo{year}{2012}), \eprint{1205.3392}.

\bibitem[{\citenamefont{Chala et~al.}(2016)\citenamefont{Chala, Nardini, and
  Sobolev}}]{Chala:2016ykx}
\bibinfo{author}{\bibfnamefont{M.}~\bibnamefont{Chala}},
  \bibinfo{author}{\bibfnamefont{G.}~\bibnamefont{Nardini}}, \bibnamefont{and}
  \bibinfo{author}{\bibfnamefont{I.}~\bibnamefont{Sobolev}},
  \bibinfo{journal}{Phys. Rev.} \textbf{\bibinfo{volume}{D94}},
  \bibinfo{pages}{055006} (\bibinfo{year}{2016}), \eprint{1605.08663}.

\bibitem[{\citenamefont{Artymowski et~al.}(2017)\citenamefont{Artymowski,
  Lewicki, and Wells}}]{Artymowski:2016tme}
\bibinfo{author}{\bibfnamefont{M.}~\bibnamefont{Artymowski}},
  \bibinfo{author}{\bibfnamefont{M.}~\bibnamefont{Lewicki}}, \bibnamefont{and}
  \bibinfo{author}{\bibfnamefont{J.~D.} \bibnamefont{Wells}},
  \bibinfo{journal}{JHEP} \textbf{\bibinfo{volume}{03}}, \bibinfo{pages}{066}
  (\bibinfo{year}{2017}), \eprint{1609.07143}.

\bibitem[{\citenamefont{Beniwal et~al.}(2017)\citenamefont{Beniwal, Lewicki,
  Wells, White, and Williams}}]{Beniwal:2017eik}
\bibinfo{author}{\bibfnamefont{A.}~\bibnamefont{Beniwal}},
  \bibinfo{author}{\bibfnamefont{M.}~\bibnamefont{Lewicki}},
  \bibinfo{author}{\bibfnamefont{J.~D.} \bibnamefont{Wells}},
  \bibinfo{author}{\bibfnamefont{M.}~\bibnamefont{White}}, \bibnamefont{and}
  \bibinfo{author}{\bibfnamefont{A.~G.} \bibnamefont{Williams}},
  \bibinfo{journal}{JHEP} \textbf{\bibinfo{volume}{08}}, \bibinfo{pages}{108}
  (\bibinfo{year}{2017}), \eprint{1702.06124}.

\bibitem[{\citenamefont{Chen et~al.}(2017)\citenamefont{Chen, Kozaczuk, and
  Lewis}}]{Chen:2017qcz}
\bibinfo{author}{\bibfnamefont{C.-Y.} \bibnamefont{Chen}},
  \bibinfo{author}{\bibfnamefont{J.}~\bibnamefont{Kozaczuk}}, \bibnamefont{and}
  \bibinfo{author}{\bibfnamefont{I.~M.} \bibnamefont{Lewis}},
  \bibinfo{journal}{JHEP} \textbf{\bibinfo{volume}{08}}, \bibinfo{pages}{096}
  (\bibinfo{year}{2017}), \eprint{1704.05844}.

\bibitem[{\citenamefont{Chung et~al.}(2013)\citenamefont{Chung, Long, and
  Wang}}]{Chung:2012vg}
\bibinfo{author}{\bibfnamefont{D.~J.~H.} \bibnamefont{Chung}},
  \bibinfo{author}{\bibfnamefont{A.~J.} \bibnamefont{Long}}, \bibnamefont{and}
  \bibinfo{author}{\bibfnamefont{L.-T.} \bibnamefont{Wang}},
  \bibinfo{journal}{Phys. Rev.} \textbf{\bibinfo{volume}{D87}},
  \bibinfo{pages}{023509} (\bibinfo{year}{2013}), \eprint{1209.1819}.

\bibitem[{\citenamefont{Huang et~al.}(2016{\natexlab{b}})\citenamefont{Huang,
  Gu, Yin, Yu, and Zhang}}]{Huang:2015izx}
\bibinfo{author}{\bibfnamefont{F.~P.} \bibnamefont{Huang}},
  \bibinfo{author}{\bibfnamefont{P.-H.} \bibnamefont{Gu}},
  \bibinfo{author}{\bibfnamefont{P.-F.} \bibnamefont{Yin}},
  \bibinfo{author}{\bibfnamefont{Z.-H.} \bibnamefont{Yu}}, \bibnamefont{and}
  \bibinfo{author}{\bibfnamefont{X.}~\bibnamefont{Zhang}},
  \bibinfo{journal}{Phys. Rev.} \textbf{\bibinfo{volume}{D93}},
  \bibinfo{pages}{103515} (\bibinfo{year}{2016}{\natexlab{b}}),
  \eprint{1511.03969}.

\bibitem[{\citenamefont{Azatov et~al.}(2015)\citenamefont{Azatov, Contino,
  Panico, and Son}}]{Azatov:2015oxa}
\bibinfo{author}{\bibfnamefont{A.}~\bibnamefont{Azatov}},
  \bibinfo{author}{\bibfnamefont{R.}~\bibnamefont{Contino}},
  \bibinfo{author}{\bibfnamefont{G.}~\bibnamefont{Panico}}, \bibnamefont{and}
  \bibinfo{author}{\bibfnamefont{M.}~\bibnamefont{Son}},
  \bibinfo{journal}{Phys. Rev.} \textbf{\bibinfo{volume}{D92}},
  \bibinfo{pages}{035001} (\bibinfo{year}{2015}), \eprint{1502.00539}.

\bibitem[{\citenamefont{de~Florian et~al.}(2016)}]{deFlorian:2016spz}
\bibinfo{author}{\bibfnamefont{D.}~\bibnamefont{de~Florian}}
  \bibnamefont{et~al.} (\bibinfo{collaboration}{LHC Higgs Cross Section Working
  Group}) (\bibinfo{year}{2016}), \eprint{1610.07922}.

\bibitem[{\citenamefont{Cline et~al.}(2011)\citenamefont{Cline, Kainulainen,
  and Trott}}]{Cline:2011mm}
\bibinfo{author}{\bibfnamefont{J.~M.} \bibnamefont{Cline}},
  \bibinfo{author}{\bibfnamefont{K.}~\bibnamefont{Kainulainen}},
  \bibnamefont{and} \bibinfo{author}{\bibfnamefont{M.}~\bibnamefont{Trott}},
  \bibinfo{journal}{JHEP} \textbf{\bibinfo{volume}{11}}, \bibinfo{pages}{089}
  (\bibinfo{year}{2011}), \eprint{1107.3559}.

\bibitem[{\citenamefont{Chiang et~al.}(2018)\citenamefont{Chiang, Lee, and
  Senaha}}]{Chiang:2018gsn}
\bibinfo{author}{\bibfnamefont{C.-W.} \bibnamefont{Chiang}},
  \bibinfo{author}{\bibfnamefont{Y.-T.} \bibnamefont{Lee}}, \bibnamefont{and}
  \bibinfo{author}{\bibfnamefont{E.}~\bibnamefont{Senaha}}
  (\bibinfo{year}{2018}), \eprint{1808.01098}.

\bibitem[{\citenamefont{Di~Luzio et~al.}(2017)\citenamefont{Di~Luzio,
  Gr{\"o}ber, and Spannowsky}}]{DiLuzio:2017tfn}
\bibinfo{author}{\bibfnamefont{L.}~\bibnamefont{Di~Luzio}},
  \bibinfo{author}{\bibfnamefont{R.}~\bibnamefont{Gr{\"o}ber}},
  \bibnamefont{and}
  \bibinfo{author}{\bibfnamefont{M.}~\bibnamefont{Spannowsky}}
  (\bibinfo{year}{2017}), \eprint{1704.02311}.

\bibitem[{\citenamefont{Kribs et~al.}(2017)\citenamefont{Kribs, Maier, Rzehak,
  Spannowsky, and Waite}}]{Kribs:2017znd}
\bibinfo{author}{\bibfnamefont{G.~D.} \bibnamefont{Kribs}},
  \bibinfo{author}{\bibfnamefont{A.}~\bibnamefont{Maier}},
  \bibinfo{author}{\bibfnamefont{H.}~\bibnamefont{Rzehak}},
  \bibinfo{author}{\bibfnamefont{M.}~\bibnamefont{Spannowsky}},
  \bibnamefont{and} \bibinfo{author}{\bibfnamefont{P.}~\bibnamefont{Waite}},
  \bibinfo{journal}{Phys. Rev.} \textbf{\bibinfo{volume}{D95}},
  \bibinfo{pages}{093004} (\bibinfo{year}{2017}), \eprint{1702.07678}.

\bibitem[{ATL(2014)}]{ATL-PHYS-PUB-2014-019}
\bibinfo{type}{Tech. Rep.} \bibinfo{number}{ATL-PHYS-PUB-2014-019},
  \bibinfo{institution}{CERN}, \bibinfo{address}{Geneva}
  (\bibinfo{year}{2014}), \urlprefix\url{http://cds.cern.ch/record/1956733}.

\bibitem[{ATL(2017)}]{ATL-PHYS-PUB-2017-001}
\bibinfo{type}{Tech. Rep.} \bibinfo{number}{ATL-PHYS-PUB-2017-001},
  \bibinfo{institution}{CERN}, \bibinfo{address}{Geneva}
  (\bibinfo{year}{2017}), \urlprefix\url{http://cds.cern.ch/record/2243387}.

\bibitem[{CMS(2015)}]{CMS-PAS-FTR-15-002}
\bibinfo{type}{Tech. Rep.} \bibinfo{number}{CMS-PAS-FTR-15-002},
  \bibinfo{institution}{CERN}, \bibinfo{address}{Geneva}
  (\bibinfo{year}{2015}), \urlprefix\url{https://cds.cern.ch/record/2063038}.

\bibitem[{\citenamefont{Di~Vita et~al.}(2017)\citenamefont{Di~Vita, Grojean,
  Panico, Riembau, and Vantalon}}]{DiVita:2017eyz}
\bibinfo{author}{\bibfnamefont{S.}~\bibnamefont{Di~Vita}},
  \bibinfo{author}{\bibfnamefont{C.}~\bibnamefont{Grojean}},
  \bibinfo{author}{\bibfnamefont{G.}~\bibnamefont{Panico}},
  \bibinfo{author}{\bibfnamefont{M.}~\bibnamefont{Riembau}}, \bibnamefont{and}
  \bibinfo{author}{\bibfnamefont{T.}~\bibnamefont{Vantalon}}
  (\bibinfo{year}{2017}), \eprint{1704.01953}.

\bibitem[{\citenamefont{Baer et~al.}(2013)\citenamefont{Baer, Barklow, Fujii,
  Gao, Hoang, Kanemura, List, Logan, Nomerotski, Perelstein
  et~al.}}]{Baer:2013cma}
\bibinfo{author}{\bibfnamefont{H.}~\bibnamefont{Baer}},
  \bibinfo{author}{\bibfnamefont{T.}~\bibnamefont{Barklow}},
  \bibinfo{author}{\bibfnamefont{K.}~\bibnamefont{Fujii}},
  \bibinfo{author}{\bibfnamefont{Y.}~\bibnamefont{Gao}},
  \bibinfo{author}{\bibfnamefont{A.}~\bibnamefont{Hoang}},
  \bibinfo{author}{\bibfnamefont{S.}~\bibnamefont{Kanemura}},
  \bibinfo{author}{\bibfnamefont{J.}~\bibnamefont{List}},
  \bibinfo{author}{\bibfnamefont{H.~E.} \bibnamefont{Logan}},
  \bibinfo{author}{\bibfnamefont{A.}~\bibnamefont{Nomerotski}},
  \bibinfo{author}{\bibfnamefont{M.}~\bibnamefont{Perelstein}},
  \bibnamefont{et~al.} (\bibinfo{year}{2013}), \eprint{1306.6352}.

\bibitem[{\citenamefont{Spannowsky and Tamarit}(2017)}]{Spannowsky:2016ile}
\bibinfo{author}{\bibfnamefont{M.}~\bibnamefont{Spannowsky}} \bibnamefont{and}
  \bibinfo{author}{\bibfnamefont{C.}~\bibnamefont{Tamarit}},
  \bibinfo{journal}{Phys. Rev.} \textbf{\bibinfo{volume}{D95}},
  \bibinfo{pages}{015006} (\bibinfo{year}{2017}), \eprint{1611.05466}.

\end{thebibliography}

\end{document}